\documentclass[aps,rmp,amsmath,amssymb,twocolumn,groupaddress,longbibliography]{revtex4-1}

\usepackage{graphicx}   
\usepackage{color}      
\usepackage{epstopdf}
\usepackage{comment}
\usepackage{bm}
\usepackage{url}

\begin{document}

\title{Yield Stress Materials in Soft Condensed Matter}

\author{Daniel Bonn}
\affiliation{Van der Waals-Zeeman Institute, Institute of Physics,
University of Amsterdam, Science Park 904, 1098 XH Amsterdam, The Netherlands}

\author{Morton M. Denn}
\affiliation{Benjamin Levich Institute for Physico-Chemical
Hydrodynamics, City College of New York, New York, NY 10031, USA}

\author{Ludovic Berthier}
\affiliation{Laboratoire Charles Coulomb, UMR 5221,
CNRS and Universit\'e Montpellier, Montpellier, France}

\author{Thibaut Divoux}
\affiliation{Centre de Recherche Paul Pascal, CNRS UPR 8641 - 115 avenue Schweitzer, 33600 Pessac, France}
\affiliation{MultiScale Material Science for Energy and Environment, UMI 3466, CNRS-MIT, 77 Massachusetts Avenue, Cambridge, Massachusetts 02139, USA}

\author{S\'ebastien Manneville}
\affiliation{Univ Lyon, Ens de Lyon, Univ Claude Bernard, CNRS, Laboratoire de
Physique, F-69342 Lyon, France}

\begin{abstract}
We present a comprehensive review of the physical behavior of yield stress materials in soft condensed matter, which encompass a broad range of materials from colloidal assemblies and gels to emulsions and non-Brownian suspensions. All these disordered materials display a nonlinear flow behavior in response to external mechanical forces, due to the existence of a finite force threshold for flow to occur: the yield stress. We discuss both the physical origin and rheological consequences associated with this nonlinear behavior, and give an overview of experimental techniques available to measure the yield stress. We discuss recent progress concerning a microscopic theoretical description of the flow dynamics of yield stress materials, emphasizing in particular the role played by relaxation time scales, the interplay between shear flow and aging behavior, the existence of inhomogeneous shear flows and shear bands, wall slip, and non-local effects in confined geometries.
\end{abstract}

\maketitle

\tableofcontents

%
%
%
%

\section*{A short introduction to yield stress materials}

\begin{figure}
	\centering
		\includegraphics{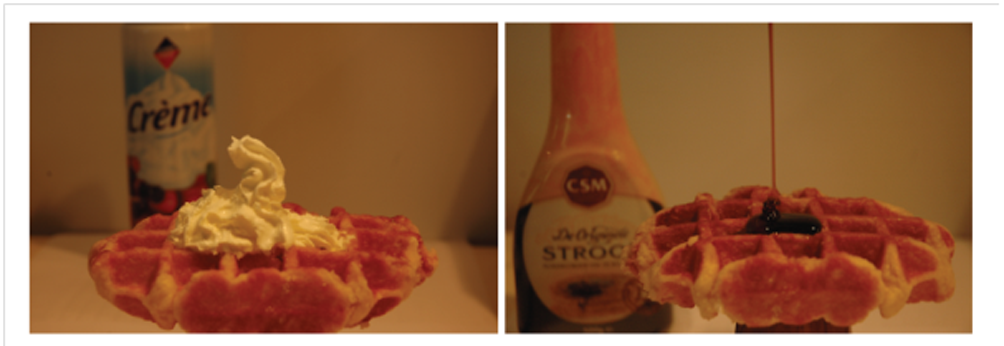}
	\caption{Which fluid is more viscous: whipped cream or thick maple syrup? Slowly stirring both materials with a spoon suggests that syrup is more viscous, while observing the flattening of piles of each material with time suggests the opposite. In fact, the question is ill-posed. The flow properties of whipped cream cannot be reduced to a single viscosity value because it is a yield stress material, whereas syrup is simply a very viscous fluid.}
		\label{fig:Ch1Fig1}
\end{figure}

Many of the materials that we encounter in our daily life are neither perfectly elastic solids nor simple Newtonian fluids, and attempts to describe these materials as being either fluid or solid often fail. Take, for instance, whipped cream and thick syrup. When moving a spoon through these two materials, one would conclude that syrup is the more viscous fluid. However, when left at rest, the syrup will readily flatten and become horizontal under the force of gravity, while whipped cream will retain its shape for a long time, suggesting that, actually, the whipped cream is more viscous than syrup (Fig.~\ref{fig:Ch1Fig1}). This paradox stems from the fact that the syrup is a Newtonian fluid, whereas whipped cream is not a simple fluid at all, and its flow properties cannot be reduced to a single number such as its viscosity. Whipped cream does not flow if the imposed stress is below a threshold value and flows rather easily after this value is exceeded. This threshold rheology is the defining feature of {\it yield stress materials}. Classical, everyday examples of yield stress materials include paints, foams, wet cement, cleansing creams, mayonnaise, and tooth paste.

Besides pharmaceutical and cosmetic applications, yield stress materials are also used in the oil industry, where estimating the minimum pressure required to restart a gelled crude-oil pipeline is crucial \cite{Chang:1999}. The yield stress is also relevant to the concrete and dairy product industries, where its value is related to the size of air bubbles that may remain trapped in the material and directly affect its properties \cite{Aken:2001,Kogan:2013}. In all these fields, it is of paramount importance to characterize as quantitatively as possible the force threshold needed to make the material flow, i.e., {\it the yield stress}.

We review recent progress concerning the fundamental understanding of the yield stress as well as the physical processes relevant to experimental studies of the yielding transition in a broad range of materials across soft condensed matter. The existence of a threshold for flow suggests that these materials respond in a highly nonlinear manner, which has a dramatic impact on their dynamical properties under flow, which we also discuss extensively. Yield stress phenomena are of key importance both from a fundamental point of view and for practical situations involving amorphous solids, spanning a wide range of materials and spanning the fields of hard and soft condensed matter physics.

There are a number of excellent topical reviews available dealing with specific aspects of yield stress materials \cite{Coussot:2005,Moller:2006,Denn:2011,Mansard:2012,NeilJ.Balmforth2014,Coussot:2014}. 
In addition, very recently a collection of relevant  papers appeared 
in a special issue celebrating the anniversary of the first paper by 
Bingham describing yield stress 
fluids~\cite{Coussot2017,Coussot2017b,Malkin2017,Dinkgreve2017,Ewoldt2017,Saramito2017,Mitsoulis2017,Frigaard2017,Cloitre2017}.
The present review attempts to give a concise overview of the physics of yield stress materials taking a very broad perspective encompassing fundamental, experimental and practical issues, along with a discussion of some important open questions.

The review is organised as follows. In Sect.~\ref{overview}, we give a general overview of the various physical concepts and issues raised by the existence of a yield stress, with emphasis on model systems and theoretical approaches. In
Sect.~\ref{yieldstressdetermination}, we provide a critical review of the experimental issues that arise due to the yield stress and of their physical causes, with emphasis on the ubiquitous phenomenon of apparent slippage of yield stress materials at the walls. Sect.~\ref{flowdyn} is devoted to the most recent developments and emerging topics regarding flow dynamics of yield stress fluids, including time-dependence and shear banding, as well as the effects of confinement and transient fluidization behaviors. We close our review with a brief summary in Sect.~\ref{Conclusion}.

\section{General concepts about yield stress fluids}
\label{overview}

\subsection{Popular rheological models for yield stress materials}

\label{sec:popular}

Quantifying the steady-state flow properties of a non-Newtonian fluid requires the measurement of its full {\it flow curve} as the shear viscosity is not a unique number. For a simple shear geometry, the flow curve is a representation of the dependence of the shear stress, $\sigma$, on the shear rate, $\dot{\gamma}$. For a Newtonian fluid, these functions are linearly related, $\sigma = \eta \dot{\gamma}$, where $\eta$ is a constant viscosity. For a yield stress material, the viscosity becomes formally a function of the shear rate, $\sigma = \eta(\dot{\gamma}) \dot{\gamma}$, and the flow curve $\sigma = \sigma(\dot{\gamma})$ is not a simple straight line crossing the origin. As discussed below, for a number of materials, the viscosity also becomes a function of the entire measurement procedure, resulting in a complex time dependence which can make practical measurements challenging.

The most elementary model capturing the existence of a finite yield stress is the Bingham model~\cite{bingham:1922}:
\begin{eqnarray}
	\sigma & < & \sigma_y \Rightarrow \dot{\gamma} = 0, \\
	\sigma & \geq  & \sigma_y \Rightarrow \sigma = \sigma_y + \eta_p\dot{\gamma},
\end{eqnarray}
where $\sigma_y>0$ is the yield stress, and $\eta_p$ a model parameter describing the slope of the flow curve in the fluid region, which is defined by $\dot{\gamma} > 0$. The Bingham model is equivalently described by an effective viscosity which is asymptotically equal to $\eta_p$ at large stresses, and diverges continuously as the stress decreases towards the yield stress: $\eta_{\rm eff}(\dot{\gamma}) \equiv \sigma / \dot{\gamma} = \eta_p + \sigma_y/\dot{\gamma}$. Its simplicity stems from the fact that it uses only a single material-dependent number, the yield stress $\sigma_y$, to describe complex, nonlinear behavior incorporating a threshold force.

\begin{figure}
	\centering
		\includegraphics{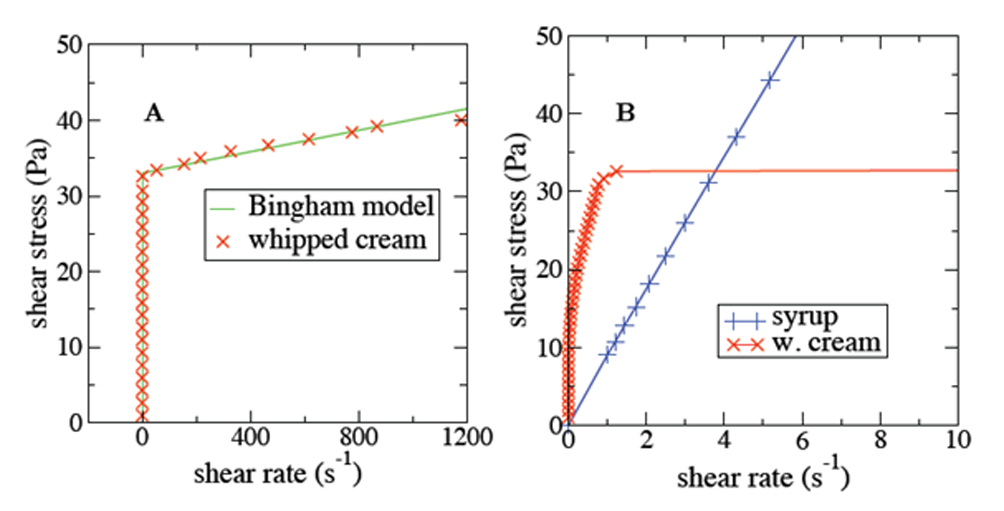}
	\caption{(a) The Bingham model (solid line) provides a reasonable fit to the experimental flow curve of whipped cream (crosses), with yield stress $\sigma_y \approx 33$~Pa. (b) The flow curves of syrup (plusses) and whipped cream (crosses) at low shear rates. Data points are connected by lines as guides to the eye. For stresses above 33~Pa, whipped cream flows more easily than syrup, while the opposite is true below 33~Pa. Using this enlarged scale, one can see that the flow curve of whipped cream below the yield stress is in fact not well described by Eq.~(1) of the Bingham model, which simply predicts zero shear rate all the way up to $\sigma_y$.}
		\label{fig:Ch1Fig2}
\end{figure}

In Fig.~\ref{fig:Ch1Fig2} we return to the examples of whipped cream and syrup, showing flow curves for both materials. Fig.~\ref{fig:Ch1Fig2}(a) illustrates that the Bingham model gives a reasonable description of the measured flow curve of whipped cream with a yield stress of about $\sigma_y \approx 33$~Pa, while Fig. \ref{fig:Ch1Fig2}(b) shows that it makes little sense to compare the viscosity of these two ``fluids'' as, in fact, only one of them is really a fluid with a constant viscosity.

Fig.~\ref{fig:Ch1Fig2} also illustrates that, whereas the Bingham model appears to be an excellent fit to the flow curve of whipped cream [Fig.~\ref{fig:Ch1Fig2}(a)], it actually fails at low shear rates, which becomes obvious once the resolution is improved [Fig. \ref{fig:Ch1Fig2}(b)]. From the latter plot, one would conclude that the yield stress is about $\sigma_y \approx 10 \, {\rm Pa}$, rather than 33~Pa. This highlights one of the many practical problems encountered when working with complex fluids: before a question about the flow properties of a complex material can be satisfactorily answered, one needs to carefully consider the exact experimental protocol as well as the range and resolution of shear rates/stresses over which the data are analyzed.

Two popular generalizations of the Bingham fluid model in shear flow are the Herschel-Bulkley \cite{Herschel:1926} and Casson equations, given respectively as
\begin{eqnarray}
	 {\rm Herschel-Bulkley:} \quad 		\sigma
& = & \sigma_y + K\dot{\gamma}^{n}, \quad		\sigma \geq \sigma_y,
\label{eq:HB} \\
	{\rm Casson:} 	\quad 						\sigma^{1/2}  & = & \sigma_{y}^{1/2} + (\eta_{p} \dot{\gamma})^{1/2},	\quad	\sigma \geq \sigma_y, \, \quad \label{eq:Casson}
\end{eqnarray}
where $K$ and $n$ are additional parameters. Obviously, the Bingham model is a specific instance of the Herschel-Bulkley equation, obtained by imposing $n=1$.

The Herschel-Bulkley model is very popular as it offers more flexibility for fitting experimental data than the Bingham model. It describes both the yield stress regime, $\sigma \approx \sigma_y$, at low shear rate, and a power-law shear-thinning behavior, $\sigma \approx K \dot{\gamma}^n$, with $n<1$ for larger shear rates.
Across a large variety of systems, the shear thinning exponent $n$ is found to have a value in the range $n = 0.2-0.8$, rather than the $n=1$ value imposed in the Bingham model. Frequently, $n$ changes very little with either the density or the temperature of the material \cite{Vinogradov:1978}, so that it appears to be a relevant ``material parameter'', for instance for microgels \cite{Roberts:2001,Oppong:2006,Gutowski:2012,Nordstrom:2010}, emulsions \cite{Mason:1996a,Becu:2006}, and foams \cite{Princen:1989,Hohler:2005,Pratt:2003,Gilbreth:2006}. As such, its determination (or prediction) has become a question of theoretical interest as well, as discussed below.

The crossover between the yield stress and the shear thinning regimes in the Herschel-Bulkley model occurs for a typical shear rate $\dot{\gamma}^\star \approx (\sigma_y/K)^{1/n}$. It is therefore tempting to interpret the corresponding time scale $1/\dot{\gamma}^\star$ as a relevant microscopic time scale for the material \cite{Bonnecaze:2010}. The Herschel-Bulkley equation also predicts the existence of a diverging time scale $\tau$ governing the relaxation to steady state in stress-controlled experiments in the vicinity of the yield point, i.e. for $\sigma \gtrsim \sigma_y$, since one gets: $\tau \sim \dot{\gamma}^{-1} \approx [K / (\sigma - \sigma_y)]^{1/n}$, which readily suggests an interpretation of the yielding transition observed in steady-state simple shear flows in terms of a critical point~\cite{Divoux:2012,Chaudhuri:2013}. This topic will be discussed extensively in Sect.~\ref{hottopics}.

\subsection{Physical origin of the yield stress in soft materials}
\label{physicalorigin}

To elucidate the physical origin of yield stress rheology in a given material, ideally one would like to know under what conditions the material exhibits a yield stress, what microscopic mechanisms are responsible for the emergence of a yield stress, and whether general rules can be formulated to predict the actual value of the yield stress for instance as a function of the composition and structural organization (constituents, interactions) of the material. The emergence of a finite yield stress is frequently referred to as a ``jamming transition'' \cite{Liu:2001,vanHecke:2010}: a broad range of dense amorphous materials (from foams and grains to dense liquids) share the important similarity that they do not flow unless a large enough shear stress is applied.
This idea was popularised via a schematic
jamming phase diagram in an
influential paper by Liu and Nagel \cite{Liu:1998}.

However, the existence of a similar type of transition between fluid and amorphous solid states does not imply that a single physical mechanism should be at work: soft condensed materials may become solid by crossing a variety of phase transitions, and the ``jamming transition'' is now
understood as being only one of them~\cite{Liu:2010,vanHecke:2010}.
In the following, we describe three important classes of yield stress materials whose solid behavior originates from qualitatively different types of phase transitions (or sharp dynamical crossovers) which are usually described by different types of theoretical approaches as well.

\subsubsection{Simple colloidal systems: soft glassy materials}

\label{glassymaterials}

Suspensions of nearly-hard-sphere colloidal particles are among the most studied experimental systems in soft condensed matter \cite{Pusey:1986,Hunter:2012}, as they represent good model systems to study a large variety of physical phenomena also occurring in atomic and molecular systems, from first-order crystallisation to glassy dynamics \cite{Royall:2013}. For colloidal particles, thermal fluctuations and Brownian motion play key roles since they ensure that the system can reach thermal equilibrium. However, when the volume fraction $\phi$ of colloidal hard spheres is increased, the system undergoes a colloidal glass transition that shares important similarities with the glass transition observed upon decreasing the temperature in molecular supercooled liquids \cite{Pusey:1987}. Experimentally, above a ``glass transition'' packing fraction of about $\phi_G \approx 0.58-0.60$ (in three-dimensional suspensions), the equilibrium relaxation time of the colloidal suspension becomes so large that the particles do not significantly diffuse over a typical experimental time scale and the system is effectively dynamically arrested \cite{Brambilla:2009}. At packing fractions above $\phi_G$, colloidal particles simply perform localized back-and-forth ``vibrational'' motion inside the cage formed by their neighbors. This empirical definition of the glass transition density demonstrates that its actual location is not very well-defined experimentally, in the sense that deciding whether a material is ``solid'' or simply ``very viscous'' depends on the observation time scale or the explored range of shear rates in steady-state flow curves.

\begin{figure}
\centering
\includegraphics[width=8.5cm]{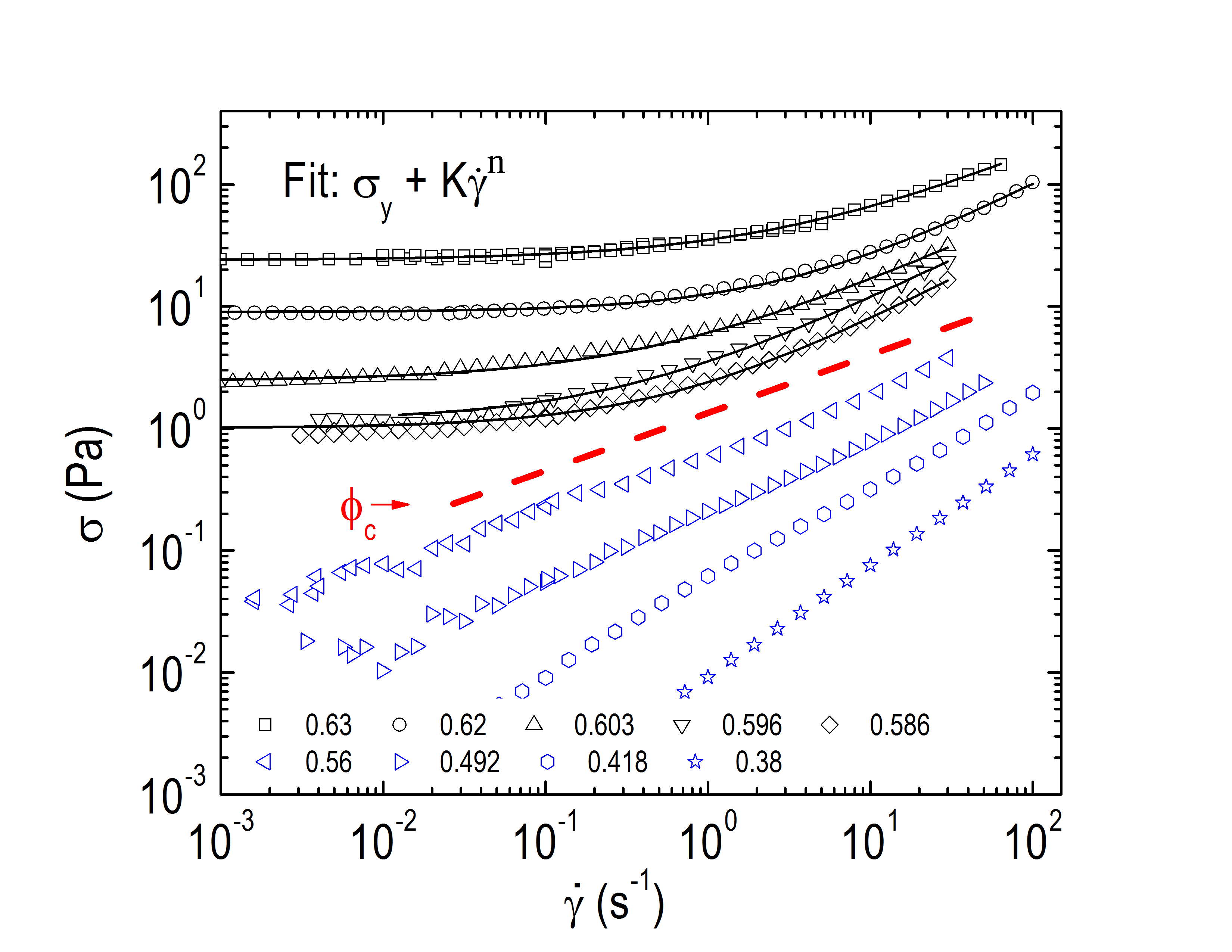}
\includegraphics[width=8.5cm]{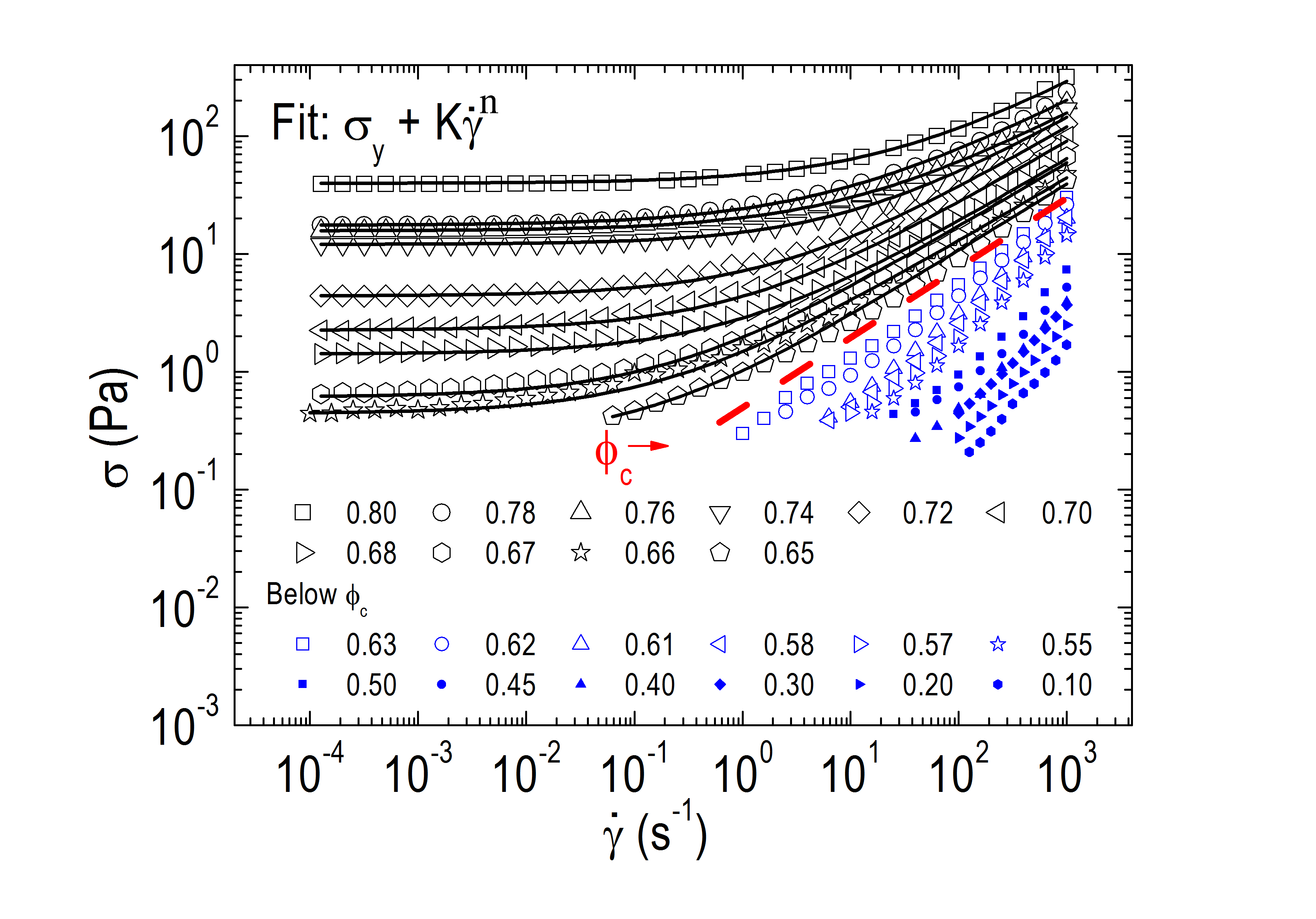}
\caption{(a) Soft glassy rheology. Evolution of the flow curves across the thermal colloidal glass transition for a suspension of PMMA hard spheres of size $a \approx 200$~nm. Extracted from \cite{Petekidis:2004}. 
(b) Jamming rheology. Evolution of the flow curves for an oil-in-water emulsion with droplet size $a \approx 3.2$~$\mu$m
across the athermal jamming transition. Extracted from \cite{Paredes:2013b}. 
In both cases, a yield stress appears above a certain critical 
density $\phi_c$ (marked with a dashed line), which corresponds 
to the glass transition $\phi_G$ for thermal systems in (a),
and to the jamming transition $\phi_J$ for athermal particles in (b). 
Although the emergence of solid behavior in both cases is conceptually 
very different, the flow curves of both materials are surprisingly similar.}
\label{fig:Ch1Fig1Ludo}
\end{figure}

The rheological consequences of the glass transition are readily observed in the flow curves shown in Fig.~\ref{fig:Ch1Fig1Ludo}(a) \cite{Petekidis:2004}. An extended Newtonian regime is observed for $\phi < \phi_G$, which defines a density-dependent viscosity, $\eta(\phi)$ that is seen to increase very rapidly as the density increases towards $\phi_G$. A finite yield stress $\sigma_y$ emerges as the glass transition is crossed for $\phi > \phi_G$, which increases as the colloidal glass concentration is increased further. In the vicinity of the glass transition, $\phi \approx \phi_G$, a shear-thinning regime is observed, where $\sigma \simeq \dot{\gamma}^n$ with $n<1$, illustrating the general fact that the rheology of glassy suspensions occurs mostly out of the linear regime. In fact, accurate measurements of the linear viscosity in hard-sphere suspensions are scarce, and often limited to a modest dynamic regime \cite{Cheng:2002}, precisely because it is challenging to access the linear rheological regime. In the glass phase, the flow curves are typically well described by the Herschel-Bulkley law, which efficiently incorporates both the yield stress and shear-thinning behaviors in a single empirical model.

Similar flow curves are observed in many systems undergoing a glass transition, from dense molecular liquids \cite{Berthier:2002} to colloidal suspensions with soft and hard repulsion between the particles \cite{Petekidis:2004,Nordstrom:2010,Siebenburger:2012b}. In all these systems, a finite yield stress emerges when the shear viscosity becomes so large (upon changing density or temperature) that the system cannot flow anymore on experimentally accessible time scales. Physically, the yield stress results from the fact that particles move too slowly and cannot rearrange the structure fast enough to relax the stress introduced by an external deformation. Therefore, a simple criterion for the emergence of a yield stress is when the time scale for the spontaneous equilibrium relaxation,  usually called ``alpha-relaxation'' time scale, $\tau_\alpha$, becomes larger than the time scale of the external deformation, given by $1/\dot{\gamma}$. In the regime where $\tau_\alpha \dot{\gamma} \gg 1$, spontaneous relaxation cannot occur over the rheologically relevant time window and the system appears solid. Empirically, $\tau_\alpha$ closely follows the behavior of the Newtonian viscosity, $\tau_\alpha \propto \eta(\phi)$, which explains why the linear regime
$\tau_\alpha \dot{\gamma} \ll 1$ becomes very difficult to study near the glass transition where the viscosity increases dramatically.

In such glassy materials, the yield stress is typically a function of temperature and density. This dependence simplifies considerably for the hard-sphere model, because the hard-sphere potential contains no energy scale. In that case, the relevant stress scale controlling solidity is $\sigma_T = k_B T / a^3$, where $k_B$ is Boltzmann's constant, $T$ the temperature and $a$ the particle diameter, so that the yield stress can be rewritten as $\sigma_y = \sigma_T f(\phi)$, where $f(\phi < \phi_G) = 0$. This behavior emphasizes the entropic origin of the solidity in colloidal hard spheres, and therefore the crucial role played by thermal fluctuations in the emergence of a yield stress in colloidal particles with purely repulsive interactions~\cite{Petekidis:2004,Ikeda:2012}.

Finally, when the colloidal glass is compressed far above the glass transition, the interparticle distance decreases and particles eventually come into near-contact as the ``random close packing'' packing fraction
is approached~\cite{Bernal:1960}. For rheology, this critical packing
fraction is more commonly called the ``jamming'' density~\cite{Liu:2010}.
As a consequence, the colloidal glass becomes stiffer when density increases. For pure hard spheres, this results in a strong increase of the yield stress, which appears to diverge as a power law, $\sigma_y \sim \sigma_T (\phi_J - \phi)^{-\gamma}$, with an exponent $\gamma \approx 1$ and where the jamming density $\phi_J > \phi_G$. This functional form shows that the yield stress vanishes for fully non-Brownian suspensions of hard particles due to the entropic prefactor $\sigma_T = k_B T / a^3$ which vanishes when the particle size becomes macroscopic,
$a \to \infty$. Therefore, suspensions of non-Brownian hard particles
such as granular particles do not belong to the family of
yield stress materials.  The density dependence of mechanical properties
is much smoother for particles with non-hard sphere interactions, such as
soft repulsive particles \cite{Koumakis:2012b,vaart:2013},
for which the concept of a sharp jamming transition cannot be
defined in the presence of thermal fluctuations~\cite{Ikeda:2013b}.

\subsubsection{Non-Brownian suspensions: Jammed materials}

\label{sec:jammed}

When the typical size $a$ of colloidal particles increases, Brownian motion becomes negligible and thermal fluctuations are less relevant. This is because the typical time scale for a Brownian particle to diffuse over a distance comparable to its own size scales as $a^2/D_0$, where $D_0$ is the single-particle diffusion constant. For an observation time scale of the order of 1 second, the crossover typically occurs for a particle diameter of about $a \approx 1~\mu{\rm m}$.

Non-Brownian suspensions of soft particles, such as foams and large emulsion droplets, become solid when the density is increased above a critical packing fraction, which corresponds to a genuine ``jamming transition''; in this case, glassy dynamics are not observed. For soft, repulsive spherical particles in three dimensions, the transition takes place near the random close-packing density, $\phi_J \approx 0.64-0.66$. Apart from experimental difficulties, an important source of the uncertainty concerning the jamming density is size polydispersity. It is empirically found that $\phi_J$ increases systematically with the size polydispersity of the sample~\cite{Torquato:2010,Hermies:2009}.

In contrast with the glass transition, thermal fluctuations play strictly no role in this process, and the emergence of solid behavior can be obtained in model systems directly at $T=0$. If the packing fraction is large enough, non-Brownian particles come into contact, and possibly deform, therefore supporting local stresses. The key concept for jamming is the existence of a sufficiently large number of contacts between the particles such that mechanical equilibrium can be maintained throughout the sample~\cite{vanHecke:2010}, but the detailed nature of this geometrical transition is different from a simple percolation transition.

A second major difference with the glass transition is that the jamming transition can in principle be defined and located with arbitrary precision, as its definition does not rely on an observation time scale, although, of course, additional experimental difficulties might intervene~\cite{vanHecke:2010,Ikeda:2013b}. The reason is that the emergence of solidity does not result from the competition between an equilibrium relaxation time scale defined at rest and a finite shear rate, as for glasses, because non-Brownian suspensions have no spontaneous dynamics at rest. The jamming transition and existence of a yield stress in soft materials can therefore be described as ``static'' transitions resulting from a sharp qualitative change in the microstructural properties of the material~\cite{Parisi:2010}.

Another important consequence of the absence of thermal fluctuations is the fact that the jamming transition, unlike the glass transition, necessarily takes place far from thermal equilibrium. In particular, this implies that the preparation protocol of the non-Brownian packings in the vicinity of the jamming transition becomes a relevant parameter controlling the location of the transition~\cite{Donev:2004,Berthier:2009b} but, quite importantly, not its physical nature and properties~\cite{Chaudhuri:2010}.

In experiments and model systems studied in computer simulations, it is found that the yield stress emerges continuously with increasing packing fraction past the jamming transition~\cite{Durian:1995}. This has been reported for foams and emulsions, which are well-described (at least near the transition) by simple models of soft repulsive spheres interacting via truncated harmonic or Hertzian potentials of the form
\begin{equation}
V(r<a) = \frac{\epsilon}{\alpha} (1-r/a)^\alpha,
\label{eq:potential}
\end{equation}
where $\epsilon$ is an an energy scale governing the mechanical property (essentially, the softness) of the particles; the potential is zero when particles
are not in contact, $V(r>a)=0$.
In that case, the relevant stress scale controlling the behavior of the yield stress is of energetic (rather than entropic) nature: $\sigma_0 = \epsilon / a^3$. As a result, the yield stress can now be written $\sigma_y = \sigma_0 g(\phi)$, where $g(\phi < \phi_J)=0$.  A robust finding for the behavior of the yield stress above the jamming transitition is a power-law behaviour: $\sigma_y = \sigma_0 (\phi-\phi_J)^\Delta$ for $\phi \geq \phi_J$~\cite{Durian:1995,Olsson:2007}. The exponent $\Delta$ can be seen as a critical exponent characterizing the rheology of jammed materials; we will now discuss whether such scalings can be retrieved in experiments, and how universal these would be.

Emulsions are systems for which the packing fraction can be changed relatively easily, without changing other system parameters much. Probably the first systematic study of flow curves across a range of volume fractions was performed by Mason {\it et al.} \cite{Mason:1996a}, using a droplet size which is however not quite large enough for thermal fluctuations to be fully irrelevant. Fig.~\ref{fig:Ch1Fig1Ludo}(b) shows similar data taken over a broader range of parameters and larger droplets so that thermal effects are fully irrelevant \cite{Paredes:2013b}. The similarity with the soft glassy rheology in Fig.~\ref{fig:Ch1Fig1Ludo}(a) is striking, as the material crosses over from a Newtonian fluid at low enough density and shear rate to a yield stress solid above jamming, where the flow curves are again well-described by the Herschel-Bulkley model with a shear-thinning exponent $n<1$. Exactly at the jamming density, a power-law shear-thinning behavior is observed. A detailed discussion of the exponent appearing in the Herschel-Bulkley law can be found in \cite{Olsson:2012}. A careful determination of $n$ requires a power-law fit of $(\sigma - \sigma_y)$ as a function of $\dot{\gamma}$. It is found in simple numerical models that such a plot actually displays two distinct power-law regimes with two different exponents: $n$ at small shear rates, and $n'$ at larger shear rates~\cite{Olsson:2012,Olsson:2011,Lerner:2012,Kawasaki:2014}. It is likely that the fitting of experimental flow curves is dominated by the second of these two exponents, and comparison to theory is thus somewhat delicate. Additionally, at finite shear rates, it is also possible that other ingredients, such as friction between particles~\cite{Bonnecaze:2010,Katgert:2008,Katgert:2009} or energy dissipation of the interstitial liquid (or in Plateau borders for foams~\cite{Schwartz:1987}) start to play a significant role and also affect the value of the shear-thinning exponent.

It is interesting to consider the limit of infinitely hard non-Brownian ({i.e., granular}) suspensions, which are also often described as possessing a yield stress. As should be clear from the above discussion, a yield stress can only exist in non-Brownian repulsive objects {if} they can be compressed strictly above the jamming density, $\phi> \phi_J$. This is by definition not possible when particles are truly hard, such as in granular suspensions that only exist in the fluid state, $\phi < \phi_J$ \cite{Andreotti:2013}. Careful measurements on suspensions of spherical particles \cite{Fall:2009, Dijksman:2013} have indeed revealed that if the particles and suspending liquid are carefully density-matched, there is no yield stress up to random close packing, where all the particles start to touch each other. However, as soon as there is the slightest density mismatch, the particles cream or sediment, so that $\phi \to \phi_J$, which is indeed the only density where hard particles can be fully arrested. This makes non-density-matched suspensions similar to dry granular systems: a sand pile has a clear, finite angle of repose, which is equivalent to stating that it has a yield stress \cite{Behringer}. As for sedimenting suspensions, this is due to the gravitational forces that push the grains together and in this way activate the frictional contacts between the grains \cite{Behringer}.

A direct consequence of the hard-particle limit is then that no time scale can be constructed from the particle interaction, and the flow curves obtained at constant density therefore simplify considerably and become fully Newtonian.
Very different from the flow curves shown in Fig.~\ref{fig:Ch1Fig1Ludo}, the rheology of this regime is described by simpler constitutive laws (with no yield stress) that have been very carefully studied and validated by many experiments in the community of granular media~\cite{Andreotti:2013,GDR:2004}.

\begin{figure}
\centering
\includegraphics[width=8.5cm]{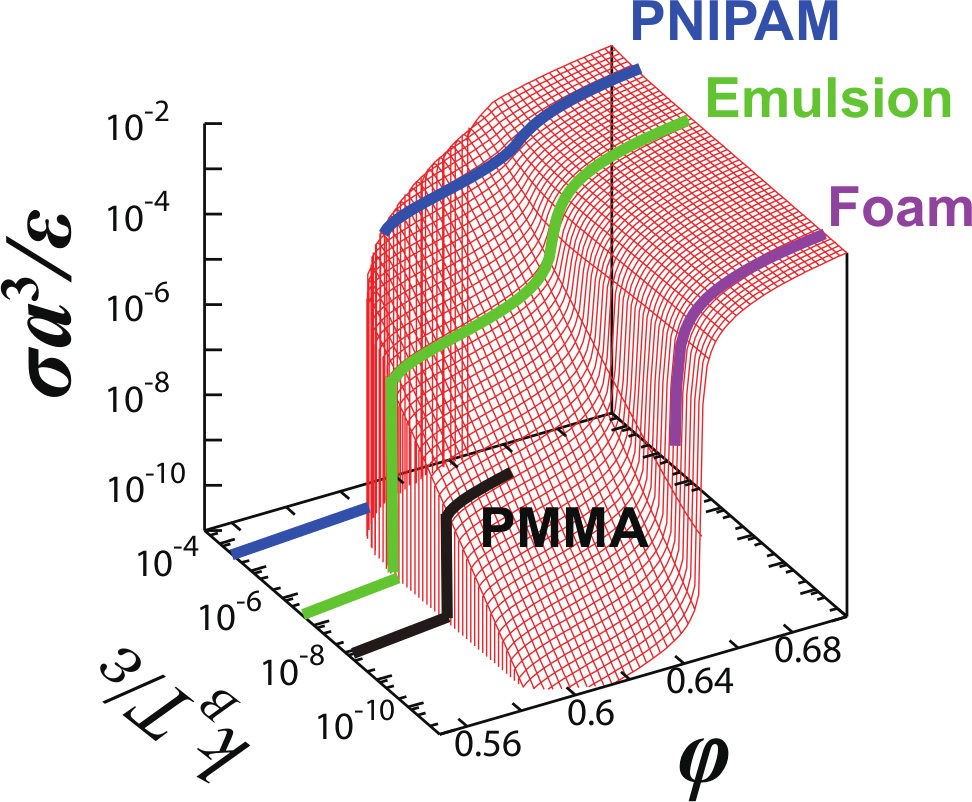}
\caption{Three-dimensional ``jamming phase diagram'' showing the reconstructed yield stress surface from numerical simulations as a function of the thermodynamic parameters temperatures and density in a dimensionless representation (particle softness $k_BT/\epsilon$, volume fraction $\varphi$, and stress $\sigma a^3/\epsilon$) for a model of soft harmonic particles~\cite{Ikeda:2013b}.
The thick lines represent the location of typical experimental measurements in various materials:
foams (rightmost line) are mainly sensitive to jamming physics; PMMA hard spheres (black line) to glass physics; emulsions display an interesting interplay between glass and jamming transitions; PNIPAM microgels (leftmost line) undergo a colloidal glass transition far from the jamming limit with no 
particular signature across the jamming density.}
\label{fig:Ch1Fig2Ludo}
\end{figure}

In Fig.~\ref{fig:Ch1Fig2Ludo} \cite{Ikeda:2013b} we summarize the evolution of yield stress with temperature $T$ and packing fraction $\phi$ for several types of three-dimensional assemblies of harmonic repulsive particles, determined using computer simulations. It demonstrates the emergence of a yield stress in thermalized colloidal assemblies at a packing fraction $\phi_G$, which depends weakly on the particle softness. This softness is quantified by the adimensional temperature scale $k_B T / \epsilon$, which compares thermal energy to particle repulsion.
Colloidal PMMA hard spheres are typically characterized by $k_B T /\epsilon \sim 10^{-8}$, whereas soft microgels are usually much softer, $k_B T /\epsilon \sim 10^{-4}$, emulsions being typically intermediate, $k_B T /\epsilon \sim 10^{-6}$. All these systems display a yield stress above the colloidal glass transition,
and the yield stress increases with density in the glass phase. For hard spheres, it diverges at the jamming transition, for emulsions it shows a strong crossover behaviour, and it has a smooth density dependence for soft microgels. 
In soft systems such as foams, $k_B T / \epsilon \sim 10^{-8}$ is again small because thermal fluctuations become irrelevant for such large particles, and the emergence of the yield stress is associated with the jamming transition, with no influence of thermal fluctuations on the rheology. The jamming transition controls the $T \to 0$ limit of the jamming phase diagram in
Fig.~\ref{fig:Ch1Fig2Ludo}.
Such a diagram is experimentally useful as it allows one to locate
systems such as microgels, emulsions, foams and colloidal hard spheres
on the same graph, and to elucidate the origin of the yield stress
observed in rheological experiments.

\subsubsection{Role of attractive forces: Colloidal gels}

In the previous section, particle systems with solely repulsive forces were described from a theoretical perspective, in which case temperature can only compete  with the particle softness, and the main control parameter is the packing fraction.
The situation becomes more complex when attractive forces come into play. Adhesion and attractive forces are relevant for a large number of model systems and real materials. For instance, dense liquids do not interact via hard-sphere
potentials, but typically also possess longer-range attractive forces, modeled for instance via a Lennard-Jones potential \cite{Hansen:2006}. In colloidal systems, attractive forces can be easily induced and tuned, using for instance colloid-polymer mixtures \cite{Royall:2013}. Many real-material systems, such as clay suspensions or more generally colloidal gels, are Brownian systems with attractive interactions between the colloidal particles \cite{Larson:1999}.

Regarding the glass transition phenomenon in simple systems, attractive forces only weakly affect the physics, in the sense that they contribute quantitatively to the relaxation dynamics and details of the phase diagram, but do not change the physical behavior qualitatively \cite{Berthier:2009,Berthier:2011}.

Attractive forces in simple liquids start to change the physics when they are strong enough to induce a nontrivial
dynamical arrest in a regime that would otherwise be characterized by a simple fluid behavior. The simplest case is when very strong bonds are present, which might result in a percolating particle network that can sustain a finite stress, very much as in chemical gels \cite{Larson:1999}. Here, a yield stress emerges and coincides with a percolation transition. When the gel is dense enough, such a network can confer a macroscopic elasticity to the system and hence be responsible for a yield stress. However, if the thermal energy is sufficient to break and reform bonds within the network, for a small applied stress the system will eventually flow at long time scales, and the system is simply viscoelastic (it is a ``transient'' gel).

Percolation only represents one of the possible routes to the production of physical gels; several other examples have been studied in recent years \cite{Zaccarelli:2007}. Here, we briefly mention three examples.

\paragraph{Nonequilibrium gels.}

A well-described example concerns colloidal gels that are formed by increasing the strength of short-ranged adhesive depletion forces, starting from an initially purely repulsive system. It has been empirically found that ``nonequilibrium gels'' can be formed over a broad range of densities as the adhesion between particles is increased \cite{Manley:2005,Lu:2008,Royall:2008}. These gels are heterogeneous, dynamically arrested structures, which thus behave mechanically as soft solids. The current understanding of the gelation process is that adhesion induces the analog of a liquid-gas phase separation in the colloidal system, which may phase-separate into colloid-rich and colloid-poor phases. However, because the attraction is very short-ranged, the coexistence curve on the colloid-rich region at large density may hit the colloidal glass transition. The emergence of slow, glassy dynamics may be able, in some cases, to slow down dramatically and even fully arrest the kinetics of the phase-separation process \cite{Lu:2008,Foffi:2005,Testard:2011}. At long times, the system may thus acquire a percolating bicontinuous structure which is mechanically rigid and does not flow if a small shear stress is applied. A consensus has been reached regarding the formation of these non-equilibrium gels as reviewed in \cite{Zaccarelli:2007}, whose structure can be controlled by tuning the flow cessation dynamics \cite{Ovarlez:2013,Koumakis:2015,Helal:2016}. However, the steady-state rheology of attractive gels is still a topic of intense research \cite{Zia:2014,Romer:2014,Helgeson:2014,Capellmann:2016}. 
Indeed, colloidal gels show a pronounced time-dependent response \cite{Ovarlez:2008b}, and a strong propensity to wall slip that appears to be non-trivially coupled to spatially heterogeneous flows \cite{ Gibaud:2008,Grenard:2014}, which makes it difficult to measure flow curves, and even questions the very existence of a unique constitutive equation. The transient and steady-state rheology of these systems are discussed in more details in Sec.~\ref{flowdyn}.

\paragraph{Attractive glasses.}

In addition to gels of repulsive colloids, the behavior of dense assemblies of attractive colloids has also attracted a large experimental interest in the recent decade \cite{Sciortino:2005,Puertas:2009}. In this situation, a complex physical interplay is to be expected, due to the competition between the non-equilibrium kinetic arrest arising at moderate densities (leading to nonequilibrium gelation) and the glassy physics emerging at large densities without adhesive interaction (leading to glass formation). Early studies have advocated that this competition produces a novel state of arrested matter, named ``attractive glass'' \cite{Fabbian:1999,Dawson:2000,Pham:2002,Sciortino:2002}, actively studied both numerically and experimentally
\cite{Pham:2002,Pham:2004,Sciortino:2003,Zaccarelli:2005,Puertas:2005}. For a discussion on how to experimentally differentiate a glass from both a gel and an attractive glass, see \cite{Bonn:1999,Tanaka:2004}.
Because the dynamics is controlled by at least two microscopic lengthscales (the adhesion range responsible for initiating phase separation, and the cage size responsible for the glassy dynamic arrest), complex relaxation patterns have been predicted \cite{Fabbian:1999,Dawson:2000} and observed \cite{Pham:2004,Zaccarelli:2005}, including in
rheological studies \cite{Koumakis:2011}. Whereas early interpretation relied on the existence of an underlying peculiar form
of glass singularity predicted by mode-coupling theory \cite{Fabbian:1999,Dawson:2000}, additional work has shown that such singularity is not needed for complex time dependences to occur \cite{Chaudhuri:2010b}. The existence of a genuine attractive glass phase has also been called into question \cite{Zaccarelli:2009,Royall:2015}, and indeed the idea of a specific type of attractive glass
does not seem needed to interpret the rheology of all concentrated
attractive glasses, see e.g. \cite{Datta:2011} for the example of an
attractive emulsion. From a practical (rather than fundamental)
viewpoint, the idea that glasses with different types of frozen-in disorder 
may exist in models with adhesive interections remains 
valuable~\cite{Pham:2004}.

\paragraph{Athermal adhesive systems.}

Finally, the role of attractive forces in non-Brownian suspensions is also relevant but necessarily has a different nature, as the
adhesive forces, by construction, cannot compete with thermal fluctuations. A few studies have explored the emergence of solidity in athermal adhesive particle systems, to understand in particular how the jamming transition is affected when adhesion is present \cite{Lois:2008,Chaudhuri:2012b,Irani:2014}. This point
is relevant for instance in the context of humid granular materials.

In particular, because adhesion creates bonds between particles, it seems physically clear that adhesive forces can only enhance solidity above jamming, and in this dense regime, adhesion acts as a small perturbation.  On the other hand, it appears that solid behavior can be maintained in a density range even below the jamming transition, $\phi < \phi_J$, which opens a novel regime for solid behavior which has no analog for purely repulsive systems. In particular, a recent numerical study suggests that a small amount of attractive force is indeed able to generate a material with a finite yield stress below the jamming transition, with a potentially interesting interplay between the imposed shear flow and the microstructure of the system, eventually giving rise to large-scale flow inhomogeneities \cite{Irani:2014}. This points to a possible mechanism for shear
banding, which is a topic that we discuss further below.

\subsection{Is the yield stress real?}

\subsubsection{A historical debate}

\begin{figure}
	\centering
		\includegraphics[width=8.5cm]{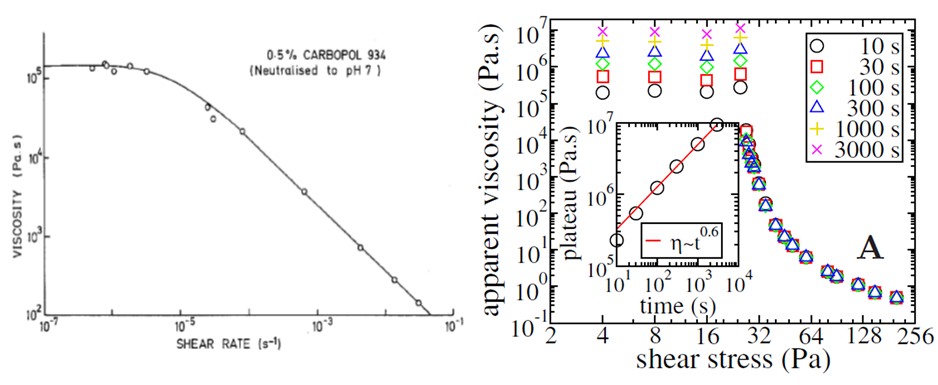}
	\caption{Viscosity vs. shear stress in Carbopol. (a)
From \cite{Barnes:1985}, (b) subsequent study of the same system by \cite{Moller:2009a}. The latter showed that the values of the low-stress viscosity plateau increase with measurement time (from 10 s to 3000 s; colored symbols). The insets show that the plateau value increases as a power law with time, with
exponent $\approx 0.6$, indicating that the measured viscosities do not correspond to steady-state shear flows
for shear stresses below the yield stress. }
\label{fig:Ch1Fig3}
\end{figure}

For many years, there has been a controversy about whether the yield stress marks a transition between a solid and a fluid state, or between two fluid states with drastically different viscosities \cite{Barnes:2007,Barnes:1999,Evans:1992,Hartnett:1989,Schurz:1990,Astarita:1990,Spaans:1995}. Numerous experimental studies argued that yield stress materials actually flow like very viscous Newtonian liquids at low stresses \cite{Barnes:1999,Macosko:1994}. \cite{Barnes:1985} presented data on Carbopol microgels to demonstrate the existence of a finite viscosity at very low shear stresses (Fig.~\ref{fig:Ch1Fig3}), rather than an infinite viscosity below the yield stress, and later published a review with numerous flow curves suggesting that yield stress materials should rather be described as Newtonian fluids with a very large viscosity \cite{Barnes:1999}.

\cite{Moller:2009a} reproduced the experiments used to demonstrate Newtonian limits at low stresses and indeed observed finite apparent viscosities at low stresses, see Fig.~\ref{fig:Ch1Fig3}. However, while all measurements collapse at high stresses, below the yield stress they no longer do, and the apparent viscosity depends in fact on the delay time $t$ between the application of the stress and the viscosity measurement. Each individual curve resembles the curves of Barnes and others, but if all the points that do not seem to correspond to a steady state are removed, one is left with a simple Herschel-Bulkley material with a well-defined yield stress. The large viscosity values obtained at low stress in fact correspond to shear rates of the order of $10^{-6} \, {\rm s}^{-1}$ or less, which are not only reaching the accuracy limits of ordinary rheometers, but also show that within a reasonable experimental measurement time, no steady state is reached where the deformation increases linearly with time, and the total deformation imposed on the sample remains well below unity. Consequently, the instantaneous shear rate cannot be interpreted as representative of a well-defined steady-state viscosity.

Besides this fundamental problem, debates about the existence of a yield stress demonstrate that the most ubiquitous practical problem encountered by scientists and engineers dealing with everyday materials such as food products, powders, cosmetics, crude oils, or concrete is that the yield stress of a given material is very difficult to determine experimentally \cite{Barnes:1999,Moller:2006,Mujumdar:2002}. Indeed, \cite{James:1987,Zhu:2001,Nguyen:2006} demonstrated that a variation of the yield stress of more than one order of magnitude can be obtained, depending on the way it is measured. This cannot be attributed to different resolution powers of different measurement techniques, but hinges on more fundamental complexities resulting from the physical processes
responsible for the flow of yield stress materials.

\subsubsection{Theoretical considerations about the existence of a yield stress}

Can theory and simulations shed light on the debate regarding the existence of a true yield stress in amorphous materials?
This is a difficult question which cannot have a simple generic answer, as it amounts to asking first whether genuine amorphous solid states exist, and second whether such states can support a finite shear stress without flowing over
arbitrarily-long time scales. Moreover, as detailed above, different materials exhibit solid properties for distinct fundamental reasons under various experimental conditions and due to various particle interactions.
Let us disentangle all these issues.

Clearly, the existence of a ``real'' yield stress in materials undergoing a glass transition is at least as ambiguous
as that of a genuine fluid-to-glass phase transition, which remains an open fundamental question. There exist theoretical approaches and simple models which describe the glassy phase of matter as a genuine thermodynamic singularity accompanied by a diverging viscosity. However, there are competing theoretical perspectives based on the opposite idea that the glass region is accessed by a dynamic crossover, and where the equilibrium relaxation time scale does not truly diverge \cite{Berthier:2011b}. Therefore, the existence of glassy phases with truly infinite viscosity is not settled theoretically, or, for that matter, experimentally.

Of course, this fundamental question is not very relevant  {\it in practice},
as glassy phases are experimentally produced by going through a dynamic crossover in a non-equilibrium manner, as explained in Sect.~\ref{glassymaterials}.
As a consequence, in the vicinity of the experimental glass transition, flow curves might display an apparent yield stress value when measured over a given window of shear rates, even though the material might eventually flow at much longer timescales. Deeper in the glassy region, when the relaxation time has become larger than any relevant experimental time scale, the distinction between a slowly flowing fluid and a kinetically arrested material is essentially irrelevant, and the question of the existence of a genuine glassy phase may appear rather academic.

In glassy materials, a system prepared in the glass region slowly ages with time because thermal fluctuations allow for a slow exploration of its complex free-energy landscape \cite{Berthier:2011b}. Importantly, this also implies that the rheological properties of glasses might depend on the time scale used to perform the measurements. For instance, the yield stress of the system has been observed to increase logarithmically with the preparation time in model systems \cite{Varnik:2004}. Additionally, the aging behavior observed in glasses at rest might be affected in a nontrivial manner by an imposed shear flow, possibly resulting in a steady-state situation where aging is prevented by the external shear flow, a situation coined ``shear rejuvenation'' \cite{Cloitre:2000,Viasnoff:2002,Bonn:2002a,Ianni:2007}.
The roles played by the preparation protocol and by the aging dynamics are similarly crucial for colloidal gels that might be formed through nonequilibrium processes, such as kinetically arrested phase separation. In that case, it is unclear how such a nonequibrium competition is affected by an externally imposed shear stress, which could for instance either ``mix'' the material or break the bicontinuous structure and accelerate the phase separation.

Assuming that genuine amorphous {\it phases} exist (where for instance ergodicity is truly broken and the Newtonian viscosity is infinite), is it necessarily obvious that such phases should display a finite yield stress? To answer this question one should ask whether there exists a physical dynamical process allowing the system to relax and flow on a finite time scale after a finite shear stress has been imposed. This problem was addressed in \cite{Sausset:2010}. Using a simple nucleation-type argument, a stress-dependent free-energy barrier for relaxation was constructed, which could then be crossed using thermal fluctuations. By connecting the constructed activation time scale to the imposed stress, a limiting flow curve $\sigma(\dot{\gamma})$ was obtained, which in three spatial dimensions is of the form
\begin{equation}
\dot{\gamma} = \frac{\sigma}{G \tau_0}
\exp \left[ - c \left(\frac{\sigma_0}{\sigma} \right)^4 \right],
\label{eq:sausset}
\end{equation}
where $G$ is the elastic shear modulus, $c$ a constant, $\tau_0$ a characteristic relaxation time, and $\sigma_0$ a temperature-dependent stress scale.
Eq.~(\ref{eq:sausset}) implies the existence of a lower bound for the resulting shear rate for a finite shear stress $\sigma > 0$, which in turn suggests that the {\it shear rate should actually be finite at any imposed shear stress} even in the ``solid'' phase. This result is not inconsistent with the existence of measured flow curves with an apparent yield stress, as it predicts that the shear stress decreases logarithmically (very) slowly with the shear rate. It might therefore be very difficult to detect such behavior in an experiment and to discriminate it, for instance, from a Herschel-Bulkley functional form with a finite yield stress (where the yield stress value is approached algebraically with decreasing $\dot{\gamma}$). Interestingly, the result is not specific to amorphous materials, but applies equally to ordered systems such as crystalline materials. This discussion shows that despite the translational symmetry breaking observed during the formation of the crystal, which contrasts with the absence of such a symmetry breaking in amorphous solids, a yield stress is conceptually not better defined in ordered systems. Therefore, the absence of a real yield stress is not due to the ``messy''  nature of soft amorphous materials, but has a more profound origin.

The reasoning leading to Eq.~(\ref{eq:sausset}) and the conclusion that a finite yield stress cannot exist even in dynamically arrested phases rely heavily on a barrier-crossing argument, and therefore on the presence of thermal fluctuations. Therefore, the situation might be different in non-Brownian suspensions undergoing a jamming transition at zero temperature. For soft jammed particle systems such as foams and large-droplet emulsions, the transition to the jammed phase is not a dynamic crossover
and the ``solidity'' is thus not destroyed by (non-existing) thermal fluctuations. In this case, there is \textit{a priori} no deep theoretical argument against the existence of a finite yield stress, so that the flow curves shown in Fig.~\ref{fig:Ch1Fig1Ludo}(b) might be true examples of genuine yield stress materials. Of course, as mentioned several times above, these experimental results do not seem to differ dramatically from measurements performed in thermal materials, which suggests that the experimental debate regarding the existence of yield stress presumably revolves around very practical issues, with little connection to the present discussion putting forward more fundamental arguments.

\subsection{Thixotropy in yield stress fluids}
\label{thixo}

\begin{figure}
	\centering
		\includegraphics{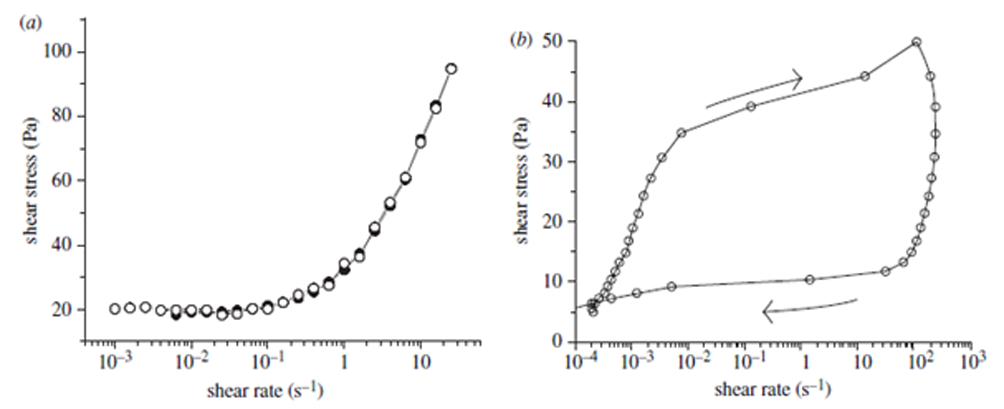}
	\caption{(a) The behavior of 0.1\% wt Carbopol microgel under increasing and decreasing shear stresses shows that this material is non-thixotropic (filled circles, up; open circles, down). (b) Thixotropy of a 10\% wt bentonite solution under an increasing and then decreasing stress ramp.}
		\label{fig:Ch1Fig4}
\end{figure}

Most yield stress fluids have an underlying microstructure that confers a macroscopic elasticity to the system. This microscopic structure may be (partly) destroyed by the flow, causing a reversible decrease of the viscosity with time, in which case the system is said to be \textit{thixotropic} \cite{Mewis:2009}. The yield stress will be different following flow application, with a value that may be dependent on the rest time prior to shearing, during which the structure may also reform. It is therefore useful in practical terms to distinguish between thixotropic and simple (non-thixotropic) yield stress fluids:

\begin{itemize}
	\item
	``Simple'' yield stress fluids: shear stress (and hence the viscosity) depends only on the imposed shear rate. Examples include non-adhesive emulsions, foams and Carbopol
microgels \cite{Bertola:2003,Becu:2006,Moller:2009b,Ovarlez:2013b}.
	\item
	Thixotropic yield stress fluids: yield stress and viscosity depend on the shear history of the sample. Examples include particle and polymer gels \cite{Moller:2008}, attractive glasses \cite{Moller:2009b}, ``soft" colloidal glasses \cite{Bonn:2002b}, adhesive emulsions \cite{Ragouilliaux:2007}, non-Brownian gels \cite{Kurokawa:2015}, pastes \cite{Huang:2005}, and hard-sphere colloidal glasses \cite{Moller:2009a}.
\end{itemize}

The distinction is straightforward, at least in principle: one can measure the flow curve by using up and down stress ramps, for instance, and check for reproducibility (Fig. \ref{fig:Ch1Fig4}). In Fig.~\ref{fig:Ch1Fig4}(b), we show that if the material is thixotropic, in general the flow will have significantly ``liquified" the material at high stresses, and the branch obtained upon decreasing the stress is significantly below the one obtained while increasing the stress.
Hysteresis is mostly negligible for simple yield stress fluids,
see Fig. \ref{fig:Ch1Fig4}(a).
The response of a thixotropic yield stress fluid will depend on the rate at which the stress is ramped up and down, and the rest time in between subsequent sweeps.

\begin{figure}
	\centering
		\includegraphics{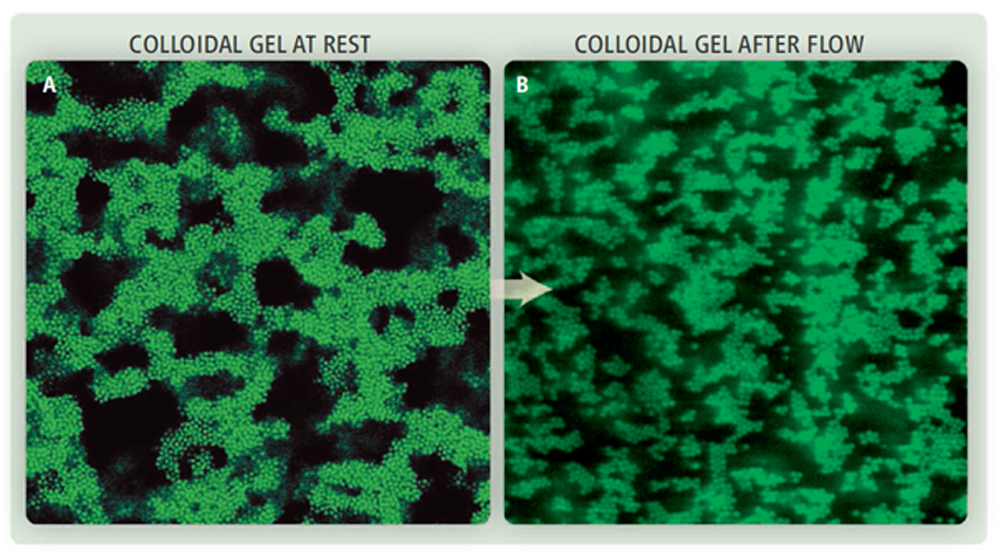}
	\caption{A colloidal gel at rest, with a percolated structure and a yield stress of 5~Pa (A), and just after flow, with individual flocs and no measurable yield stress (B). The gel is made up of 1.3 $\mu$m fluorescent PMMA particles and 3$\cdot10^7$ Mw polystyrene in a mixture of decalin and
cyclohexyl bromide. From \cite{Bonn:2009}.}
		\label{fig:Ch1Fig5}
\end{figure}

Figure \ref{fig:Ch1Fig5} shows a direct qualitative observation of the effect of stress-dependent structural organization in a colloidal gel. At rest (A), the gel exhibits a percolated structure and exhibits a yield stress of about 5 Pa. Just after flow (B), the gel has broken up into individual flocs and there is no measurable yield stress. Detailed images of the shear-induced breakup of two-dimensional colloidal gels at interfaces for different values of the shear rate and strain were shown by \cite{Masschaele:2011}, who quantified the effect of surface coverage and deformation on the morphology (i.e., transient networks or individual deformed aggregates); the undeformed structures in these experiments undoubtedly exhibit a yield stress, but direct mechanical measurements are not available.

The distinction between the two main families of ``simple'' yield stress fluids and thixotropic yield stress fluids is at present mostly driven by empirical considerations. It would be interesting to understand if it can also be rationalised at a more fundamental level. Experimentally, it would be useful to develop model systems allowing both types of behaviors to be observed and controlled, for instance by devising materials that are only weakly thixotropic and where ``simple'' yield stress behavior can be continuously recovered in
some well-controlled limit.

\subsection{Theoretical descriptions of yield stress materials}
\label{theory}

\subsubsection{Why a theory of yield stress solids is difficult}

While properly defining and measuring a yield stress is a debated issue from an experimental point of view, as emphasized throughout this review,
in theoretical work one usually identifies the yield stress as the shear stress measured in steady-state shear flow in the limit where the deformation rate goes to zero:
\begin{equation}
\sigma_y = \lim_{\dot{\gamma} \to 0} \sigma(\dot{\gamma}).
\label{eq.ysdef}
\end{equation}
Thus, the challenge for theoreticians does not lie in the practical definition of the yield stress or its best quantitative
determination, but in the conceptual difficulty to describe the nonlinear mechanical properties of disordered complex solids.

An additional difficulty can be appreciated by comparing the situation of disordered materials to that of crystalline solids.
Crystals are formed through a phase transition across which translational invariance is broken. Because the broken symmetry is easily identified, it is not difficult to recognize the associated defects (such as dislocations) directly from the structure of an imperfect crystalline system. It is well established that nonlinear flow and mechanical deformation in crystalline materials are mostly driven by these defects, so that an understanding of the flow defects of crystals is indeed the key to understanding their rheology. So, we are led to ask what the ``defects'' are in an amorphous material that is formed without breaking any obvious symmetry. Are there at least equivalent localized structures allowing us to efficiently describe flow and mechanical deformation in amorphous solids? These are two long-standing questions in the area of amorphous material rheology, which have received some constructive answers in the last decades, mostly from
numerical and experimental studies \cite{Barrat:2011}.

\begin{figure}
	\centering
		\includegraphics[width=8.5cm]{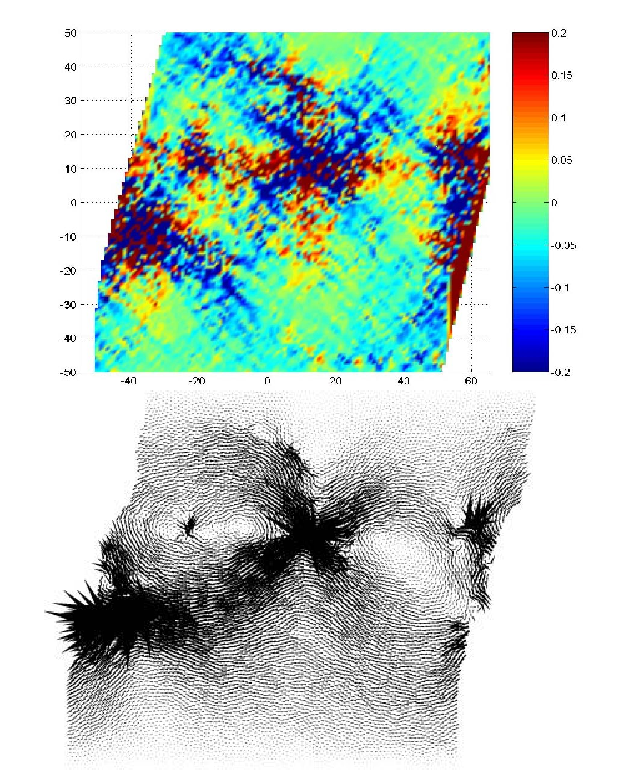}
	\caption{Changes in the local shear stresses (as indicated by
the color coding) during a
localized plastic event (top); the color coding gives the amplitude of the stress changes, and associated displacement field (bottom) observed in the numerical simulation of a quasi-statically sheared model of atomic glass. Adapted from \cite{Tanguy:2006}.}
\label{fig:Tanguy}
\end{figure}

It has been demonstrated in many different studies that flow in amorphous materials occurs at the microscopic scale in very localized ``zones'',
sometimes identified as ``shear transformation zones'' \cite{Falk:1998}. These zones are best observed in studies of amorphous systems which are sheared so slowly that individual events can be resolved in space and time, such as computer simulations in quasi-static shear conditions \cite{Maloney:2006} or confocal microscopy experiments on slowly deformed colloidal glassy systems \cite{Schall:2007}. It has been observed that flow occurs mostly near zones comprising a small number of particles (say, 5 to 10) undergoing the largest irreversible rearrangements. However, because the material is globally an elastic solid, these local plastic events additionally induce a long-range redistribution of the stress field in their
surroundings \cite{Picard:2004}, which, in turn, can couple to a different zone, or trigger further relaxation elsewhere in the system. An example of such an event detected in the numerical simulation of a slowly sheared glass model \cite{Tanguy:2006} is shown in Fig.~\ref{fig:Tanguy}.
Notice however that both computational studies and colloidal experiments
are performed on disordered systems that are prepared in physical conditions
that are vastly different from the ones relevant for molecular and
polymeric glasses (i.e. ``hard'' glasses),
for which these ideas remain to be experimentally validated.

\subsubsection{Theoretical approaches}

The previous section suggests that theory still has trouble describing the
transition between a fluid and an amorphous solid (glass, gel, and jammed
states), and that describing the rheology of these materials requires
in addition a description of a non-linear response to flow, which
is typically accompanied by strong spatial fluctuations
and localized flow defects that may induce long-range correlations,
intermittent relaxations, and even catastrophic responses with
complex time dependencies. It should therefore come as no surprise
that no complete, well-accepted, first-principle theoretical
approach exists that can account for all aspects
of the rheology of yield stress materials.
Instead, several layers of (potentially complementary) theoretical descriptions are found in the literature. In the following, we distinguish two main types of theoretical approaches.

\paragraph{Mode-coupling theories and trap models.}

In a first category of theoretical models, the
focus is primarily on a detailed description of the rheological consequences
of the existence of a fluid-to-amorphous solid phase transition. In particular, numerous theoretical approaches to the description of the glass transition in dense fluids and colloidal systems have now been extended to account for the mechanical properties in the vicinity of the glass transition \cite{Berthier:2011b}, such as for instance mode-coupling theories \cite{Gotze:2008} and Bouchaud's trap model \cite{Bouchaud:1992}. Even in this restricted context, these approaches
differ widely in their approach.

On the one hand, mode-coupling theories were developed as truly microscopic or ``first-principle'' approaches to understand the dynamics of simple liquids near a glass transition~\cite{Bengtzelius:1984,Gotze:2008}.
A large amount of work has been performed to develop tractable equations of motion that can attack complex flow histories while retaining aspects of the driven dynamics of the microscopic degrees of freedom.

On the other hand, trap models correspond to more phenomenological descriptions of the glass phenomenon, and have attracted a lot of attention in particular in the context of aging phenomena inside glassy phases. The rheological trap model is called the ``soft glassy rheology model'' (SGR) and has been studied extensively \cite{Sollich:1998,Sollich:1997}, both in steady-state conditions, in the context of rheological aging, and in even more complicated time-dependent situations, with interesting connections to the physics of thixotropic materials \cite{Fielding:2000}. By introducing spatial dependences, the SGR model has also been studied to give insight into spatially inhomogeneous
flows \cite{Fielding:2009,Moorcroft:2011,Moorcroft:2013}.

\paragraph{Shear-transformation zones and elasto-plastic models. }

A second family of theoretical models actually postulates from the start that a solid amorphous state exists, which is characterized by a finite yield stress. These models are then able to explore in more detail how such a solid system might flow under an applied shear stress larger than the yield stress.

The starting point for these models is the observation
that flow occurs in a spatially inhomogeneous manner, and occurs
mostly at localised shear transformation zones as illustrated in
Fig.~\ref{fig:Tanguy}. This empirically well-established observation
made in different systems suggests a theoretical pathway
to model the mechanical properties of yield stress amorphous solids.

A well-studied model constructed in this manner is the
shear transformation zone model, pioneered by Falk and
Langer \cite{Falk:1998,Falk:2011}. Building upon their numerical observations,
they devised a set of minimal equations of motion for the dynamic evolution
of a sparse population of shear transformation zones. In later refinements
and theoretical reformulations of the model, spatio-temporal aspects
were introduced in the original mean-field version of the model,
allowing it to attack a large variety of physical situations,
from simple and time-dependent flows to shear-banding phenomena and
fractures in amorphous materials \cite{Manning:2007b,Manning:2009}.
The shear transformation zone model has also been used to understand the
thermodynamic properties of sheared amorphous solids \cite{Langer:2004,Bouchbinder:2009a,Bouchbinder:2009b,Bouchbinder:2009c}, and has been generalised to include also the effects of thermal
fluctuations \cite{Falk:2004,Manning:2007}. The model
continues to be actively developed
to more complex situations, see e.g. \cite{Rycroft:2012,Hinkle:2016}.

An alternative modeling effort gaining increasing attention
builds on the observation of localised shear transformation zones to construct ``mesoscopic''
elasto-plastic descriptions of the rheology of amorphous
materials \cite{Baret:2002,Picard:2002,Picard:2005,Bocquet:2009,Rodney:2011,Cheddadi:2011}. These models are coarse-grained descriptions in the sense that no attempt is made to describe the microscopic origins of the yield stress. Instead, they assume that a yield stress exists and directly explore the consequences of deforming a solid material. The clear advantage of such models is that they open up the possibility to explore large-scale consequences of the dynamics of shear transformation zones. For instance, numerical simulations have revealed that elastic deformation in the vicinity of a local rearrangement induces long-range spatial correlations, which may induce correlations between plastic events \cite{Vandembroucq:2011,Martens:2012}. These correlations have been observed to lead to system-spanning avalanches in quasi-static deformations that are sometimes also described as precursors for the formation of permanent shear bands or strong flow localization \cite{Barrat:2011,Falk:2011,Falk:2004,Shi:2007,Maloney:2006}. The obvious drawback is that no information can be gained about the dependence of the yield stress on external control parameters, but these models can more efficiently explore the consequences of nonlinear flow curves, and might be able to describe in a relevant manner more complex situations such as shear bands, kinetic heterogeneties under flow, fractures, time-dependent phenomena, or flow in confined geometries, as discussed in more detail in Sect.~\ref{flowdyn}.

\subsubsection{Theoretical flow curves}

In the preceding section, we have described two broad classes of methods to describe the fluid-amorphous solid transition. In the following we ask how we can quantitatively describe and compare their outcomes. We consider mode-coupling and soft glassy rheology-type ``trap models" separately, and therefore consider three families of theoretical paradigms to analyze steady-state flow curves in yield stress materials. These approaches go beyond (or in some cases justify) the popular Herschel-Bulkley model described in Sect.~\ref{sec:popular}, which provides an efficient fitting model but is essentially empirical.

It should be noted that flow curves in steady-state simple shear flows only represent one of the many aspects of the rheology of yield stress materials, and some models also make detailed predictions for, e.g., time-dependent flows or more complex geometries. Reviewing model predictions for all these phenomena would however require a review article on its own \cite{Voigtmann:2014}.

\paragraph{Soft glassy rheology.}

The soft glassy rheology (SGR) model is a direct extension of Bouchaud's trap model \cite{Bouchaud:1992} that incorporates mechanical degrees of freedom in a minimal manner to describe the interplay between glassy dynamics and shear deformation \cite{Sollich:1997}.
The original trap model was mainly devised to study the physics of the glass transition and the aging dynamics in systems quenched suddenly into a glassy phase \cite{Bouchaud:1992,Bouchaud:1995}. The SGR model provides an evolution equation for the probability distribution of the system in terms of energy and stress variables. In the presence of a constant
shear stress, steady-state flow curves can be predicted, with a behavior that is governed by the only control parameter of the model, namely the ``temperature'' $T$. Whereas the initial trap model for aging glasses explicitely refers to $T$ as the temperature of a thermal bath coupled to the system, the SGR model differs somewhat on the precise interpretation of the temperature and uses the words ``effective temperature'', in order to include athermal materials such as foams or emulsions in the same framework. The temperature $T$ is then thought as quantifying the strength of ``mechanical noise'' triggered by the flow itself. A more detailed discussion of effective temperatures in driven materials can be found elsewhere:
\cite{Berthier:2001,Cugliandolo:1997,Sollich:2012,Bouchbinder:2009a,Bouchbinder:2009b,Bouchbinder:2009c}. Recent work has critically revisited the
properties of the mechanical noise triggered by shear transformation
zones \cite{nicolas:2014},
offering in particular a detailed comparison between the SGR model and an
alternative mean-field modeling proposed by H\'ebraud and Lequeux
\cite{Hebraud:1998},
where a Langevin dynamics is studied in which noise is directly
related to the amount of plastic deformation generated in the material

Despite its simplicity, the SGR model offers a rich variety of possible flow curves, depending on the considered temperature regime \cite{Sollich:1998}.
In the absence of a flow, the system undergoes a glass transition at some critical temperature $T_c$, below which
ergodicity breaking occurs. With an imposed shear flow, three temperature regimes are observed:
\begin{itemize}

\item First, when $T > 2T_c$, the system exhibits a Newtonian flow, as expected for a simple fluid state.

\item A second, somewhat unexpected regime occurs when $T_c < T < 2 T_c$, where the system displays a pure power-law rheology of the form $\sigma \approx \dot{\gamma}^{n}$, with a ``shear-thinning" exponent $0 < n = T/T_c - 1 < 1$. This regime is peculiar as it corresponds to a ``solid" system with an infinite viscosity at rest when $\dot{\gamma} \to 0$, but with no yield stress.
When the shear-thinning exponent $n$ becomes small, it might be difficult to distinguish this behavior from a Herschel-Bulkley functional form. A peculiarity of this regime is the infinite shear viscosity for temperatures that are strictly above the glass transition temperature where the system actually reaches thermal equilibrium. See \cite{Lequeux:2001} for a detailed discussion of this
curious issue.

\item In the third regime, for temperatures below the critical temperature, $T<T_c$, the rheology can be described by the Herschel-Bulkley model, $\sigma \approx \sigma_y(T) + \dot{\gamma}^{n}$ and
the shear-thinning exponent obeys $0 < n = 1-T/T_c < 1$. A
temperature-dependent yield stress $\sigma_y(T)$ emerges continuously at the glass temperature, with a linear onset $\sigma_y(T \lesssim T_c) \approx 1-T/T_c$
and a smooth approach to a finite limit at zero tempeature,
$\sigma_y(T \to 0) > 0$.
\end{itemize}

Overall, within the SGR model, the behavior of the flow curves is smooth at the transition temperature, $T=T_c$, where the system has no yield stress but the shear-thinning exponent vanishes---a situation that could easily be confused experimentally with a finite yield stress.

Moreover, since all the characteristic exponents of the model are temperature-dependent quantities, they carry no deep physical meaning but simply reflect the complex interplay between the broad distribution of relaxation times in the equilibrium model and the external mechanical forcing in the presence of thermal fluctuations. This remark implies, in particular, that no particular scaling form
is predicted to describe the flow curves derived within the SGR model in any of the temperature regimes, or even in the close vicinity of the critical temperatures of the model.

\paragraph{Mode-coupling theories.}

The mode-coupling theory of the glass transition is
now understood as a building block of a larger theoretical construction
to understand the physics glassy materials called random first-order transition theory, which aims at describing dynamic and thermodynamic aspects of the
statistical mechanics of materials undergoing a fluid-to-glass
transition \cite{Lubchenko:2007,Berthier:2011b}.

The mode-coupling approach itself is not a unique theory, and several related lines of research coexist which differ in their microscopic starting point but often provide similar predictions. A few of these approaches
were extended to also include mechanical degrees of freedom,
much in the spirit of the SGR model. There are
at present two main starting points for solving the dynamics in the amorphous solid state.

\begin{itemize}

\item A first approach \cite{Cugliandolo:1997b,Berthier:2001} consists of solving exactly the driven dynamics of simple, but rather abstract, glass models that are known to exhibit an equilibrium dynamics that is in the same universality class as other mode-coupling approaches, like for instance the $p$-spin glass
models or other disordered models \cite{Kirkpatrick:1987}.

\item  A second line of work starts from microscopic equations of motion for particles in a dense fluid, and develops mode-coupling approximations to derive closed, but approximate, dynamical equations for microscopic correlation functions based on density fluctuations \cite{Fuchs:2002,Miyazaki:2002,Brader:2007,Fuchs:2009}.

\end{itemize}

Both approaches have been extended to include external driving forces
and shear flows in order to study
the interplay between glassy dynamics and rheology.
In mode-coupling theories, the equilibrium dynamics (without shear flow) is characterized by a critical temperature $T_c$ where the (alpha) relaxation time $\tau_\alpha(T)$, diverges as a power law. Near the glass transition, time-correlation functions develop a two-step decay with an intermediate plateau reflecting the transient caging of the particles in the dense fluid. In rheological terms, this simply signals that very viscous fluids near a glass transition are viscoelastic and behave as solids at intermediate time scales, and flow at long times,
associated to complex frequency spectra for the linear rheological response.
The approach to, and departure from, this plateau regime involve non-trivial power-laws for time correlation functions, which are characteristic signatures
of mode-coupling theories \cite{Gotze:2008}. A known limitation
of the theory is that the algebraic divergence that it predicts for the
equilibrium relaxation time is not observed in experiments,
where it is replaced by a smooth crossover. The current view is that
mode-coupling theories describe the initial regime of slow
dynamics in glassy materials well, but fail closer to the glass transition.
Interestingly, this ``mode-coupling''
regime coincides with the physically relevant one
for colloidal systems \cite{Gotze:2008,Siebenburger:2012b},
which justifies why the mode-coupling approach is included in this soft-matter review article.

In both mode-coupling approaches, the glass transition is destroyed
by the imposed shear flow, and the
microscopic relaxation time scale is never infinite in the presence of a finite
driving force, but rather becomes dependent on the imposed shear rate
$\dot{\gamma}$. However, the resulting flow curves differ somewhat in their
details as we discuss in the following.

In the first class of models, namely
schematic mean-field models, the rheology exhibits Newtonian behavior at
temperatures above $T_c$ at very low shear rates, but the dynamics become
strongly dependent on $\dot{\gamma}$ when the ``dressed" P\'eclet number,
$P_e \equiv \tau_\alpha \dot{\gamma}$, becomes larger than unity.
Since the shear flow accelerates the microscopic structural relaxation,
the viscosity decreases as $\dot{\gamma}$ increases, a shear-thinning
behaviour. As a result, the following scaling form for the flow curves,
\begin{equation}
\eta(\dot{\gamma},T) = \frac{\eta_0(T)}{[ 1 + \dot{\gamma}/
\dot{\gamma}_0(T) ]^{1-n}},
\label{eq:flowresc}
\end{equation}
where $\eta_0(T) \sim \tau_\alpha(T)$ is the Newtonian viscosity,
and $\dot{\gamma}_0(T)$ a critical shear rate separating Newtonian from shear
thinning regimes, and $n$ the usual shear-thinning exponent, whose value is $n=1/3$ in the specific family of models studied in \cite{Berthier:2001}.

At the glass transition, a pure power-law rheology is thus obtained, $\sigma(\dot{\gamma},T=T_c) \sim \dot{\gamma}^n$,
whereas a temperature-dependent power-law rheology is obtained in the glass phase: $\sigma(\dot{\gamma}, T<T_c)
\sim \dot{\gamma}^{n(T)}$ with a shear-thinning exponent decreasing from $n(T=T_c)=1/3$ to $n(T \to 0)=0$, but with no finite yield stress. Therefore, the rheology of the glass phase is very similar to the intermediate temperature
regime of the SGR model with the difference that here the viscosity divergence
coincides with the equilibrium glass transition of the model.

The absence of a yield stress is natural in the context of mean-field approaches whose aging dynamics in the glass phase is well understood \cite{Cugliandolo:1993}. In the absence of external flow, the system slowly relaxes along the flat or ``marginal" regions of its free-energy landscape, but does not penetrate deeper free-energy minima. This is a general feature of mean-field glassy
dynamics \cite{laloux:1996}. The power-law
rheology found in the glass phase directly results from
this marginal dynamics, and non-mean-field effects
are believed to manifest themselves by the emergence of a
finite yield stress, as explored in \cite{Berthier:2002pp}.

\begin{figure}
	\centering
		\includegraphics[width=8.5cm]{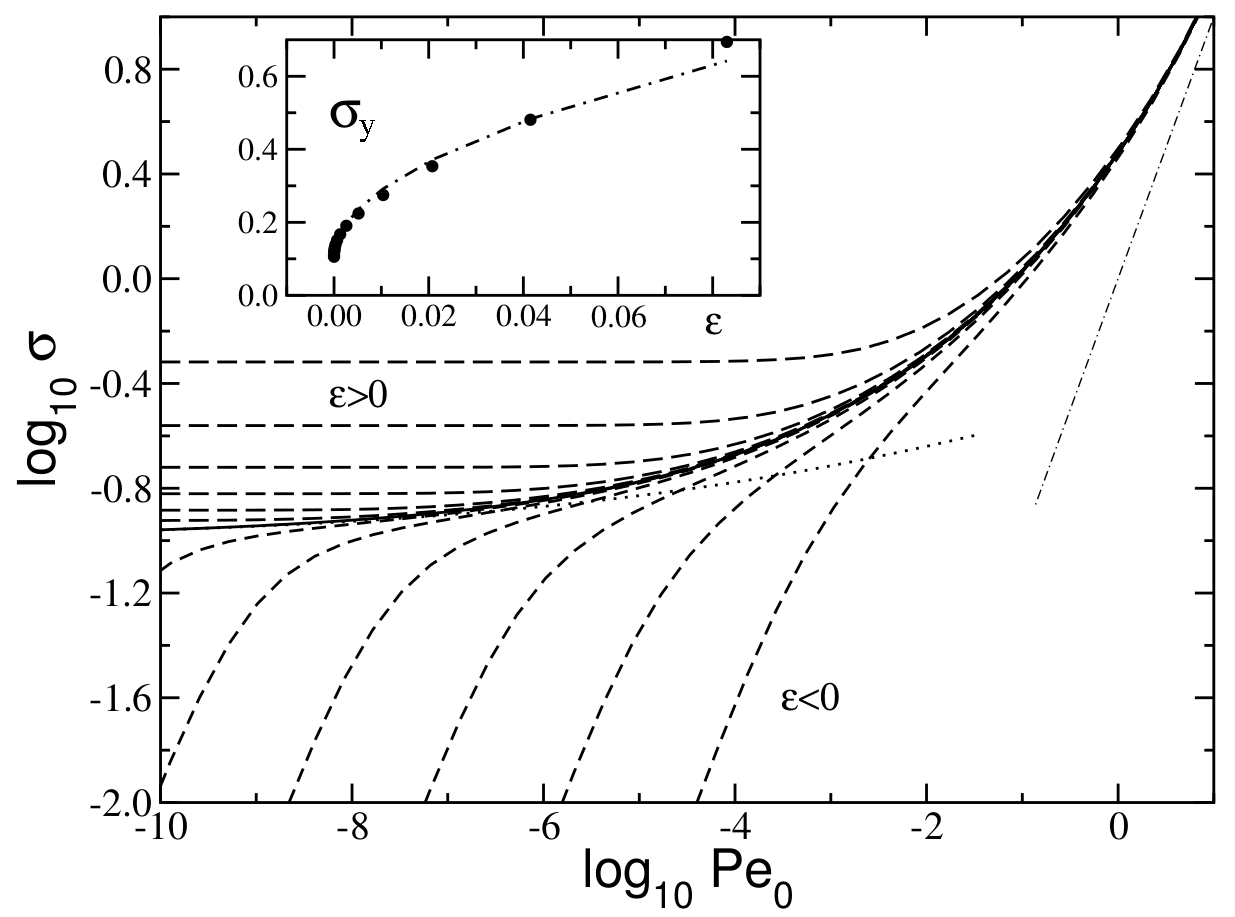}
	\caption{Flow curves predicted for a range
of temperatures $T$ across the mode-coupling critical temperature $T_c$;
$\epsilon = (T_c-T)$ is the distance to the critical temperature,
and the shear rate is rescaled by a microscopic time unit $\tau$
to form a P\'eclet number $Pe_0 = \dot{\gamma} \tau$. The inset shows the
discontinuous emergence of the yield stress at $T_c$.
From \cite{Fuchs:2003}.}
\label{fig:fuchs}
\end{figure}

Using liquid-state theory to derive mode-coupling equations for glassy
fluids under flow results in a set of dynamical
equations that reduce to the usual mode-coupling phenomenology described above for the equilibrium dynamics. However, the driven dynamics under shear flow provides a set of predictions that differ somewhat from the schematic mean-field models, for reasons that are more technical than physical and presumably stem
from the application of different types of ``mean-field'' approximations.
These mode-coupling equations have been derived in a number of ways that are technically quite involved, but all derivations essentially provide similar predictions for the flow
curves \cite{Fuchs:2002,Fuchs:2009,Miyazaki:2002}.
The predicted flow curves in the vicinity of the glass transition closely reflect the complexity of the time regimes observed
for time correlation functions, as illustrated in Fig.~\ref{fig:fuchs}.

Specifically, the flow curves
predicted by this second class of models in the fluid region exhibit a
Newtonian regime at sufficiently small $\dot{\gamma}$ followed
by a strong shear-thinning regime for large Peclet numbers
$P_e = \dot{\gamma} \tau_\alpha$, as
found in the mean-field and SGR models.
The flow curve at the critical temperature
$T=T_c$ obeys a Herschel-Bulkley functional form with a finite yield stress, $\sigma_y(T=T_c) > 0$, and a shear-thinning exponent that takes a non-universal value [specific approximations give $n \approx 0.15$ \cite{Fuchs:2003}].
The power-law approach to a finite yield stress closely mimicks the power-law approach to a finite plateau found for time-correlation functions.

Just below the glass transition, the yield stress increases algebraically with decreasing temperature, $\sigma_y(T \lesssim T_c) \approx \sigma_y(T_c) + c \sqrt{T_c - T}$, which again mimicks the temperature behavior of the plateau in time-correlation functions (indeed the two are intimately connected within the theory). Note, however. that the yield stress emerges discontinuously at
the critical temperature $T_c$, a prediction that seems unique
to this approach. This abrupt emergence of a yield stress
can however not exist in experiments
where a true mode-coupling transition is not observed. In practice 
it is replaced
by a crossover between flow curves where the Newtonian regime of the flowing
liquid slowly shifts outside the experimental time window, so that
by construction the ``first'' measurable value of the yield stress
must indeed be a finite number \cite{Varnik:2006,wittmer:2013}).
Therefore, the question of the (dis)-continuous nature of the
emergence of a yield stress at the glass transition is ill-posed.
Of course, when analysing experimental and numerical flow curves
(which do not have a real transition)
within the framework of the mode-coupling approach
(which has a real transition), the discontinuous emergence 
of a yield stress is needed~\cite{Voigtmann:2011,Siebenburger:2012b}.

In the glass regime, the flow curves are again well described by a Herschel-Bulkley functional form. The limit of low temperatures is, however, problematic within the theory as it makes the unphysical prediction that the yield stress eventually vanishes in the $T \to 0$ limit~\cite{Ikeda:2013}. This implies that the theory is actually not well suited to describe the yield stress of glassy systems deep in the glass phase, which is perhaps not surprising, as its starting point is actually an equation of motion for the fluid.

It should thus be kept in mind that mode-coupling theories are to be
used to describe the interplay of glassy dynamics
and shear flow in the immediate vicinity of kinetic arrest,
over a modest window of shear rates.
All the detailed predictions of the theory have been tested in great detail
in both numerical and experimental studies
\cite{Brader:2010,Siebenburger:2012b,Amann:2013,Ballauff:2013,Amann:2015}.
We emphasize that despite the presence of a genuine critical temperature
in the mode-coupling approach and the existence of power laws controlling the
divergence of the viscosity and the discontinuous
emergence of a finite yield stress, no specific ``critical data collapse''
of the flow curves is obtained within the theory.

The theoretical limitations of mode-coupling approaches are
fully understood in the broader context of random first-order transition theory, where the structure and dynamics of the glass phase are treated analytically using a completely different method based on an approximate treatment (involving replica calculations \cite{Parisi:1988}) to describe the complex free-energy landscape characterizing glassy materials \cite{Yoshino:2010}. Recent progress in this direction has been substantial \cite{Parisi:2010,Charbonneau:2014}, as the nature of the equilibrium glass transition has been analytically elucidated for particle systems in the (abstract) limit of a large number of spatial dimensions.
This approach opens new ways to treat analytically the nature of the glass phase, of the dynamics of the viscous liquid near the glass transition, and potentially of its rheological properties. Currently, the theory is being developed to treat mechanical properties, such as the shear modulus \cite{Yoshino:2014}. Very recently, stress-strain curves in quasi-static deformation protocols have been obtained analytically \cite{Rainone:2014}, thus pushing the theory closer to being able to describe the yielding transition in glassy solids \cite{urbani:2017}. 
Reconciling these thermodynamic replica calculations to dynamic equations derived within mode-coupling theories remains an open issue \cite{Szamel:2010}. Another promising route is the possibility to perform 
a systematic treatment of non-mean-field effects, thus paving the way for a generalisation of mode-coupling approaches that do not suffer from the shortcomings described above. 

\paragraph{Jamming rheology.}

In Sect.~\ref{sec:jammed}, we provided a qualitative description of the flow curves obtained from simple computational models undergoing an ideal jamming transition, in connection with the experimental results displayed in Fig.~\ref{fig:Ch1Fig1Ludo}(b) for emulsions with sufficiently large (i.e. non-Brownian) droplets. In the vicinity of the jamming transition, these flow curves can display a number of scaling features that are fully specific to non-Brownian assemblies of particles.
Upon compression towards $\phi_J$,
the system exhibits a Newtonian viscosity that diverges algebraically, accompanied by a power-law shear-thinning behavior. Above $\phi_J$, a finite yield stress emerges continuously at the transition, and its increase with packing fraction is also described by a power law.

We emphasize that the presence of these
power law behaviors is unique to athermal rheology, and that the situation differs qualitatively from the behaviors observed in Brownian systems sheared across their glass transition. There has been some confusion in the literature
about the distinction between the two types of yield stress rheology.
The scaling behavior proposed for athermal systems has
for instance been incorrectly applied to Brownian and thermal systems
as well. As mentioned, the distinction
is readily made by looking at adimensional shear rates (Peclet numbers)
and stress scales \cite{Ikeda:2012,Ikeda:2013b,Ikeda:2016}.

The scaling properties of the jamming rheology near the zero-temperature
jamming transition have been fully elucidated in computer simulations
of soft repulsive potential, such as harmonic or Hertzian pair potentials,
see Eq.~(\ref{eq:potential}).
These flow curves have now been
characterized numerically in great detail
\cite{Olsson:2007,Olsson:2012,Olsson:2012b,Hatano:2010,Ikeda:2012,Vagberg:2014}.

An approximate scaling form similar to Eq.~(\ref{eq:flowresc}) is obtained below the jamming transition in the non-Brownian suspension regime, where the Newtonian viscosity diverges as $\eta_0(\phi) \sim (\phi_J - \phi)^{-m}$, with $m \approx 1.5-2.5$ \cite{Andreotti:2012,Boyer:2011}. A series of recent large-scale numerical studies for non-frictional particles report $m \approx 2.55$~\cite{Kawasaki:2014,Vagberg:2014}, but it should be noted that this power law only holds extremely close to the jamming density, with strong corrections further away from the critical point, which presumably explain the large spread in literature values for the exponent $m$ of the viscosity divergence. Because the Newtonian regime is reached at very low shear stresses where particles barely overlap, the
particle softness does not
affect the value of $m$, which thus remains pertinent
to describe the hard sphere limit.

In the jammed region, the density increase of
the yield stress is well described by a power law,
$\sigma_y(\phi) \sim (\phi - \phi_J)^{\Delta}$,
where $\Delta$ is a critical exponent. Because
this exponent describes the solidity of
a compressed assembly of soft overlapping particles, it is not surprising
that it is found to depend on the chosen form of the pair repulsion between
the particles. In particular, simulations show if $\alpha$ is the exponent
describing the pair repulsion $V(r)$ in Eq.~(\ref{eq:potential})
(with $\alpha =2$ for harmonic sphere,
and $\alpha=5/2$ for Hertzian potential), then $\Delta$ is very close to
the value $\Delta = \alpha - 1$, with small but measurable deviations from
this estimate (for instance $\Delta \approx  1.15$ for harmonic
spheres \cite{Olsson:2012b}). This estimate
is reasonable as $(\alpha - 1)$ is also the
exponent controlling the increase of
the pressure in compressed packings, as predicted by dimensional
analysis \cite{OHern:2003}.
Systematic deviations, $\Delta \gtrsim \alpha-1$,
have now been reported in several numerical
studies \cite{Olsson:2011,Olsson:2012,Olsson:2012b,Hatano:2010,Kawasaki:2014}.

Finally, above jamming, the flow curves are well described by a
Herschel-Bulkley model, with a shear-thinning exponent $n\approx 0.38$ which
is also independent of the form of the soft potential \cite{Olsson:2012b}.
Exactly at the jamming transition, a pure power-law rheology is
obtained, $\sigma \sim \dot{\gamma}^{n'}$, with another non-trivial
shear-thinning exponent, $n'$, which depends on the form of the pair
potential and is thus not universal \cite{Olsson:2012b}.
We emphasize that a precise determination of the various scaling regimes and
the precise values of all these critical exponents ($m$, $\Delta$,
$n$, $n'$)
is a difficult numerical task, which are in addition plagued by strong
finite size effects \cite{Vagberg:2010,Kawasaki:2014}.
These difficulties also suggest that direct comparison
to experimental results should be done with some caution.

An important consequence of these multiple scaling regimes is that
despite the presence of power laws in the rheology of model assemblies of
soft particles, the flow curves measured over a large domain of densities
and shear rates cannot be rescaled onto master curves, as initially
proposed in \cite{Olsson:2007}. These deviations have been described
with great analytic precision as a form of correction to scaling
in a series of studies \cite{Vagberg:2010,Vagberg:2014};
see \cite{Kawasaki:2014} for a
specific illustration of a ``failed'' data collapse for a fully athermal
assembly of soft particles.

In addition, there is currently a large theoretical activity to better understand the physical origin of these exponents and to relate them to more microscopic quantities characterizing the structure of athermal packings in the vicinity of the jamming
transition \cite{Tighe:2010,Lerner:2012,DeGiuli:2014,Yoshino:2014}.

\begin{figure}
	\centering
		\includegraphics[width=8.5cm]{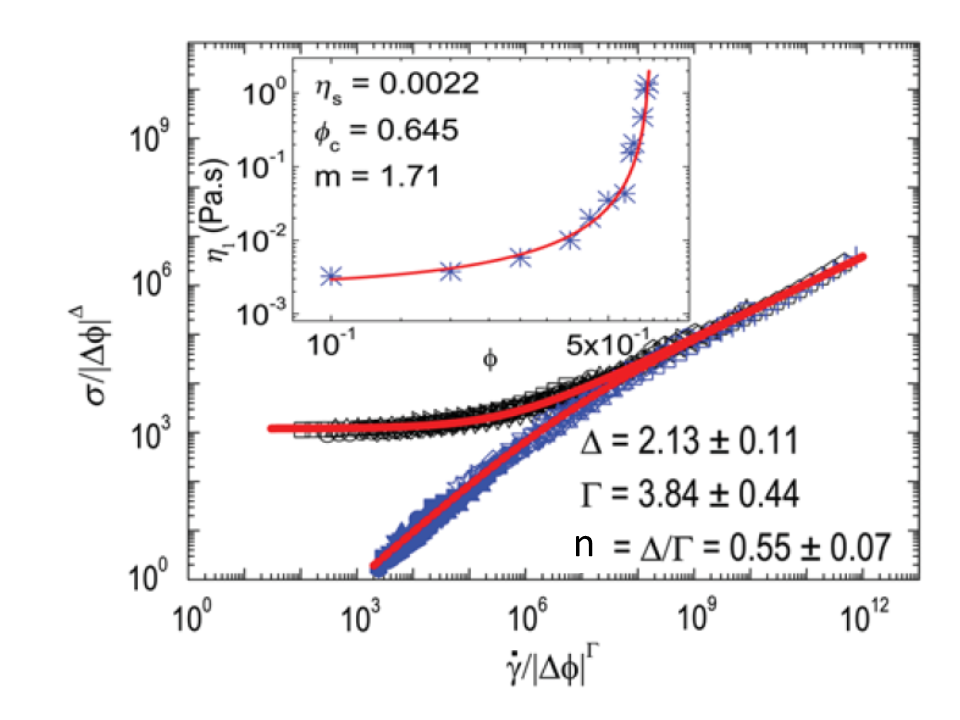}
	\caption{Master curves showing a good collapse of
the flow curves onto two branches, one for samples with $\phi < \phi_J$   and one for $\phi > \phi_J$, when stress and shear rate are rescaled with the distance to jamming to a certain power. The lines are supercritical and subcritical branches representing empirical fits of the master curve, respectively. The inset shows a fit of the the low-shear viscosity to a power-law divergence \cite{Paredes:2013}. Flow curves were obtained for emulsions prepared with different volume fractions of the dispersed phase.}
\label{fig:rescalingflowcurves}
\end{figure}

In experiments, these scaling forms have also been used to analyze flow curves measured in a variety of systems. Using the exponents defined above and assuming that power laws hold for the entire range of explored densities and shear rates, the flow curves measured for different volume fractions of a given system can be collapsed onto two master curves (one below and one above jamming) by rescaling both the stress and the shear rate with appropriate powers of the distance to the jamming transition $(\phi - \phi_J)$ \cite{Nordstrom:2010,Paredes:2013b,Dinkgreve:2015}, as shown in Fig.~\ref{fig:rescalingflowcurves}. This type of data
collapse is empirically useful, as it organizes the experimental data around
the critical density $\phi_J$, while using simple, but
reasonable functional forms for their density dependence.
This strategy was first employed in numerical work \cite{Olsson:2007},
for which it is now understood to be only approximately correct.

In all published cases \cite{Nordstrom:2010, Paredes:2013b, Dinkgreve:2015}, the rescaling appears to work well; $\Delta \approx 2$ and the exponent for the rescaling of the shear rate axis is $\Gamma \approx 4$. The observation that the rescaling collapses the Herschel-Bulkley flow curves above jamming then immediately implies that the shear-thinning exponent is $n=\Delta/\Gamma \approx 1/2$. The collapse of the Newtonian flow regime curves at low shear rates below jamming implies in turn that the exponent for the divergence of the viscosity is $m=\Gamma-\Delta \approx 2$ \cite{Paredes:2013b}, see the inset of Fig. ~\ref{fig:rescalingflowcurves}. This rescaling with very similar exponents has now been observed for soft polymer particles (PNIPAM and Carbopol), emulsions with mobile and immobile surfactants and foams \cite{Nordstrom:2010,Paredes:2013b,Dinkgreve:2015}, suggesting that either these systems have very similar interactions, or that the exponents ($\Delta$, notably) do not sensitively depend on the interactions, in contrast to theoretical predictions \cite{Tighe:2010}. Another possibility for the difference between experiments and simulations could be that the simulations and experiments use quite different regimes to determine the critical exponents (in general, in the simulations, one is much closer to the jamming transition), so that experimentally determined values could represent ``effective'' values.
Moreover, some of the analyzed systems (notably, microgels) are not fully athermal and should perhaps be described by exponents characteristic of the glass transition (if exponents exist for this situation) which may be different from the exponents from the thermal jamming transition.

Finally, it is worth mentioning that other scaling analysis have
been proposed that are also based on experimental data,
e.g. for the evolution of the yield stress or the shear
modulus \cite{Mason:1996a,Scheffold:2014,Scheffold:2016,Yodh:2014,Mohan:2013}.
These remain interesting open questions and we refer to a recent
publication \cite{Dinkgreve:2015} for a compilation of the different jamming
exponents in theory, experiment and numerics.

%
%

\section{Physical insights from yield stress measurements}
\label{yieldstressdetermination}

While the definition of the yield stress from a theoretical point of view, i.e., Eq.~(\ref{eq.ysdef}), looks very simple, its practical determination is known to raise challenging experimental problems. As discussed above, aging and time-dependences --most generically, thixotropy-- have led to long-standing controversies in the rheology community. Other issues include instrument artifacts or slippage of the material at the walls of the measuring device. Such problems with the measurement of yield stress have been reviewed from an engineering point of view for example in:  \cite{Uhlherr:2005,Nguyen:2006,Moller:2009b,NeilJ.Balmforth2014,Coussot:2014}. Here, we try to clarify what experimentalists call ``the yield stress,'' what they exactly measure and what physical mechanisms they actually probe in the various classical techniques. We will ignore techniques that involve complex geometries such as squeeze flows \cite{Rabideau:2009}, penetrometry tests \cite{Boujlel:2012} or stop flows on inclined planes \cite{Kee:1990,Coussot:1995} in order to focus only on techniques that rely on the drag flow produced by a rheometer, and show how their diversity proves relevant to address specific fundamental questions pertaining to yield stress behavior.

\begin{figure}
	\centering
		\includegraphics[width=8.5cm]{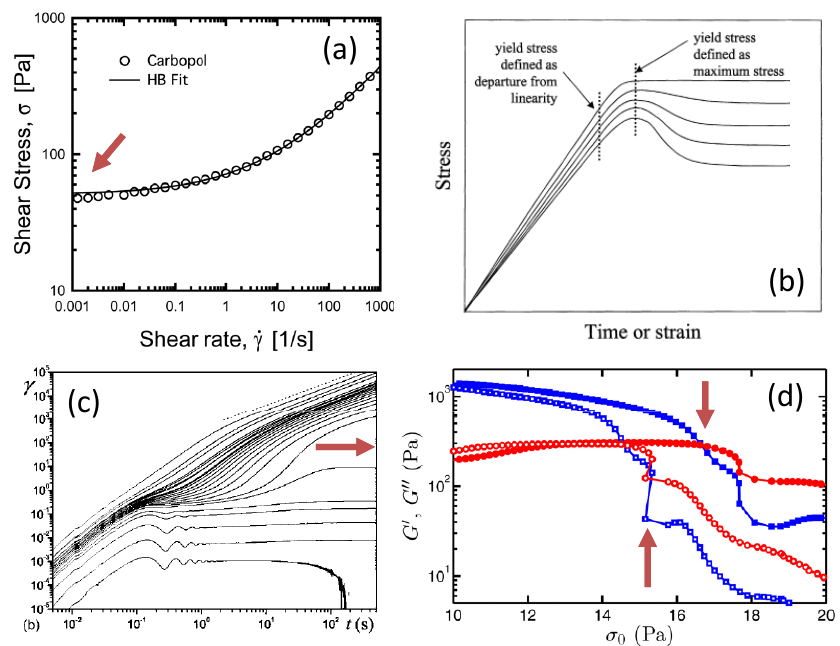}
	\caption{Various methods to determine a yield stress experimentally. (a) Extrapolation of the flow curve in the limit of vanishing shear rates. Experiments performed on a Carbopol microgel using roughened cone and plate fixtures. The black line is the best Herschel-Bulkley fit. Extracted from \cite{Dimitriou:2013}. (b) Sketch of the stress response to a shear start-up experiment. The yield stress can be defined as the stress corresponding to the end of the linear regime, as the stress maximum, or as the equilibrium stress. Extracted from \cite{Barnes:2001}. (c) Strain response to  step stress experiments for various stresses ranging from 0.22 to 220~Pa. Extracted from \cite{Coussot:2006}. (d) Oscillatory stress sweep experiment performed on a 6\% wt carbon black gel at two different sweep rates: 7 (open) and 34 (filled) mPa.s$^{-1}$. Here, the yield stress, defined as the intersection of $G'$ (blue/dark grey) and $G''$ (red/light grey), depends on the sweep rate. Extracted from \cite{Perge:2014b}.}
		\label{fig:Ch2Fig3}
\end{figure}

In rotational shear rheometry a shear stress $\sigma$ is imposed and the corresponding shear rate $\dot\gamma$ (or strain $\gamma$) is recorded, or vice-versa. Typical geometries used for performing this type of measurement include concentric cylinders, plate-plate and cone-plate geometries \cite{Larson:1999,Barnes:1989}. In the following we first review methods that involve liquid-to-solid transitions to determine the yield stress and then those based on solid-to-liquid transitions (see Fig.~\ref{fig:Ch2Fig3}).

\subsection{Experiments probing the liquid-to-solid transition}

\subsubsection{Extrapolating the flow curve in the limit of vanishing shear rates}

The experiment matching the definition of Eq.~(\ref{eq.ysdef}) consists of measuring the flow curve ($\sigma$ vs $\dot{\gamma}$) by applying a \textit{steady shear} and progressively ramping down the shear rate to reach the limit $\dot{\gamma}\rightarrow 0$. The material, liquid-like at first, is thus progressively brought into a solid-like state, ideally through a series of steady states. The extrapolation of the stress in the limit of vanishing shear rates points towards a stress value that is generally referred to as the {\it dynamic yield stress}. In practice, however, this extrapolation can be problematic as it requires the establishment of a steady shear flow at arbitrarily low shear rates. An alternative is to fit the flow curve to a rheological model, such as the Bingham, Herschel-Bulkley or Casson models [see Fig.~\ref{fig:Ch2Fig3}(a)] \cite{Nguyen:1992}. As noted above, the Herschel-Bulkley model is observed to fit the experimental data properly over several decades in the case of dense assemblies of soft particles, such as emulsions, microgels and foams \cite{Ovarlez:2013b}, and to provide a reproducible yield stress value. For this model, the most convincing representation of the flow curve is to plot the {\it viscous} stress, namely the difference between the stress and the yield stress, $\sigma-\sigma_y$, vs the shear rate $\dot\gamma$, which should show pure power-law behavior, as reported for instance by \cite{Katgert:2009,Mobius:2010,Tighe:2010,Fall:2010b,Shaukat:2012}.

However, the above methodology suffers from several important limitations. First, wall slip can affect the flow at low shear rates, an issue that will be discussed in more details in Sect.~\ref{wallslip}. Second, time-dependent phenomena such as thixotropy cause the shape of the flow curve and therefore its extrapolation in the limit of vanishing shear rates to depend on the rate at which the shear rate is swept \cite{Divoux:2013}. For some materials such as various attractive colloidal gels \cite{Ovarlez:2013}, this may even be a subtle function of the previous flow history. The dynamic yield stress obtained for a time-dependent material thus depends on the details of the experimental protocol. Furthermore, from a theoretical viewpoint it is unclear whether Eq.~(\ref{eq.ysdef}) strictly holds even for simple types of glassy materials, whereas very little is known for
physical and
non-equilibrium gels from computer simulations. Even for athermal systems
characterized by a genuine jamming transition, extracting the
yield stress using extrapolations to vanishing shear rates
requires rheololgical measurements over a very broad time window.

\subsubsection{Determining the residual stress after flow cessation}

\begin{figure}
 \includegraphics[width=7.5cm]{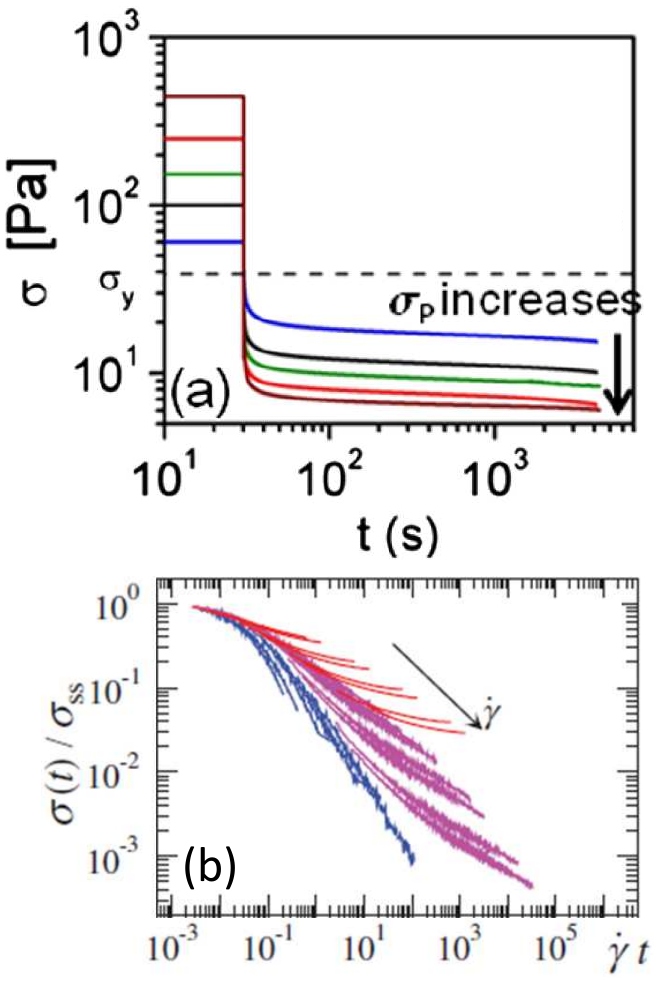}
	\caption{(a) Stress relaxation upon flow cessation: experiments with microgels for different preshear stresses (from 60 to 443~Pa). The internal stress $\sigma_r$, defined by linear extrapolation of the stress measured over a short time interval ($<50$~s) after flow cessation, is larger for smaller preshear stress and becomes quite significant for pre-shear stresses approaching the yield stress. From \cite{Mohan:2013}.
(b) Evolution of the stress after flow cessation normalized by the stress prior to flow cessation ($\sigma_{ss}$) as a function of $\dot \gamma$ for a hard-sphere colloidal suspension. The curves correspond to various imposed shear rates $\dot \gamma$ prior to flow cessation, and different packing fractions. Glass states are shown in red (light gray), liquid states in blue (dark grey). From \cite{Ballauff:2013}.}
	\label{fig:Chapter2FigureFlowCessation}
\end{figure}

In order to minimize the influence of previous flow history and thixotropy, some researchers prefer switching off the flow rather than progressively decreasing the shear rate. In such a flow-cessation experiment, the sample is sheared at a given shear rate long enough to reach steady state. Then the shear rate is suddenly set to zero and the initially liquid-like material turns into a solid while the stress decreases towards a constant {\it residual} or {\it internal} stress $\sigma_r$, see Fig.~\ref{fig:Chapter2FigureFlowCessation}. Historically, this stress value has also been coined a {\it yield stress} by several authors \cite{Michaels:1962,Tiu:1974,Nguyen:1983,Magnin:1990} but it was soon recognized that this residual stress was always much smaller than the dynamic yield stress \cite{Keentok:1982}. In fact, the residual stress decreases for increasing values of the shear rate applied prior to flow cessation \cite{Osuji:2008,Lidon:2016}. Therefore $\sigma_r$ is not a material constant but rather gives access to a history-dependent frozen-in quantity that accounts for the microstructural anisotropy imprinted to the material by previous shear.

Still, recent years have seen a renewed interest in internal stresses triggered by various theories for soft glasses and their predictions of the dynamics upon flow cessation. In dense assemblies of soft particles, the stress has been shown to relax through two distinct steps. A rapid relaxation, interpreted as the ballistic motion of the particles in the framework of a micromechanical model \cite{Seth:2011} is followed by a slower relaxation of the elastic contact forces between the jammed particles \cite{Mohan:2013} [see Fig.~\ref{fig:Chapter2FigureFlowCessation}(a)]. Whereas such a slow relaxation due to aging dynamics is expected for the Brownian particles studied in these experiments, no such slow relaxation should exist for fully athermal soft particles, as there is no mechanism to induce fluctuations that would allow for a slow exploration of the complex free-energy landscape of the material. Simulations of the behavior of non-Brownian particles
after shear is suddenly stopped confirm the rapid convergence of the residual stress to a finite value \cite{Chaudhuri:2012b}, with no slow
relaxation involved in that relaxation process.

In hard-sphere colloidal glasses, the stress relaxes as a power-law as predicted by the SGR model \cite{Cates:2004}, and is associated with subdiffusive motions of the particles \cite{Ballauff:2013} [see Fig.~\ref{fig:Chapter2FigureFlowCessation}(b)]. This is not surprising because the model
was initially devised as a rheological model to
study the interplay of aging dynamics and shear flow in glassy
materials \cite{Sollich:1998,Fielding:2000}.
By contrast, the mode-coupling approach developed in \cite{Fuchs:2009}
does not include aging effects, and so it cannot describe the slow
relaxation of the stress after flow cessation or the sub-diffusive particle
displacements observed in experiments \cite{Ballauff:2013,Fritschi:2014},
and predicts instead a fast convergence to an arrested state with a finite
residual stress.

Finally, aging laponite clay suspensions display a sigmoidal stress relaxation upon flow cessation with a characteristic time that scales inversely with the quench rate \cite{Negi:2010}. The latter behavior contrasts with the simpler relaxation reported in dense systems with strong aging and remains to be interpreted from microscopic and/or theoretical points of view.

\subsection{Experiments probing the solid-to-liquid transition}

\subsubsection{Analyzing the transient stress response during shear start-up}

\label{shear-start-up}

When an external shear rate $\dot\gamma$ is imposed on a soft solid at time $t=0$, and is kept constant thereafter, the stress $\sigma(t)$ first increases linearly with the strain $\gamma=\dot\gamma t$ which is indicative of elastic response [see Fig.~\ref{fig:Ch2Fig3}(b)]. It then departs from linearity at intermediate strains, typically $\gamma \sim 0.1$ for bentonite suspensions \cite{Nagase:1986} and $0.2$ for microgels \cite{Divoux:2011b}. Although a characteristic stress associated with departure from linearity can be inferred from this early-time stress response \cite{Lin:1985,Nagase:1986}, such a ``yield stress'' involves an arbitrary definition of how far from linearity the system should be. More importantly, this behavior may relate to local yielding events rather than to global yielding of the material. Nonetheless, this crossover from linear to nonlinear behavior is an interesting phenomenon for which experiments can be confronted to theories and simulations perhaps more easily than for larger strains.

Upon entering the fully nonlinear regime, $\sigma(t)$ in general goes through a maximum before decreasing toward its steady-state value. Such a {\it stress overshoot} is observed in a large number of yield stress fluids such as foams \cite{Khan:1988}, emulsions \cite{Papenhuijzen:1972,Batista:2006}, microgels \cite{Divoux:2011b}, clays \cite{Nagase:1986} and attractive gels \cite{Liddell:1996,Koumakis:2011}. The maximum value of the stress reached during shear start-up has been widely used as an estimate of the yield stress. However, it does not coincide with the definition of Eq.~(\ref{eq.ysdef}) and it is now referred to as the {\it static yield stress} \cite{Varnik:2004} in order to clearly distinguish it from the dynamic yield stress inferred from flow-curve measurements measured in the flowing regime. In particular, as they are performed at a finite shear rate, start-up experiments introduce the additional time scale $1/\dot\gamma$ and the subsequent nonlinear stress response is generally not a function of $\gamma$ only but also depends on $\dot\gamma$.
Although the static yield stress is not a material constant, the stress overshoot phenomenon still raises important fundamental questions: Does it have any simple microstructural interpretation? Can it be predicted from theory? The influence of the various experimental control parameters on the stress maximum, reviewed below, might give some clues.

\begin{figure}
	\centering
	\includegraphics[width=8.5cm]{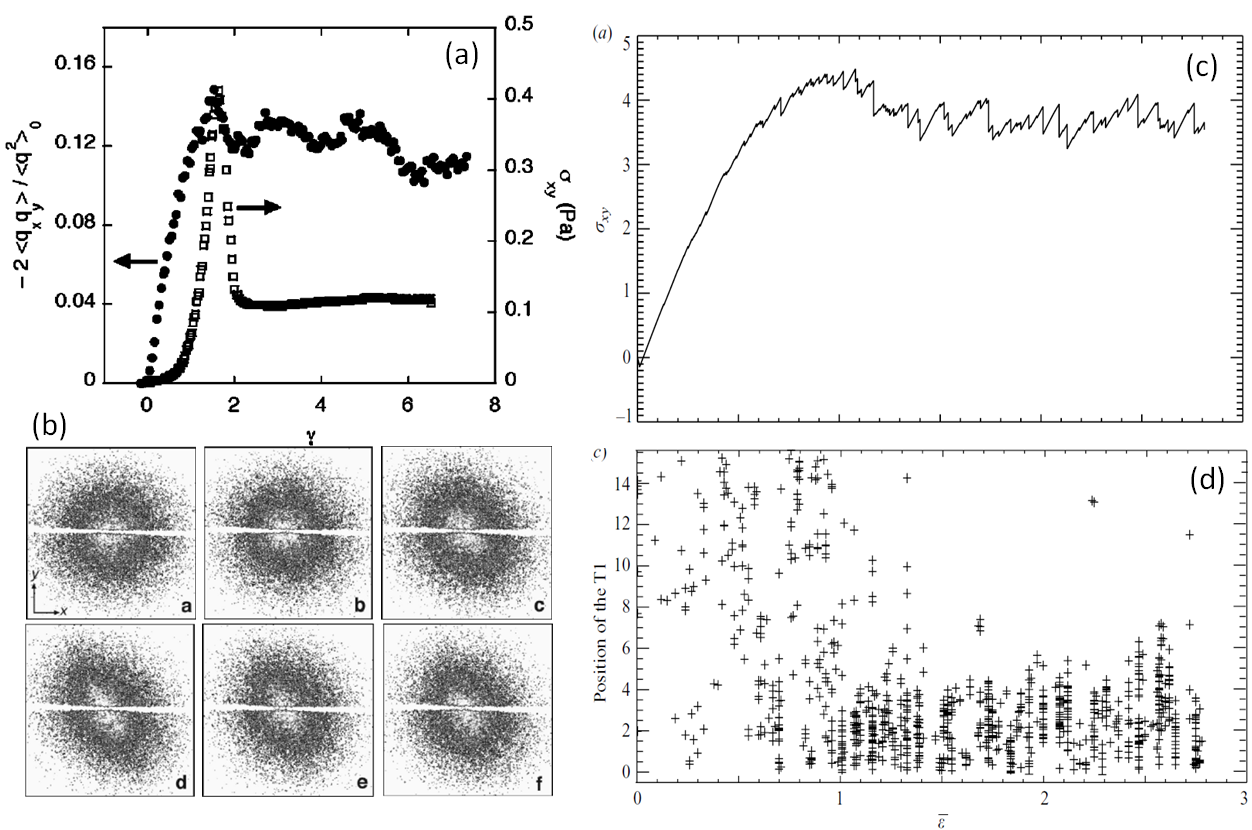}
	\caption{(a) Normalized off-diagonal component
of the second moment tensor of the dimensionless scattering
vector $\hat{q}$ weighted by the structure factor:
$<\hat{q}\hat{q}>$, and shear stress $\sigma_{xy}$ vs strain $\gamma$ during a shear start-up experiment ($\dot \gamma = 0.17$~s$^{-1}$) for a DLCA polystyrene gel ($\phi=10^{-3}$). (b) Contour plots of a representative cascade of scattering patterns collected during a start-up experiment ($\dot \gamma = 0.56$~s$^{-1}$)  with $t=0.1$ [a], 1.1 [b], 2.2 [c], 3.5 [d], 6.3 [e], and 8.3 s [f]. Maximum anisotropy is observed at $t\simeq3.5$~s. (a,b) extracted from \cite{Mohraz:2005}. (c) Evolutions of the shear stress and (d) positions of the T1 events in a foam sheared in a 2D Couette cell as a function of the applied strain during a shear start-up experiment. Data from numerical simulations, extracted from \cite{Kabla:2007}.}
	\label{fig:Chapter2FigureOvershoot}
\end{figure}

First, if the effect of boundaries and the possibility of wall slip (see Sect.~\ref{wallslip}) are ignored, the stress overshoot mainly depends on the value of $\dot\gamma$ \cite{Nguyen:1983}. Experiments performed on stabilized suspensions of silica particles \cite{Derec:2003}, Carbopol microgels \cite{Divoux:2011b} and attractive gels \cite{Koumakis:2011} report a power-law increase of the stress maximum with external shear, with an exponent $\nu$ in the range 0.1-0.5. This power-law scaling is captured by fluidity models \cite{Derec:2003}, Stokesian simulations \cite{West:1994} and Brownian dynamics simulations of particle gels \cite{Whittle:1997,Park:2013}, although the microscopic parameters controlling the exponent $\nu$ are still unclear. Power laws also contrast with the logarithmic increase reported for bidisperse Lennard-Jones mixtures for which the increase of the stress maximum can be interpreted in the framework of the Ree-Eyring viscosity theory \cite{Varnik:2004,Rottler:2005} and appears quite natural in the context of aging studies of glassy materials, in which slow aging dynamics very often leads to logarithmic time dependences.

Second, for a given applied shear rate, the stress maximum increases with the ``sample age,'' i.e. the waiting time $t_w$ between the preshear used to reset the fluid memory and the start-up of shear. The overshoot eventually disappears for waiting times shorter than $1/\dot \gamma$ \cite{Derec:2003,Letwimolnun:2007,Divoux:2011}. Such a behavior is well captured by the SGR model \cite{Fielding:2000} and the fluidity model \cite{Moorcroft:2011}, although both models predict a logarithmic increase of the stress maximum with $t_w$, whereas experimental results rather point to a weak power-law dependence.

Finally, regarding the local behavior of the fluid during shear start-up, recent experimental and numerical studies have shed new light on the nature of the stress maximum. In Brownian colloidal systems, the stress maximum coincides with the maximum structural anisotropy \cite{Mohraz:2005,Koumakis:2012} [see Fig.~\ref{fig:Chapter2FigureOvershoot}(a,b)]. For attractive gels, the stress maximum corresponds to the rupture of the gel network, while for dense hard-sphere-like systems, individual colloids experience an (apparent) superdiffusive motion as they are being pushed out of their cage by shear \cite{Zausch:2008,Koumakis:2012},
which can be readily interpreted in terms of a delayed onset of diffusive
behavior.
In the case of a jammed assembly of soft particles, the deformation is almost elastic and only a few rearrangements that are uniformly spatially distributed have been reported in foams \cite{Kabla:2007} [see Fig.~\ref{fig:Chapter2FigureOvershoot}(c,d)], while linear velocity profiles have been observed in microgels up to the stress maximum \cite{Divoux:2011}. These recent local approaches clearly show that shear start-up and more specifically the stress overshoot phenomenon are powerful tools to finely distinguish between various types of yield stress materials. Theoretical understanding of the full transient scenario is however still far from reach.

\subsubsection{Creep experiments}

In a creep experiment, a constant shear stress $\sigma$ is applied from time $t=0$ and the strain response $\gamma(t)$ is monitored. Although also
a shear start-up experiment, this protocol does not necessarily fluidize the
material which may remain solid, and the results may be qualitatively
distinct from those discussed in Sect. \ref{shear-start-up}.

For stresses applied above the yield stress, the material eventually flows, i.e. $\gamma(t)$ increases linearly with time, whereas for stresses lower than the yield stress, the material behaves as a solid and $\gamma(t)$ tends toward a constant. Equivalently, the shear rate $\dot\gamma(t)$ reaches a non-zero steady-state value in the former case, while it vanishes in the latter case.
Following the discussion in Sect. \ref{shear-start-up}, the yield stress
measured by this approach should again
provide an estimate of the static yield stress.

This ``bifurcation'' between a finite steady-state viscosity and an apparently infinite viscosity in principle provides a well-defined estimation of the yield stress as the critical stress separating these two regimes \cite{Coussot:2002b,Coussot:2002c,DaCruz:2002,Moller:2006,Coussot:2006} [see Fig.~\ref{fig:Ch2Fig3}(c)]. This method is cumbersome, however, as the yield stress is obtained by dichotomy, and for each experiment the time for the material to flow increases as the applied shear stress gets closer to the yield stress \cite{Moller:2009a}. The question of deciding whether a steady state is reached and whether the system eventually flows or not becomes even more important in the case of very long transients and of so-called ``delayed yielding,'' where no apparent flow can be detected for long times before the material finally yields \cite{Uhlherr:2005,Magda:2009,Gibaud:2010,Chaudhuri:2013}.

 \begin{figure}[t]
\centering
\includegraphics[width=8.5cm]{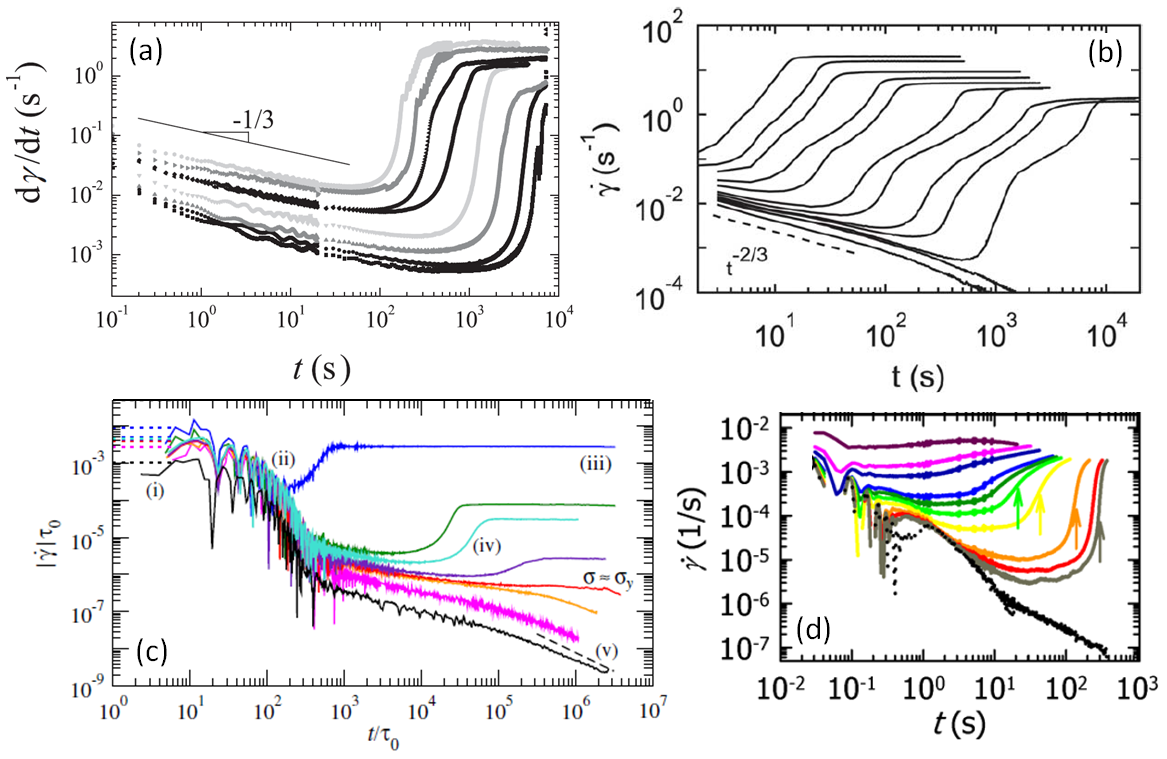}
\caption{Shear rate responses vs time for creep experiments at different imposed shear stresses in various materials: (a)~polycrystalline hexagonal columnar phase, extracted from \cite{Bauer:2006}, (b)~Carbopol microgel, extracted from \cite{Divoux:2011b}, (c)~core-shell PS-PNIPAM particle glass, extracted from \cite{Siebenburger:2012a}, and (d)~carbon black gel at 8~\% wt, extracted from \cite{Sprakel:2011}.}
\label{Chapter3FigureCreep}
\end{figure}	

Besides the determination of the yield stress, the transient strain or shear rate responses also provide potentially rich information on the physical processes at play in soft solids under constant stress. In particular, a robust feature of creep responses prior to fluidization is a power-law decrease of the shear rate [see Fig.~\ref{Chapter3FigureCreep}] that strongly resembles the ``Andrade creep'' reported for hard solids \cite{Andrade:1910}, which has been attributed to collective dislocation dynamics \cite{Miguel:2002,Csikor:2007,Miguel:2008}. Power-law creep has been reported for cellulose gels \cite{Plazek:1960} and more recently for various amorphous soft solids such as polycrystalline surfactant hexagonal phases \cite{Bauer:2006}, Carbopol microgels \cite{Divoux:2011b,Lidon:2016}, core-shell p-NIPAM colloidal particles \cite{Siebenburger:2012a}, thermo-reversible protein gels \cite{Brenner:2013} and colloidal glasses \cite{Sentjabrskaja:2015}. Yet, Andrade-like creep remains mostly unexplored in soft materials such as yield stress fluids. Local velocimetry suggests that the strain field remains macroscopically homogeneous during this first regime \cite{Divoux:2011b,Grenard:2014}. Still, characterizations at finer --ideally microscopic-- scales are needed to unveil the presence of plasticity or microcracks during the initial loading phase and to make a clear link between the physical mechanisms at play in the creep of ordered solids and of disordered soft materials. New insights can also be gained by adapting recent numerical models to creep situations \cite{Fusco:2014,Colombo:2014}.

Finally, for stresses above the yield stress, the initial power-law creep is followed by a gradual acceleration up to an abrupt fluidization of the material that later reaches a steady state. The dynamics associated to fluidization will be discussed in Sect.~\ref{hottopic_time scales} together with the characteristic time scales involved in the yielding process. For stresses below the yield stress, the interplay between creep deformation and aging leads to long-time strain responses that are more complex than pure power-laws and strongly depend on the sample age, as reported for laponite clay suspensions \cite{Negi:2010c,Baldewa:2012} and star glassy polymers \cite{Christopoulou:2009}.

\subsubsection{Large-amplitude oscillatory shear experiments}

So far, the yielding transition has only been considered from the point of view of a {\it steady} external shear. Yet, the solid-like vs liquid-like behavior of a complex material can also be quantified through {\it oscillatory} shear experiments. By imposing a sinusoidal shear strain of amplitude $\gamma_0$ and pulsation $\omega$, given by $\gamma (t) = \gamma_0 \; \sin(\omega t)$, and measuring the corresponding stress response $\sigma(t)$, the storage ($G'$) and loss ($G''$) moduli can be defined from the amplitudes of the stress response that are respectively in phase and in quadrature with $\gamma (t)$ at the excitation pulsation $\omega$ \cite{Ferry:1980}. The solid-to-liquid transition of a yield stress material can thus be probed by progressively increasing the amplitude $\gamma_0$ of the oscillatory strain. At low strain amplitudes, the solid-like material is elastically deformed and the storage modulus $G'$ remains roughly constant and much larger than the loss modulus $G''$. This corresponds to the linear regime of deformation referred to as ``small-amplitude'' oscillatory shear (SAOS). At larger strain amplitudes, the material response gets nonlinear. Under such ``large-amplitude'' oscillatory shear (LAOS), $G'$ typically decreases, then crosses $G''$ and becomes much smaller than $G''$ as the material becomes liquid-like [see Fig.~\ref{fig:Ch2Fig3}(d)].

\begin{figure}[t!]
	\centering
	\includegraphics[width=8.5cm]{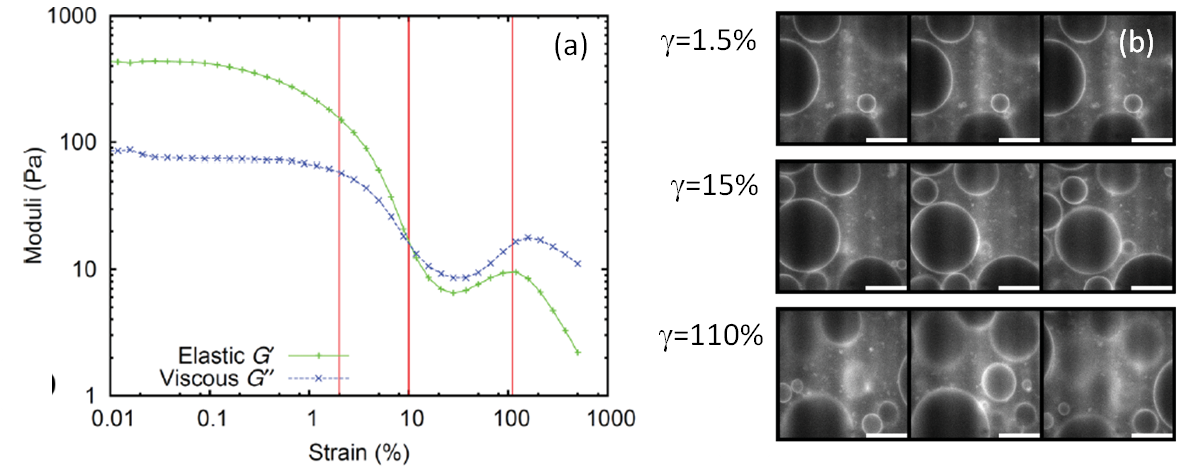}
	\caption{(a) Evolution of the shear moduli of a Pickering emulsion stabilized by silica colloids during a LAOS strain amplitude sweep. The volume fraction of the oil is 65\%. (b) Confocal images of the emulsion during shear taken 40 mm into the sample to avoid wall effects and obtained at different strains during the strain sweep. Scale bars correspond to 20~$\mu$m. For $\gamma_0<0.10$, the droplets slide along each other but remain trapped in the cages formed by their neighbors. For $\gamma_0\simeq 0.10$, the moduli intersect and the droplets can be seen to move irreversibly, although their displacement over a period is much less than their diameter. For $\gamma_0>0.30$, $G'$ and $G''$ increase due to jamming, which results in apparent shear-thickening, and the droplets move rapidly during each period, over distances larger than their own diameter. Extracted from \cite{Hermes:2013}.}
	\label{fig:Chapter2FigureLAOS}
\end{figure}

LAOS experiments allow for a number of different estimations of the yield point. By plotting $G'$ and $G''$ as functions of either the strain amplitude $\gamma_0$ or the shear stress amplitude $\sigma_0$, the stress amplitude $\sigma_{0y}$ at which the material yields may be given by the point at which $G'= G''$, which has been called ``characteristic modulus'' by \cite{Larson:1999} [see Figs.~\ref{fig:Ch2Fig3}(d) and \ref{fig:Chapter2FigureLAOS}(a)], or by the intersection of power-law fits of the moduli behaviors well above and well below the yielding point \cite{Rouyer:2005}. Alternatively, $\sigma_{0y}$ can be estimated by plotting $\sigma_0$ vs $\gamma_0$ from the intersection between a linear behavior with slope $G'$ at low strains and a power-law fit at high strains \cite{Mason:1996a,SaintJalmes:1999}.

Clearly, contrary to steady-shear measurements, all LAOS estimates of the yield stress as $\sigma_{0y}$ involve the additional time scale $1/\omega$ and thus do not comply with the definition of Eq.~(\ref{eq.ysdef}) unless vanishingly small frequencies are considered. Moreover, as already addressed in several rheology reviews \cite{Wilhelm:2002b,Hyun:2011}, the response to LAOS is intrinsically nonlinear and needs to be analyzed considering the full spectrum of strain or stress harmonics rather than the sole fundamental frequency through $G'$ and $G''$ only. The various estimates of $\sigma_{0y}$ should depend on both $\omega$ and the harmonic content of the stress or strain response, and there is no particular reason why they should coincide and correspond to the dynamic yield stress inferred from steady-state measurements. Finally, in the case of strongly time-dependent materials, the estimate of $\sigma_{0y}$ is most likely to depend on the details of the LAOS ramp protocol, as illustrated in Fig.~\ref{fig:Ch2Fig3}(d). Wall slip and/or bulk heterogeneous flows may complicate yielding under oscillatory shear even more \cite{Walls:2003,Gibaud:2010,Perge:2014b,Gibaud:2016}.

Interest in LAOS has grown, leading to a surge in the number of  experimental and theoretical studies over the last decade. First, LAOS has been used to unveil a striking difference between attractive and repulsive colloidal glasses. Whereas the elastic modulus decreases monotonically in dense hard-sphere-like systems, attractive glasses display a two-step yielding, which results from the existence of two distinct microscopic length scales in the system: the adhesion range (responsible for initiating phase separation) and the cage size in the dense glassy phase (responsible for the dynamic arrest) \cite{Pham:2006,Laurati:2011,Chan:2012,Koumakis:2013}. Such a two-step scenario has been observed through strain-step experiments \cite{Koumakis:2011}, but its interpretation still requires full confirmation from direct local investigations. Of course, this two-step yielding process immediately leads to the question whether a yield stress or even two yield stresses should be defined.

Second, physical insights into the microscopic dynamics under LAOS have been gained by coupling oscillatory shear to other characterization techniques, such as structural measurements or local tracking of particle motion. The ``light scattering (LS) echo'' technique has allowed the quantification of the global amount of irreversible rearrangements \cite{Hebraud:1997,Petekidis:2003,Laurati:2014}. Direct optical imaging of the microstructure has recently been used to assess the transition to irreversibility with emphasis on the physical properties of rearrangements such as their correlation length \cite{Nagamanasa:2014}, the presence of dynamical heterogeneities \cite{Knowlton:2014} and cage breaking in repulsive vs attractive systems \cite{Hermes:2013} [see Fig.~\ref{fig:Chapter2FigureLAOS}(b)]. Along the same lines, time-resolved neutron and x-ray scattering now allows one to follow the evolution of shear-induced anisotropy in colloidal gels during one single LAOS cycle \cite{MinKim:2014,Kim:2014,Rogers:2014}. Such measurements come as a crucial complement to recently proposed nonlinear analyses of rheological data during one oscillation cycle \cite{Ewoldt:2008,Rogers:2011a,Ewoldt:2013,Dimitriou:2013,Dimitriou:2014}, and can be used as an additional tool to study the yielding behavior and hence the value of the yield stress.

Third, from a more theoretical point of view, recent advances in modeling and simulation of LAOS flows have also led to significant progress in unveiling the physical importance of caging effects in the yielding of both hard-sphere glasses \cite{Brader:2010,Koumakis:2013} and dense assemblies of soft particles \cite{Mohan:2013b}, and this situation was analysed within a mode-coupling 
approach~\cite{Seyboldt:2016}.
The idea has emerged that the yielding transition corresponds to a change in the dynamics at the microscopic scale between reversible particle trajectories at small applied stress, and a chaotic dynamics beyond the yielding point \cite{Fiocco:2013,Regev:2013,Kawasaki:2015}.
Whereas early numerical work seemed to predict a continuous phase transition between the two regimes characterized by power law divergences \cite{Brader:2010,Perchikov:2014}, more recent work \cite{Kawasaki:2015}
put forward the idea that the transition is
indeed sharp but discontinuous, akin to a non-equilibrium first-order
phase transition.
In experiments, contrasting evidences have been reported on this point
\cite{Hermes:2013,Knowlton:2014,Denisov:2015}.
Therefore, the nature of the yielding transition under oscillatory shear
remains to be fully elucidated.

\subsubsection{Non-viscometric flows}

Many flows encountered in practice are not simple viscometric flows. A typical example is that of a sphere falling through a yield stress fluid; this has often been used as a benchmark, e.g. for numerics. The fluid around the sphere will be set in motion because of the stress exerted by the falling sphere, but the fluid far away will remain motionless; the question is where the ‘yield surface’ (i.e., the transition from the flowing to the non-flowing material) is localized in space. Experiments on the flow of yield stress fluids around falling spheres far from any boundaries have revealed the location of the yield surface, but have also shown that the usual constitutive equations are unable to describe this situation. In a number of experiments, the loss of foreaft symmetry \cite{Gueslin:2006,Putz:2008} was observed. For thixotropic fluids \cite{Gueslin:2006}, this can easily be understood: where the sphere has passed through the material, it has liquefied. However similar observations made on carbopol gels \cite{Putz:2008} cannot be explained by properly invariant 3-D generalizations of classical models like the Bingham and Herschel-Bulkley fluids \cite{Putz:2008}. Symmetry breaking may in fact be an elastic effect: the resemblance between the yield stress fluid flow around a sphere and the flow of viscoelastic polymer solutions has been noted. The notion of combining viscoelasticity and yield stress behavior has spurred Saramito \cite{Saramito:2007} and de Souza and Mendes
\cite{SouzaMendes:2009,SouzaMendes:2011} to attempt to add elasticity to the usual visco-plastic models and developed properly invariant continuum 
elasto-viscoplastic constitutive equations. There has been some success in simulating the breaking of flow foreaft symmetry in 
\cite{Cheddadi:2011,Fonseca:2013,dosSantos:2014}.
Notably, Fraggedakis and coworkers \cite{Fraggedakis:2016} 
have successfully simulated the
flow of Carbopol solutions past isolated spheres by incorporating the
plastic “back pressure” \cite{Dimitriou:2013} into the Saramito model.

\subsection{Wall slip in yield stress materials}
\label{wallslip}

As briefly mentioned above, in the vicinity of a {\it smooth} solid boundary, the velocity of a yield stress fluid ($v_{\rm sample}$) may differ from the velocity of the boundary ($v_{\rm wall}$). One may either observe $v_{\rm sample} < v_{\rm wall}$ (e.g. in the case of a shearing device driven at constant velocity $v_{\rm wall}$) or $v_{\rm sample} > v_{\rm wall}=0$ (close to a fixed surface or in capillary or channel flows). In both cases, the apparent discontinuity in velocity at the wall is caused by a thin and highly sheared region adjacent to the wall of lower viscosity than the bulk material.
This phenomenon, referred to as {\it apparent wall slip} or more often simply as {\it wall slip} in the literature, has been first described as an artifact that experimentalists should get rid of in order to avoid misinterpretating rheological measurements \cite{Barnes:1995}.
Although elucidating the exact microscopic structure of the lubrication layer remains an experimental challenge due to its small width (typically smaller than 1~$\mu$m), very high local shear rate, and proximity to a solid boundary,
it is presumably composed of pure solvent in the case of colloidal gels or suspensions \cite{Hartman:2002,Hartman:2004} or of a film of continuous phase in emulsions \cite{Princen:1985}.
The emergence of local techniques to quantify slip velocities have brought about a better understanding of the behavior of yield stress materials near boundaries, allowing the development of successful microscopic models in the case of dense assemblies of soft particles \cite{Seth:2011}. In the case of attractive gels, a recent body of evidence strongly suggests that the dynamics close to a wall may not be easily decoupled from the bulk dynamics and that wall slip is not merely a rheometric complication, as recently pointed out by \cite{Buscall:2010}.
This section addresses the various issues related to wall slip in yield stress fluids in light of such recent developments.

\begin{figure}
\centering
\includegraphics[width=8.5cm]{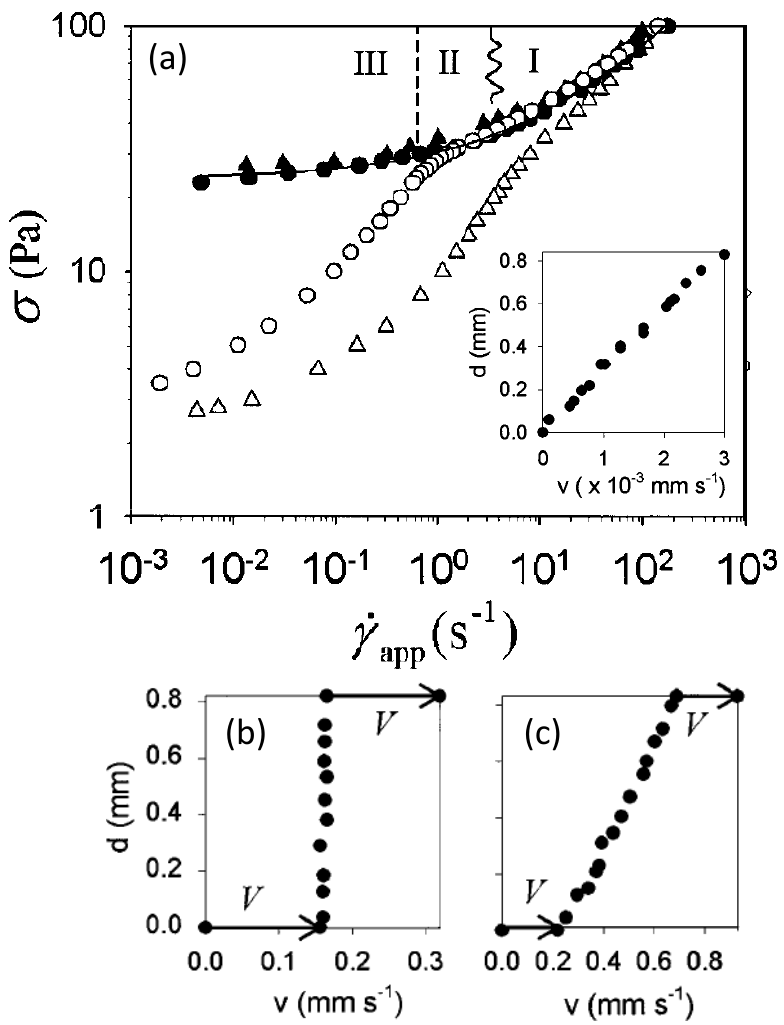}
\caption{(a) Stress $\sigma$ vs the apparent shear rate $\dot \gamma_{\rm app}$ for a microgel paste ($\circ$, $\bullet$) and an emulsion ($\bigtriangleup$,$\blacktriangle$) of packing fraction $\phi \simeq 0.77$ obtained in a cone-and-plate device for smooth (open symbols) and rough (closed symbols) surfaces. Regimes I to III refer to microgel slip behavior discussed in the text. The inset shows the velocity profile measured with rough surfaces for $\sigma/\sigma_y=1.05\pm 0.1$. (b,c) Velocity profiles measured with smooth surfaces for $\sigma/\sigma_y=0.9\pm 0.1$ (b) and $1.3\pm 0.1$ (c). Adapted from \cite{Meeker:2004a}. }
\label{Chapter3FigureWallSlip}
\end{figure}

 \subsubsection{Impact on flow curve measurements}

In the presence of wall slip, the measured {\it apparent} shear rate overestimates the {\it true} shear rate within the material (or, correspondingly, the strain indicated by a rheometer overestimates the true deformation experienced by the bulk of the material). Consequently, the apparent flow curve is shifted to higher shear rates compared to the actual constitutive equation of the bulk material. In general, at low shear rates, where wall slip effects are most pronounced, the apparent flow curve displays a kink and/or a plateau at a stress lower than the yield stress estimated in the absence of wall slip [see Fig.~\ref{Chapter3FigureWallSlip}(a) for examples]. This signature has been reported in the literature as early as 1975 in the pioneering work of \cite{Vinogradov:1975} and, since then, for a broad range of yield stress materials including colloidal gels \cite{Buscall:1993,Mas:1994}, dense Brownian suspensions \cite{Ballesta:2008b,Ballesta:2012}, emulsions and microgels \cite{Meeker:2004a,Meeker:2004b}, and foams \cite{Marze:2008}. In particular, wall slip leads to significant deviations from the Herschel-Bulkley behavior at low shear rates, and the apparent flow curve is strongly surface-dependent in this limit \cite{Seth:2008}.

 \subsubsection{Physical origin of wall slip in yield stress fluids}

Direct flow visualization coupled to rheology has made it possible to go one step further in interpreting the apparent lower stress plateau in the presence of wall slip. The seminal work of \cite{Magnin:1990} on Carbopol microgels, coupling rheology to direct observations of the strain field, has inspired numerous subsequent studies coupling flow visualization to standard rheology \cite{Kalyon:1993,Aral:1994,Persello:1994}. Subsequently, combinations of rheology and other local measurement techniques, such as light scattering velocimetry in a Couette geometry \cite{Salmon:2003a} and particle tracking velocimetry in cone-and-plate \cite{Meeker:2004a,Meeker:2004b,Ballesta:2008b,Ballesta:2012,Paredes:2011} and plate-plate geometries, \cite{Seth:2012} have provided quantitative measurements of slip velocities (defined as $v_s=\vert v_{\rm sample}-v_{\rm wall}\vert$) and wall slip scenarios for the different yield stress materials. Let us first discuss the case of yield stress fluids composed of soft deformable particles before turning to rigid particles.

\begin{figure}[t]
\centering
\includegraphics[width=8.5cm]{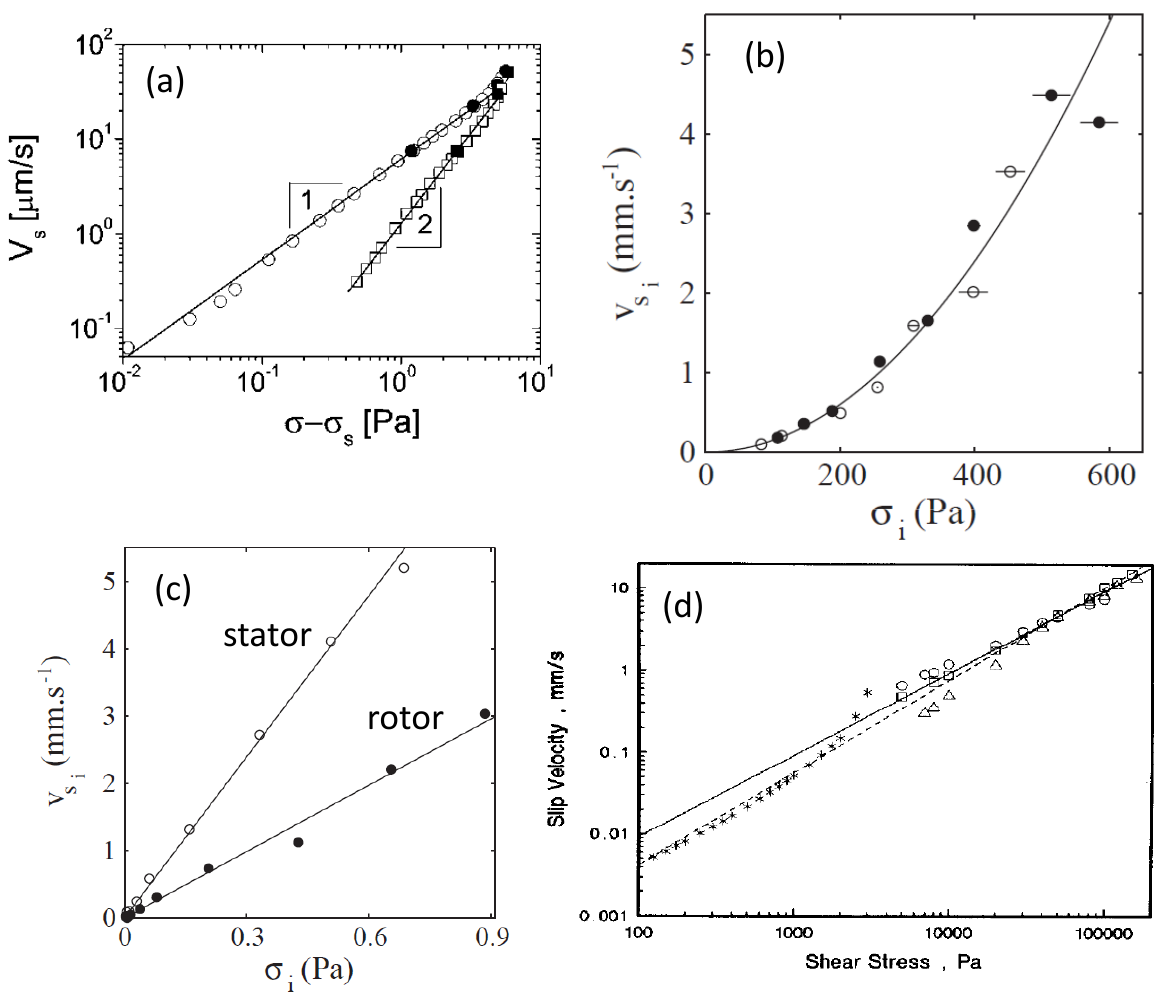}
\caption{(a)~Slip velocity vs excess stress in a dense emulsion for stresses below the yield stress in a plate-plate geometry. The top plate is either coated with a weakly adhering polymer surface ($\square, \blacksquare$) or a non-adhering glass surface ($\circ,\bullet$). The wetting properties of the boundary conditions strongly impact the behavior of the slip velocity. Extracted from \cite{Seth:2012}. (b,c) Slip velocity vs shear stress at the rotor ($\bullet$) and stator ($\circ$) for (b)~a dense emulsion $\phi=0.75$ and (c)~a dilute emulsion $\phi=0.2$. The slip velocity is linear in the dilute regime and quadratic for dense packing for stresses above the yield stress. Extracted from \cite{Salmon:2003a}.  (d)~Slip velocity vs shear stress in a suspension of ammonium sulfate particles in PBAN (poly(butadiene acrylonitrile acrylic acid) terpolymer). Data obtained in capillary flows with dies (extrusion nozzles) of various aspect ratios ($\circ,\square,\triangle$) and a plate-plate geometry ($\star$). The solid line corresponds to a linear behavior. Extracted from \cite{Yilmazer:1989}.}
\label{Chapter3FigureSlipvelocity}
\end{figure}

\paragraph{Wall slip in case of soft particles.}
In yield stress fluids made of soft particles, the solid behavior results from the tightly packed structure of deformable objects, and the lubrication layers that develop at smooth walls are intimately related to the particle deformability. For shear rates such that $\sigma < \sigma_y$ [see regime III in Fig.~\ref{Chapter3FigureWallSlip}(a)], the bulk remains unsheared and the apparent motion is entirely due to wall slip [see Fig.~\ref{Chapter3FigureWallSlip}(b)]. This situation is referred to as ``total'' wall slip or ``plug-like'' flow. In this regime, $v_s$ has been shown to increase as a power-law of the excess stress (i.e. the stress at the wall minus the apparent yield stress $\sigma_s$ inferred from the extrapolation of the flow curve to vanishing shear rates in the smooth geometry): $v_s \propto (\sigma-\sigma_s)^p$, where the exponent $p$ does not depend significantly on the packing fraction but is strongly influenced by the chemical nature of the walls. The slip velocity $v_s$ displays a nearly quadratic scaling, i.e. $p \simeq 2$, in the case of attractive/non-wetting surfaces, whereas $p = 1$ for repulsive and/or wetting walls \cite{Seth:2008,Seth:2012} [see Fig.~\ref{Chapter3FigureSlipvelocity}(a)]. Both exponents $p=1$ and $p=2$ have been successfully described at the scale of single particles by elastohydrodynamic lubrication theory as the result of a balance between bulk osmotic pressure and viscous dissipation taking place in the thin lubrication layer that separates the squeezed particles from the wall \cite{Meeker:2004b,Seth:2008}. The upper limit of this total wall-slip regime generally correlates well with the stress drop or ``kink" on steady-state macroscopic measurements.

For larger shear rates such that $\sigma > \sigma_y$, the bulk material is sheared but wall slip remains significant, at least for $\sigma \gtrsim \sigma_y$ [see regimes I and II in Fig.~\ref{Chapter3FigureWallSlip}(a) and Fig.~\ref{Chapter3FigureWallSlip}(c)]. In this partial wall-slip regime, it appears that the slip velocity scales as a power law of the slip stress only (i.e. stress at the wall). Here, the physical picture is much less clear. Both the absolute value of $v_s$ and the exponent $p$ depend strongly on the geometry. For instance, in similar systems, $p\simeq1$ has been reported for a  plate-plate geometry \cite{Seth:2012} while $p\simeq 2$ for a smooth Couette cell \cite{Salmon:2003a} [see Fig.~\ref{Chapter3FigureSlipvelocity}(b)] and for rough microchannels \cite{Geraud:2013}. We note that in this regime, the influence of the packing fraction in the glassy state and the impact of the chemical nature of the walls have not been systematically explored. Yet, as seen in Fig.~\ref{Chapter3FigureSlipvelocity}(b,c), one finds $p\simeq 2$ above jamming to $p\simeq 1$ for low packing fractions where soft particles are no longer compressed against each other and the yield stress vanishes. The exponent $p=1$ is also found in liquid-like suspensions of rigid particles as discussed in the following paragraph.
From a recent study on soft thermoresponsive particles conducted at different temperatures in a Couette cell, it appears that the scaling of the slip velocity depends mainly on the packing fraction \cite{Divoux:2015}. Nonetheless, more experiments in other geometries are required to provide a truly universal scaling to the slip velocity across the jamming transition.

\paragraph{Wall slip in case of hard particles.}
Concerning wall slip in suspensions of hard particles, a great deal of work has been done on non-Brownian systems, which corresponds to large P\'eclet numbers. Over a wide range of packing fractions, wall slip is associated with a depletion layer near the wall, which thickness depends on the particle size and decreases roughly linearly with increasing bulk packing fraction \cite{Kalyon:2005}. The latter result may not hold for polydisperse samples \cite{Soltani:1998} and appears to be affected by migration effects \cite{Jana:1995}, which shows that the detailed mechanism for slip in non-Brownian systems is not fully understood. Nonetheless, the slip velocity at the wall scales linearly with the slip stress in a remarkably robust fashion \cite{Yilmazer:1989,Aral:1994,Jana:1995,Soltani:1998} [see Fig.~\ref{Chapter3FigureSlipvelocity}(d)]. The chemical properties of both the particle surface and boundary conditions seem to affect wall slip \cite{Kao:1975} although their quantitative impact on $v_s(\sigma)$ remains to be determined.

For Brownian hard spheres, the thickness of the depletion layer depends weakly on the P\'eclet number \cite{Hartman:2004}, and decreases for increasing packing fractions \cite{Ballesta:2008b}. The slip velocity scales linearly with the slip stress for both dilute and glassy assemblies. Moreover, wall slip in glassy samples is characterized at low shear rates by a stress-kink on the macroscopic flow curve together with plug-like velocity profiles \cite{Ballesta:2012}. The latter result is strikingly similar to the one reported for soft particles in contact with non-adhering surfaces in Fig.~\ref{Chapter3FigureWallSlip}(a), suggesting that a common mechanism might be at work. Attractive colloidal gels display the same phenomenology as glassy suspensions over a broader range of packing fractions, down to very low values of $\phi$. However, both the kink and wall slip tend to disappear as the polydispersity is increased \cite{Ballesta:2013}. Particle migration is also more likely to play a major role in these yield stress fluids with low packing fractions by promoting concentration gradients and/or segregation. One can thus anticipate that wall slip in attractive gels originates from the combined effects of migration and polydispersity, with a strong dependence on the shearing geometry, including stress gradients.

 \begin{figure}[t]
\centering
\includegraphics[width=8.5cm]{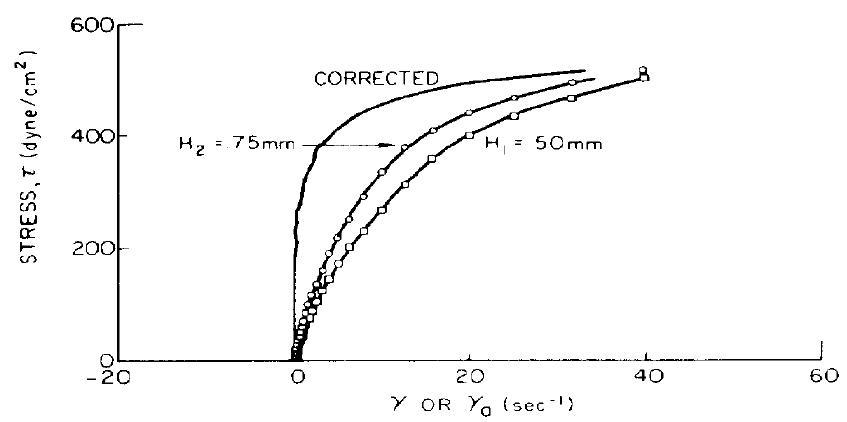}
\caption{Shear stress vs shear rate for a dense emulsion ($\phi=0.923$) measured in a plate-plate geometry with smooth boundary conditions for two different gap sizes (500 and $750~\mu$m, respectively). The stress vs the shear rate computed from the method developed by Mooney and extended by Yoshimura and Prud'homme displays a yield stress, while this is not obvious from the raw measurements. Extracted from \cite{Yoshimura:1988a}.}
\label{Chapter3FigureMooney}
\end{figure}	

\subsubsection{Dealing with wall slip in practice}

\begin{figure}[t]
\centering
\includegraphics[width=8.5cm]{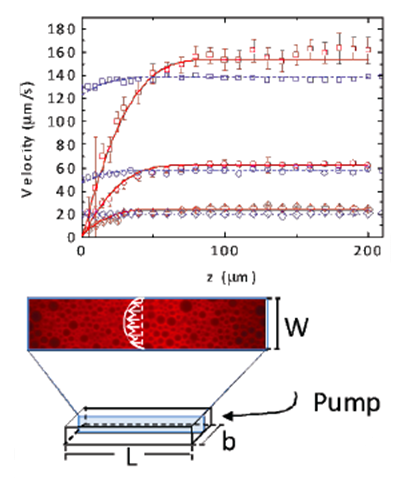}
\caption{Velocity profiles of a dense emulsion flowing in a rectangular microchannel (gap $w=400$~$\mu$m). Images obtained by confocal microscopy. Velocity profiles in blue (dashed lines) correspond to smooth boundary conditions treated with a piranha solution. The oil droplets experience wall slip. Velocity profiles in red (continuous lines) correspond to smooth, silanized boundary conditions. The oil droplets stick to the surface, creating an effectively rough boundary condition. Flow rates: 0.2, 0.5 and 1.2 10$^{-2}$~mL/min. From \cite{Paredes:2015}.}
\label{Wallsliplocal}
\end{figure}

Two types of practical approaches towards wall slip have been proposed in the literature: either to quantify the effect of wall slip from the experimentally determined flow curve, or to eliminate it. An elegant solution due to Mooney \cite{Mooney:1931} and further extended by \cite{Yoshimura:1988a}, \cite{Kiljanski:1989} and \cite{Wein:1992} consists of determining the relationship between the apparent shear stress and rate for gaps of different sizes. Combining at least two measurements and assuming that ($i$) the slip velocity is a function of stress only and ($ii$) slippage is the same at both walls, one can recover the constitutive relationship $\sigma(\dot \gamma)$ corrected for wall slip (see Fig.~\ref{Chapter3FigureMooney}). This method has been applied to various yield stress fluids, including emulsions \cite{Yoshimura:1988a}, microgels \cite{Meeker:2004b}, and dense suspensions \cite{Hartman:2002,Hartman:2004, Yilmazer:1989,Kalyon:1993,Kalyon:2005} although the two assumptions on which it relies have been verified by local measurements of slip velocities only in a few cases \cite{Salmon:2003a,Meeker:2004b}.

To prevent wall slip, the nature of the wall needs to be modified. The use of rough boundary conditions allows to properly determine constitutive equations without wall slip \cite{Vinogradov:1975}. The roughness of the wall has been tuned from a few microns to hundreds of microns by using sandblasted surfaces \cite{Buscall:1993}, grooved surfaces \cite{Magnin:1990}, serrated tools \cite{Nickerson:2005}, or by gluing waterproof sand-paper \cite{Seth:2008,Seth:2012} or a monolayer of particles on the cell walls \cite{Isa:2007}. The accepted paradigm is that the roughness of the surface should be comparable to the size of the microstructure, since a lower roughness would not be efficient, and a higher roughness, including vane cup geometries, may trigger secondary flows \cite{Ovarlez:2011}. In that respect, recent attempts to systematically explore the effect of the roughness-to-particle-size ratio \cite{Mansard:2012} look promising in order to go beyond empirical knowledge. Last but not least, the chemical nature and wetting properties of the walls can be also tuned to force the adhesion of the material even for smooth interfaces. This has been successfully achieved for colloidal silica gels at low deformations using hydrophobic boundaries \cite{Walls:2003} and for aqueous microgels and oil-in-water emulsions using silicon boundaries \cite{Seth:2008,Seth:2012} and chemical treatment of PMMA \cite{Metivier:2012} or glass surfaces \cite{Paredes:2015}, as illustrated in Fig.~\ref{Wallsliplocal}.

\section{Steady-state flow dynamics of yield stress fluids: flow curves and shear banding}
\label{flowdyn}

This section is devoted to the dynamics of yield stress materials in the case where a stress above the yield stress is applied. After briefly reviewing methods to experimentally distinguish between simple and thixotropic yield stress fluids, we examine the current interpretations and models for the steady-state shear-banding flows generally observed in thixotropic materials. We close this section with two emerging topics that have emerged within the last few years on the flow of yield stress fluids under confinement and the time scales involved in transient regimes of yield stress fluid flow.

\subsection{Flow curves of simple and thixotropic yield stress fluids}
\label{distinction}

As discussed in Sect.~\ref{thixo}, yield stress fluids can be broadly divided into ``simple'' yield stress fluids (microgels, dense emulsions and foams) and thixotropic yield stress fluids (clays, fiber suspensions and colloidal gels) \cite{Moller:2009b,Bonn:2009,Ovarlez:2009,Ovarlez:2013b}. Here, we briefly review how the distinction can be made experimentally, before turning to most recent ideas.

\subsubsection{Distinction between flow curves}

\begin{figure}
\centering
\includegraphics[width=8.5cm]{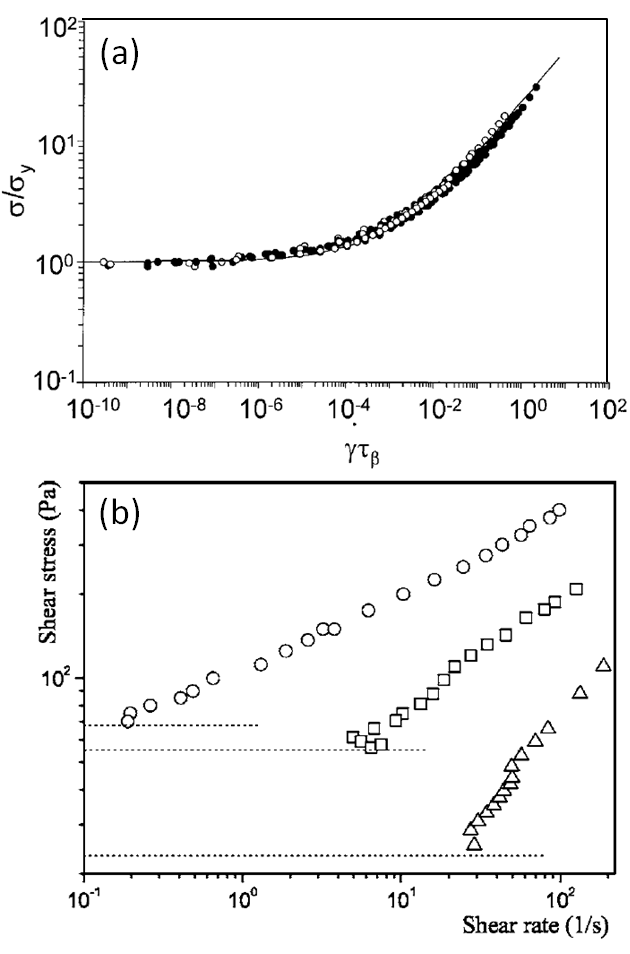}
\caption{Representative flow curves for (a) a simple yield stress fluid, here microgels of different cross-link densities and concentrations. The solid line is the equation $\sigma/\sigma_y=1+(\dot{\gamma}\tau_{\beta}/\gamma_0)^{0.45}$, where $\tau_{\beta}$ is the fluid relaxation time [extracted from \cite{Cloitre:2003}]. (b) Three different thixotropic materials. From top to bottom: a hair gel, a commercial mustard, and a bentonite suspension. Note that each of these materials displays a minimum shear rate $\dot{\gamma}_c$ below which no steady flow is possible. The horizontal dotted lines indicate the yield stresses of the different materials. Extracted from \cite{Coussot:2006}.}
\label{flowcurve}
\end{figure}

Steady-state flow curves $\sigma (\dot \gamma)$ can be used to distinguish between the two types of yield stress materials \cite{Moller:2009b}:
\begin{itemize}
\item Simple yield stress fluids exhibit a {\it continuous} and monotonic constitutive equation, which is well fitted by the phenomenological Herschel-Bulkley law $\sigma=\sigma_y+A\dot\gamma^n$ [Fig.~\ref{flowcurve}(a)]. As a consequence, whatever the applied shear rate, even in the limit of vanishing values, there is always a finite shear stress above $\sigma_y$ at which the material flows. Conversely, whatever the applied shear stress above $\sigma_y$, there is always a finite shear rate reached by the material. 
\item In contrast, thixotropic yield stress fluids are characterized by a {\it discontinuous} underlying flow curve [Fig.~\ref{flowcurve}(b)] with a pronounced time-dependence that is the definition of thixotropy. Indeed, in addition to a yield stress $\sigma_y$, these materials are also characterized by a critical shear rate, $\dot \gamma_c$, below which they cannot flow steadily in homogeneous conditions when imposing the shear rate \cite{Coussot:2002a}.
\end{itemize}

At this stage, it is tempting to draw an analogy with equilibrium phase transitions, where such a discontinuous flow curve would correspond to a first-order solid-to-fluid transition while a continuous flow curve would correspond to a second-order solid-to-fluid transition. However, in practice it may be difficult to discriminate between these two categories on the sole basis of the steady-state flow curve \cite{Dennin:2008}. 
This is because the flow may become unstable and heterogeneous at low imposed shear rates ($\dot\gamma<\dot\gamma_c$), leading to an {\it apparent} flow curve that does not necessarily reflect the unstable constitutive behavior of the material.

\subsubsection{Existence of a ``viscosity bifurcation''}

\begin{figure}
\centering
\includegraphics[width=8.5cm]{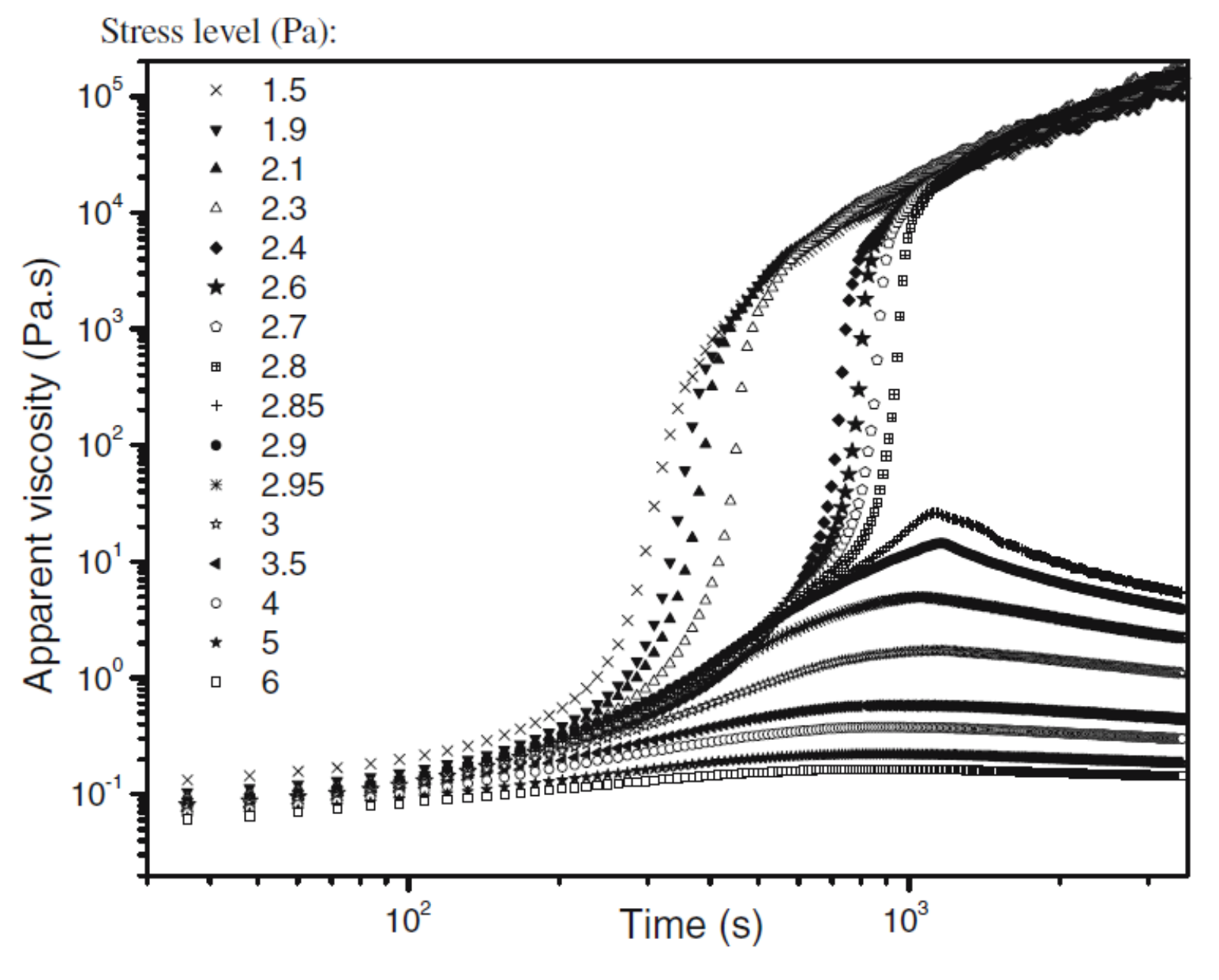}
\caption{Temporal evolution of the apparent viscosity $\eta = \sigma/\dot \gamma$ of a drilling mud for various applied shear stresses below and above the yield stress $\sigma_y\simeq 2.82$~Pa. The drastic change of behavior within a range of less than 0.1~Pa around the yield stress illustrates the viscosity bifurcation scenario. Extracted from \cite{Ragouilliaux:2006}.}
\label{viscobifurc}
\end{figure}

A practical consequence of the existence of a critical shear-rate $\dot \gamma_c$ is the striking  avalanche-like behavior of thixotropic yield stress fluids under an applied shear stress in the vicinity of the yield stress \cite{DaCruz:2002,Coussot:2002b,Coussot:2002c}. Within a narrow range of stresses of a few pascals around the yield stress, two very different macroscopic responses can be observed [see Fig.~\ref{viscobifurc}]: for $\sigma<\sigma_y$, the material deforms and progressively stops flowing as the viscosity takes up ever-increasing values, whereas for $\sigma>\sigma_y$ the material experiences an abrupt fluidization, characterized by an increase of the shear rate up to a finite steady-state value, which recalls avalanche behavior.

The above behavior is characteristic of thixotropic yield stress fluids and has been coined a ``viscosity bifurcation'' in the sense that the yield stress separates two regimes characterized by widely different values of the steady-state viscosity.
The terminology is, however, somewhat unfortunate since a divergence of the viscosity is also expected for $\sigma\rightarrow \sigma_y^+$ or $\dot\gamma\rightarrow 0$ in simple yield stress fluids. For these materials, the divergence is continuous and any (arbitrarily large) value of the final viscosity can be reached close to the yield stress without any forbidden shear rate range. In the language of bifurcations, yielding in simple yield stress fluids would therefore be analogous to a {\it supercritical} bifurcation while in thixotropic yield stress fluids it would correspond to a {\it subcritical} bifurcation.

\subsubsection{Consequences for local measurements}
\label{local}

Within the last two decades, a number of different tools have emerged to measure the local velocity field within standard rheological geometries, including particle tracking, dynamic light scattering, magnetic resonance or ultrasonic imaging \cite{Manneville:2008,Bonn:2008,Gallot:2013,Manneville:2004a,Salmon:2003b,Callaghan:2008,Besseling:2009}. As reviewed elsewhere \cite{Ovarlez:2013b}, local velocity profiles under shear allow to make a clearer distinction between simple and thixotropic yield stress materials. It has been shown that, as expected from their monotonic macroscopic rheology, simple yield stress fluids display homogeneous velocity profiles in steady state, at least in experimental geometries with small enough stress heterogeneity. In this case, the {\it local} rheology, given by the local shear stress $\sigma(\mathbf{r})$ as a function of the local shear rate $\dot \gamma (\mathbf{r})$ derived from the velocity $v(\mathbf{r})$, perfectly matches the global rheology \cite{Divoux:2012,Ovarlez:2013b}. 
In the case of wide-gap geometries, $\sigma(\mathbf{r})$ may vary so much that it falls below the yield stress. This leads to a heterogeneous flow where a solid region characterized by a pluglike flow (where $\sigma(\mathbf{r})<\sigma_y$) coexists with a flowing region (where $\sigma(\mathbf{r})>\sigma_y$). Similar plug-like flow is observed in the case of channel flows of simple yield stress fluids, where the local stress necessarily vanishes in the center of the channel \cite{PerezGonzalez:2012,Poumaere:2014}. Such a shear localization due to large stress heterogeneity does not contradict the continuous solid--fluid transition of simple yield stress fluids. It is now clearly distinguished from the intrinsic shear localization, referred to as ``shear-banding,'' observed in thixotropic yield stress fluids and discussed in the next paragraph \cite{Ovarlez:2009,Ovarlez:2013b}.
Finally, even more subtle effects were recently discovered in simple yield stress fluids that are made to flow in confined geometries or during transient regimes close to the yield stress. These effects are reviewed below, in Sect.~\ref{hottopics}.

\begin{figure}[t]
\centering
\includegraphics[width=\columnwidth]{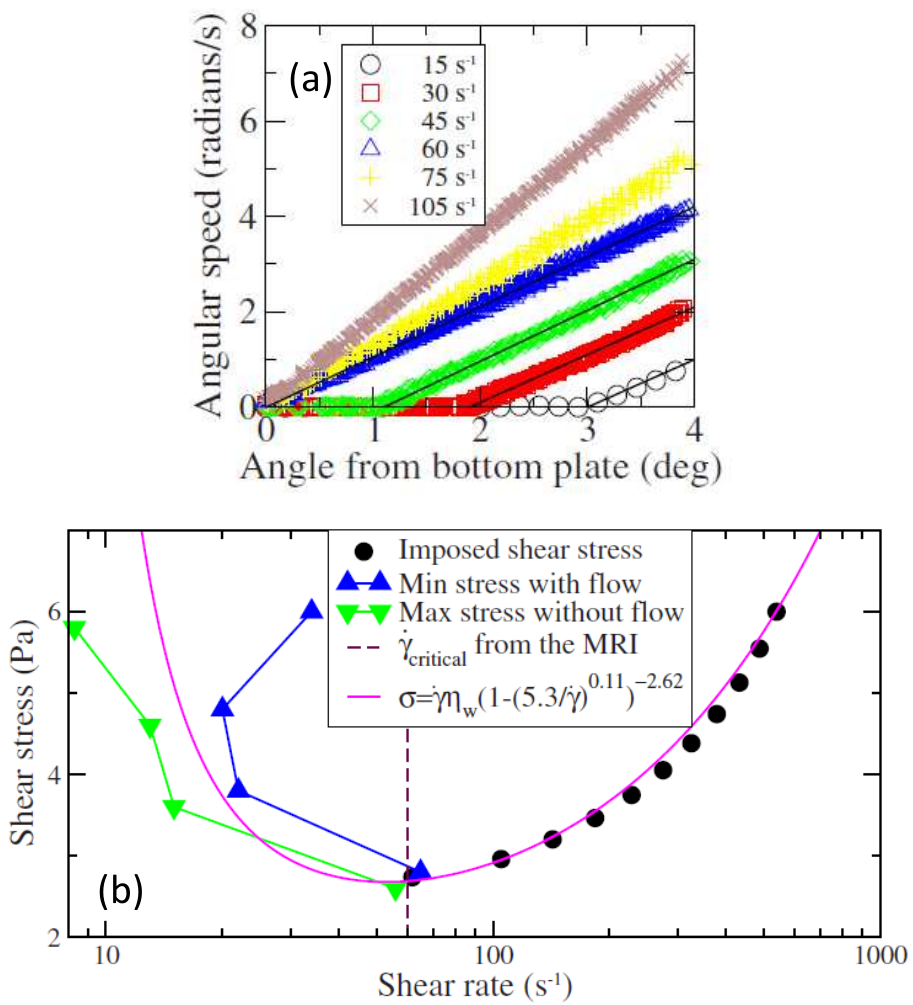}
\caption{Example of shear-banded flows. (a) Velocity profiles in a 4$^{\circ}$ cone-and-plate geometry of a colloidal silica suspension (Ludox TM-40) for shear rates ranging from 15 to 105~s$^{-1}$. (b) Steady-state flow curve determined by two different types of measurements. The branch at larger shear rates is obtained under constant external stress. The branches at lower shear rates are determined by estimating the minimum (resp. maximum) shear stress with (resp. without) flow. Extracted from \cite{Moller:2008}.}
\label{SBvelocityprofiles}
\end{figure}	

Contrary to simple yield stress fluids, thixotropic yield stress materials have been shown either to flow homogeneously (for $\dot \gamma > \dot \gamma_c$) or to display heterogeneous velocity profiles (for $\dot \gamma < \dot \gamma_c$). In the latter case, a solid-like region coexists with a liquid-like band sheared at $\dot \gamma = \dot \gamma_c$ \cite{Moller:2008} and the relative extent of both bands ensures that the average shear rate coincides with the macroscopic applied shear rate \cite{Coussot:2002a,Ovarlez:2009}.
As seen in Fig.~\ref{SBvelocityprofiles}(a) for the case of a colloidal gel, the amount of fluid-like material increases proportionally to the global applied shear rate for $0< \dot \gamma < \dot \gamma_c$. This points to an equivalent of a ``lever rule'' for solid--fluid coexistence and once again emphasizes the analogy between yielding in thixotropic materials and a first-order phase transition. It corresponds to {\it true} shear localization, i.e. to shear banding, in the sense that it is observed independently of any geometry-dependent stress heterogeneity. Table~\ref{tableSB} summarizes the distinction between shear banding (due to the existence of a critical shear rate) and shear localization (due to stress heterogeneity). It also recalls that in both cases, apparent slippage at the walls, which can be seen as an extreme kind of shear localization, may come into play as discussed in Sect.~\ref{wallslip}.

\begin{table}[h]
  \begin{tabular*}{8.5cm}{@{\extracolsep{\fill}}lclclcl}
    \hline
     Origin of shear localization & Simple YSF &  Thixotropic YSF \\
    \hline
    from critical shear rate & not possible & possible\\
    from stress heterogeneity & possible & possible\\
    wall slip & possible & possible \\
    \hline
  \end{tabular*}
  \caption{Different types of shear localization in simple
and thixotropic yield stress fluids (YSF).}
  \label{tableSB}
\end{table}

\subsection{Causes of steady-state shear banding}	
	

\subsubsection{Competition between aging and shear rejuvenation }

The existence of a critical shear rate $\dot \gamma_c$ in thixotropic yield stress materials has been explained in terms of an underlying decreasing branch of the flow curve at low shear rates \cite{Olmsted:2008,Divoux:2016a}, as first discussed for viscoelastic ``wormlike micelle'' surfactant solutions \cite{Spenley:1993}. In such a scenario, the constitutive relation of the material is actually a decreasing function for shear rates ranging from 0 to $\dot \gamma_c$. In this shear rate range, the flow is  mechanically unstable, which leads to some sort of phase separation into an arrested region that coexists with a flowing band sheared at $\dot \gamma_c$ \cite{Picard:2002}. This coexistence is expected to correspond to a flat portion of the steady-state flow curve, analog to the Maxwell plateau in first-order phase transitions, where the size of the flowing band should follow the ``lever rule'' mentioned in Sect.~\ref{local}. Transient measurements in the unstable shear rate range can be used to unveil the underlying decreasing flow curve [see Fig.~\ref{SBvelocityprofiles}(b)].

The unstable part of the flow curve is most often interpreted and modeled as the result of competition between spontaneous aging and shear-induced rejuvenation, although only indirect evidence for such a competition has been reported up to now. Aging processes arise from particle aggregation in systems with microscopic attractive interactions, e.g. clays and attractive colloidal gels, or from the thermally activated reorganization towards minimal energy in dense systems, such as dense emulsions or microgels \cite{Sollich:1997,Cloitre:2000,Viasnoff:2002,Coussot:2007}. Such {\it physical} aging may occur over a wide range of time scales and is different in nature from the {\it chemical} aging due to slow chemical reactions, such as the release of Na$^+$ ions in laponite clays, which cannot be reversed by shear \cite{Shahin:2010,Shahin:2012}.
While attractive interactions have been shown to be a sufficient ingredient to induce shear banding \cite{Becu:2006,Ragouilliaux:2007,Fall:2010b,Paredes:2011}, the minimal amount of attraction necessary to permanently form banded profiles is still an open issue.
A better understanding of the role of microscopic interactions should be gained from experiments where the attraction between microscopic constituents is continuously tuned, e.g. in a model system of colloids or deformable droplets \cite{Saunders:1999}.

The simplest phenomenological model based on the idea of a competition between aging and shear rejuvenation is the toy-model known in the literature as the ``$\lambda$-model'' \cite{Mujumdar:2002,Coussot:2002b}. The basic assumptions of this model are: (i) there exists a structural parameter, $\lambda$, that describes the local degree of interconnection of the microstructure, (ii) the viscosity $\eta$ increases with increasing $\lambda$, and (iii) for an aging system at low (or zero) shear rate $\lambda$ increases, while at sufficiently high shear rates the flow breaks down the structure so that $\lambda$ decreases to a low steady-state value. For $n>1$, this model is easily shown to predict flow curves with
a minimum at a critical shear rate $\dot{\gamma}_c$, therefore qualitatively reproducing the case of a thixotropic yield stress material showing a viscosity bifurcation and steady-state shear banding. More refined versions of the $\lambda$-model have been proposed in the literature, e.g. for fractal colloidal gels \cite{Moller:2008} and elasto-viscoplastic structured fluids \cite{SouzaMendes:2011,deSouzaMendes:2013}, leading to similar results.
The kinematic hardening model used by \cite{Dimitriou:2013}
incorporates a “back stress” that evolves dynamically and affects the mechanics in the neighborhood of yielding. This back stress can be viewed as a $\lambda$ parameter in simple shear flow and causes the location of the yield surface to adjust, depending on the deformation state.

\begin{figure}
\centering
\includegraphics[width=8.5cm]{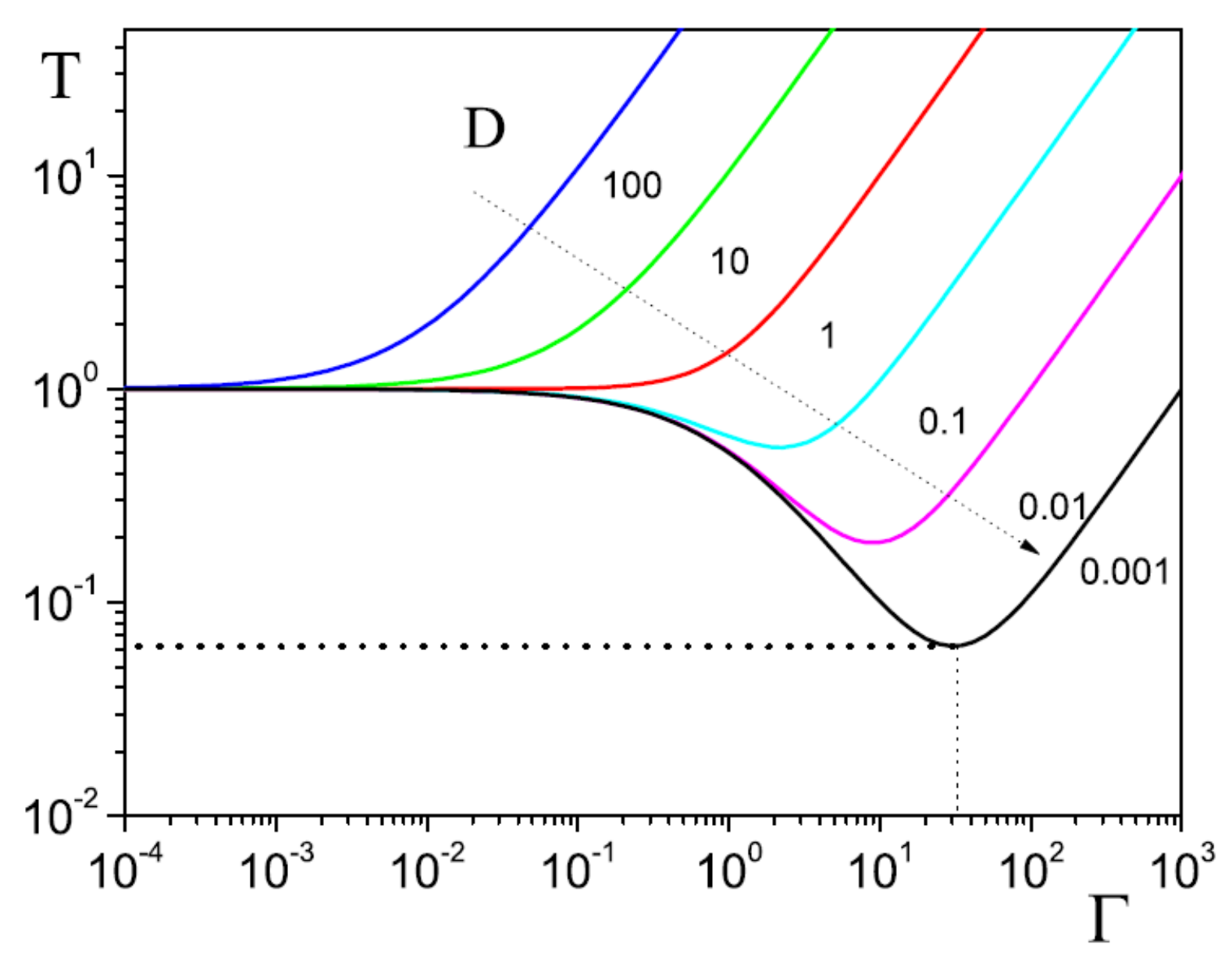}
\caption{Dimensionless flow curves (stress $T$ vs shear rate $\Gamma$) for different values of the ratio $D$ of the fluid relaxation time $\tau_{\rm rel}$ over the restructuration time $\tau_{\rm age}$, i.e., the time for a microscopic link to reform after being broken. Extracted from \cite{Coussot:2010}.}
\label{SBtoymodel}
\end{figure}

In order to achieve a more realistic picture of the effects of aging that accounts for the viscoelasticity of the material, a simplified mean-field argument has recently been proposed based on two time scales \cite{Coussot:2010}: a macroscopic relaxation time $\tau_{\rm rel}$, equivalent to the viscoelastic time which can easily be measured through step-strain or stress relaxation experiments, and a microscopic restructuring time $\tau_{\rm age}$ associated with the fluid spontaneous aging.
This model produces a simple expression for the flow curve:
\begin{equation}
\frac{\sigma}{G\gamma_c}=\tau_{\rm rel}\frac{\dot{\gamma}}{\gamma_c}+\frac{1}{1+\tau_{\rm age}\dot\gamma/\gamma_c}\,,
\end{equation}
where $G$ is the characteristic elastic modulus of elements that break above a critical strain $\gamma_c$.
Interestingly, the predicted flow curve has a minimum at a critical shear rate for $\tau_{\rm rel}<\tau_{\rm age}$, i.e. for a sufficiently long restructuration time while simple yield stress behavior sets in when restructuring becomes faster than viscoelastic relaxation, i.e. for $\tau_{\rm age}<\tau_{\rm rel}$ (see Fig.~\ref{SBtoymodel}). In this model, one can thus continuously go from a monotonic flow curve to a non-monotonic flow curve, i.e. from a simple to a thixotropic yield stress fluid, by increasing the duration of the restructuration time.

The idea of a competition between a restructuring time scale and shear flow has recently been implemented in the elasto-plastic coarse-grained modeling initiated by \cite{Picard:2004}. The influence of the time-scale competition between structural rearrangement and elastic recovery has been explored in full detail in \cite{Martens:2012}; see also \cite{Benzi:2016}. This study not only confirms the non-monotonic character of the global flow curves for a certain set of control parameters, but also explores the spatial consequences of the non-monotonicity in a realistic geometry. In particular, the emergence of a ``phase separation'' between flowing and non-flowing regions in the system, i.e. permanent shear bands, was clearly observed,
thus putting the ideas of \cite{Coussot:2010} on firmer grounds. Similar ideas involving a self-consistent dynamics following
structural reorganization have been explored in various modeling contexts, see for instance \cite{Jagla:2007,Fielding:2009,Cheddadi:2012,Maki:2012,Joshi:2015}.

Although these models all give a consistent picture of a time scale competition leading to non-monotonic flow curves in some well-chosen regimes, and thus may give rise to shear bands and viscosity bifurcation, very little progress has been made towards understanding at a more microscopic level both the physical origin of these time scales and how to control their evolution by tuning, for instance, the interaction between colloidal particles. A notable exception is recent
work exploring the athermal rheology of sticky particles in the vicinity of the
jamming transition \cite{Irani:2014}. Here, it was shown that
stickiness may promote a yield stress even below
jamming, which is however easily disrupted by a slow shear flow. At larger
shear rate, particles are pushed against each other, and therefore
repulsive forces should produce an increase of the shear stress.
For a narrow range of control parameters, this competition
produces a non-monotonic flow curve, very much
in the spirit of Fig.~\ref{SBtoymodel}.

\subsubsection{Static versus dynamic yielding}

\begin{figure}
\centering
\includegraphics[width=8.5cm]{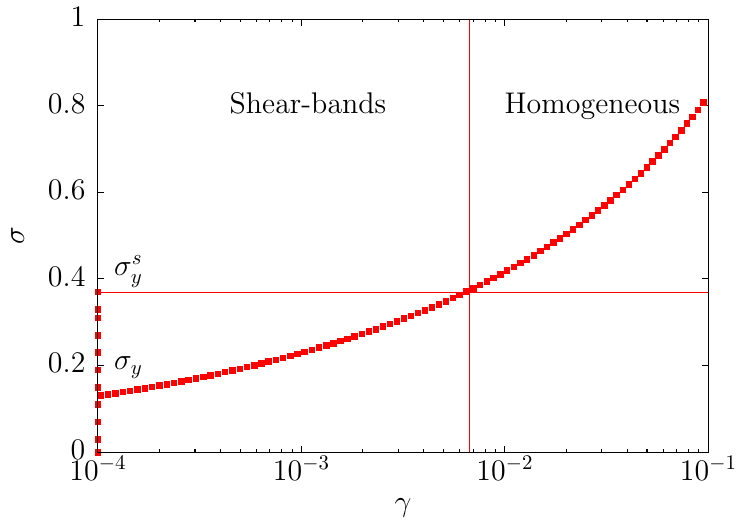}
\caption{Scenario for shear banding in yield stress materials.
A monotonic flow curve with a finite dynamic yield stress
$\sigma_y$ coexists with a static branch at $\dot{\gamma}=0$ and
$\sigma <\sigma_y^s$, where $\sigma_y^s$ is the static yield stress.
The shear rate is bi-valued for a range of shear stresses
$\sigma \in [\sigma_y, \sigma_y^s]$, which may lead to shear bands.
Adapted from \cite{Berthier:2002pp}.}
\label{fig:shearband}
\end{figure}

Whereas non-monotonic flow curves necessarily give rise to shear bands \cite{Olmsted:2008}, as observed in a variety
of complex fluids, a simpler scenario can also hold in the specific context of yield stress materials. Because the shear bands observed in a yield stress material delimit a flowing phase from an arrested phase (and not two different fluids as in more traditional complex fluids), shear banding can be explained by a simple picture where a monotonic global flow curve of the Herschel-Bulkley type with a finite {\it dynamic yield stress} $\sigma_y$ [as in Eq.~(\ref{eq:HB})] coexists with a static branch at $\dot{\gamma}=0$ existing for $\sigma < \sigma_y^s$, where $\sigma_y^s$ is a {\it static yield stress}. In that case, a
strict inequality $\sigma_y^s > \sigma_y$ directly ensures the existence of a finite stress regime, $\sigma  \in [\sigma_y, \sigma_y^s]$ where the shear rate is bi-valued, and can be either zero or finite, see Fig.~\ref{fig:shearband}.

This scenario was explored theoretically in \cite{Berthier:2002pp}, where the two phases were shown to correspond to two different families of dynamical solutions under the same external stress values in the context of a specific driven glassy model. These solutions respectively correspond to a fluid and an arrested phase. A similar explanation was shown to account for the presence of permanent (or at least very long-lived) shear bands in the computer simulation of a glass-forming
liquid in the glassy region below the glass transition temperature \cite{Varnik:2003}. There, again, a clear separation was observed between the dynamic extrapolation of the homogeneous flow curve and the direct determination of the static yield stress value. Detailed numerical studies have shown, however, that carefully measuring these two yield stress values is not an easy task \cite{Xu:2006,Peyneau:2008}.

The validity of this scenario was demonstrated in a numerical study of concentrated assemblies of soft particles where the degree of particle adhesion was tuned continuously \cite{Chaudhuri:2012b}, in analogy with the experimental investigations described above \cite{Becu:2006,Ragouilliaux:2006,Ovarlez:2008a,Fall:2010b}. In this numerical study, the emergence of flow inhomogeneity was again directly connected to a discontinuity of the flow curve at $\dot{\gamma}=0$, which was moreover observed to be strongly enhanced by adhesive forces, thus establishing a direct link between increasing the
adhesion and promoting shear-banding behavior \cite{Chaudhuri:2012b}.

Although the flow curves depicted in Figs.~\ref{SBtoymodel} and
\ref{fig:shearband} appear qualitatively distinct at first sight,
they may become
more similar in the case where the minimum of the flow curves
in Fig.~\ref{SBtoymodel} occurs at the lower end of the accessible
experimental range, in which case the remaining part of the flow curve
at small $\dot{\gamma}$ is ``compressed'' along the $\dot{\gamma}=0$ axis,
very much as in Fig.~\ref{fig:shearband}. In addition, in both cases
a shear-band may appear where a slow (or arrested) flowing phase
and a fast flowing phase coexist, and it may
be experimentally challenging to
distinguish between both scenarios unless non-banded, steady-state flows
can also be characterized at very low shear rates. The distinction could be
easier in computer simulations, where it is possible to impose
a global shear rate and follow either set of flow curves
shown in Figs.~\ref{SBtoymodel} and \ref{fig:shearband} down to arbitrary
low shear rates.

\subsubsection{Flow-concentration coupling}

In the case of suspensions of dense and rigid non-colloidal particles, shear banding may also result from volume fraction heterogeneities. As particles are denser than the surrounding fluid, there is a competition between sedimentation and shear-induced resuspension \cite{Ovarlez:2006,Fall:2009}. If shear-induced resuspension is not efficient enough, contacts between particles trigger the formation of a percolated network and of heterogeneous volume fraction profiles, leading to shear banding. Interestingly, such a flow-concentration coupling argument has also been invoked recently to account for shear banding in colloidal glasses \cite{Besseling:2010}. The underlying idea is that, despite a homogeneous stress field, minute local variations of the volume fraction $\phi$ may result in significant changes in the yield stress value, which for a homogeneous system strongly depends on the overall volume fraction $\phi$. At low applied shear rates in sufficiently dense packings, the flow may become unstable \cite{Schmitt:1995}: fluctuations trigger the jamming of a region of the material, which further turns into steady-state shear banding. This type of localization could therefore be interpreted as a precursor to shear-induced thickening \cite{Fall:2010}, although more experimental work is needed to draw an overall conclusion.

\subsection{Emerging topics: confinement and transient regimes}
\label{hottopics}

In the following we focus on two questions that have recently attracted growing interest, as examples of current challenges towards understanding the dynamics of yield stress fluids.

\subsubsection{Yield stress materials in confined geometries}
\label{hot_topic_non_local}

\begin{figure}[t]
\centering
\includegraphics[width=5.cm]{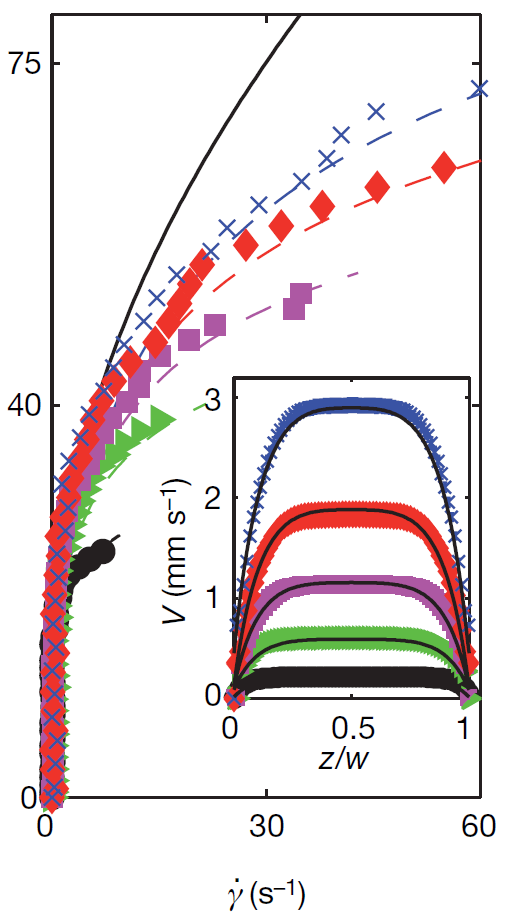}
\caption{Global and local flow curves (black solid line and symbols respectively) for a dense emulsion ($\phi=0.75$ and 20\% polydispersity). Global data are obtained in a wide-gap Taylor-Couette cell, while local flow curves are deduced from velocity profiles measured in a $w=250$~$\mu$m thick microchannel with rough surfaces, for various pressure drops ranging from 300 to 900~mbar (inset). No overlap of the local flow curves is observed. Dashed lines are predictions for the local flow curves at the given pressure drop, as obtained from the nonlocal rheological model [see Eq.~(\ref{nonlocaleq})] with a flow cooperativity length $\xi=22.3$~$\mu$m. Inset: solid lines are the velocity profiles predicted by the nonlocal rheological model. The y-axis for the main figure is the stress in Pa and the figure thus represents the flow curve; the important observation is that the flow curves for different driving pressures no longer overlap due to collective effects. Extracted from \cite{Goyon:2008}.}
\label{LocalGlobal}
\end{figure}

Flow properties of yield stress fluids have been discussed up to now in the context of ``large'' geometries, i.e. with gaps much larger than the granularity of the fluid microstructure, typically by at least two orders of magnitude. In this limit, the macroscopic behavior does not depend on the gap size. However, when the gap size becomes comparable to the mesoscopic scale characteristic of the fluid microstructure, i.e. in a {\it confined} geometry, rheological data have been reported to depend on the gap width \cite{Clasen:2004,Davies:2008,Yan:2010}. Accordingly, the local rheology in confined geometries no longer follows the Herschel-Bulkley model valid for large gaps [see Fig.~\ref{LocalGlobal}], as clearly demonstrated for emulsions \cite{Goyon:2008,Goyon:2010} and Carbopol microgels \cite{Geraud:2013} in small microchannels.

\paragraph{Cooperative effects in simple yield stress fluids.}

In such a confined geometry, the effect of shear-induced local rearrangement spans over a range larger than the single grain/drop/bubble scale  and can become comparable to the gap size, in which case finite-size effects influence the measured viscosity. The idea that the flow occurs through successive, plastic events over a certain ``cooperativity'' length $\xi$ has led to the development of so-called {\it nonlocal} models \cite{Bocquet:2009}. The simplest such model is a spatial version of the ``fluidity model'' \cite{Derec:2003} where the number of plastic events per unit time (or fluidity) taken as $f=\dot \gamma /\sigma$ is influenced both by the local contribution of the flow and by plastic events taking place at distances smaller than $\xi$. This leads to the following simple differential equation for $f$:
\begin{equation}
\xi^2 \frac{\partial^2 f}{\partial x^2} + (f_{\rm bulk}-f)=0\,,
\label{nonlocaleq}
\end{equation}
where $x$ is the direction of the stress gradient and $f_{\rm bulk}$ denotes the ``bulk'' fluidity value, i.e. the fluidity expected in a large-gap geometry in the absence of nonlocal effects. Here ``nonlocality'' stems from the double spatial derivative in $\partial^2 f/\partial x^2$ that involves the local fluidity over a typical size $\xi$. The solution to Eq.~(\ref{nonlocaleq}) successfully accounts for experimental flow profiles [see inset of Fig.~\ref{LocalGlobal}] and for dynamical arrest in confined geometries \cite{Chaudhuri:2012a}.

Recently the local fluidity has been related to the local shear rate fluctuations $\delta \dot \gamma(x)$ \cite{Jop:2012,Benzi:2014}. In particular, the
study by Jop {\it et al.} strongly suggests that the nonlocal rheology originates in the mechanical noise induced by the flow. Such a modification of the rheology due to confinement is not specific to yield stress materials since it also affects for instance the flow of surfactant wormlike micelles \cite{Masselon:2008}. In fact, the leftmost term in Eq.~(\ref{nonlocaleq}) was first introduced to model shear banding in these systems \cite{Dhont:1999,Yuan:1999}. More details on current issues raised by confinement of yield stress fluids are summarized in the recent review by \cite{Mansard:2012}. Here, we only emphasize that the question of whether cooperative effects may be at play during start-up flows should also be addressed. Indeed, in a partially fluidized material undergoing a transient regime, the fluid at rest is confined between the wall and the flowing band. This point, which raises the question of whether the --possibly slow-- dynamics of cooperative effects might be related to the diverging duration of transient regimes, is discussed in the next section.

\paragraph{Shear-induced structuration of attractive yield stress fluids.}

\begin{figure}[t]
\centering
\includegraphics[width=8.5cm]{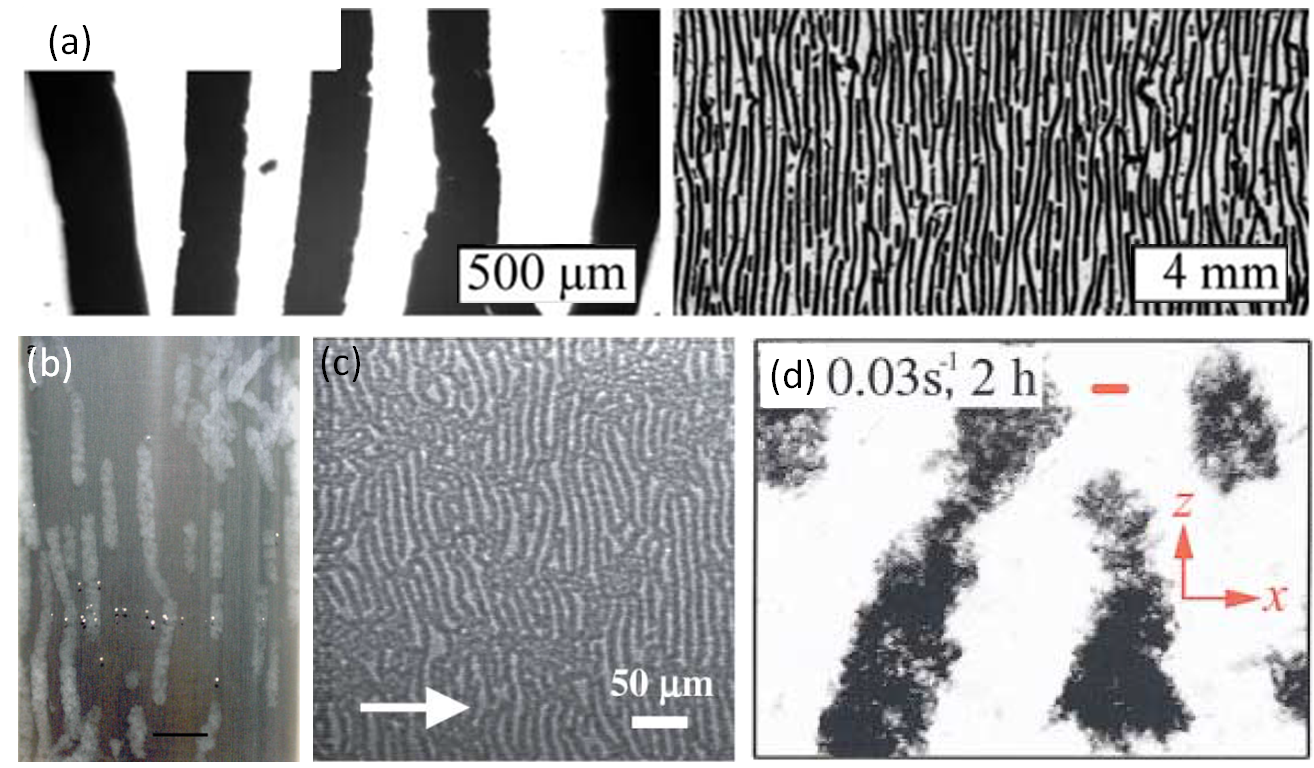}
\caption{Shear-induced patterns observed in various yield stress fluids under shear in confined geometries. (a) Carbon black gel under simple shear with a gap thickness of 173~$\mu$m as seen with optical microscopy with large and low magnifications (left and right respectively). Extracted from \cite{Grenard:2011}. (b) Suspension of micro fibrillated cellulose at 0.1~\% wt after shearing 10~min at 0.5~s$^{-1}$ in a Taylor Couette cell. Extracted from \cite{Karppinen:2012}. (c) Emulsions under simple shear for a gap thickness of 12~$\mu$m. The arrow indicates the direction of shear. Extracted from \cite{Montesi:2004}. (d) Optical micrograph of a quiescent semidilute non-Brownian colloidal nanotube suspension at 0.5~\% wt. The red scale bar is 10~$\mu$m and the gap thickness is 50~$\mu$m. Extracted from \cite{LinGibson:2004}.}
\label{Chapter3FigureStructure}
\end{figure}

Another striking effect of confinement on yield stress fluids is the spectacular shear-induced structuration observed in the case of attractive particle systems at moderate shear rates, typically $0.1<\dot\gamma<10$~s$^{-1}$. Examples include colloid polymer mixtures \cite{DeGroot:1994}, flocculated magnetic suspensions \cite{Navarrete:1996}, carbon nanotubes \cite{LinGibson:2004}, attractive emulsions \cite{Montesi:2004}, carbon black and alumina dispersions \cite{Osuji:2008,Negi:2009,Grenard:2011}, and microfibrillated cellulose \cite{Karppinen:2012} [see Fig.~\ref{Chapter3FigureStructure}]. In all these thixotropic yield stress fluids, the microstructure fully rearranges into a striped pattern of log-rolling flocs aligned along the vorticity direction, as demonstrated either indirectly through light scattering measurements \cite{DeGroot:1994} or through scanning electron microscopy \cite{Navarrete:1996} and optical microscopy \cite{LinGibson:2004,Montesi:2004,Osuji:2008,Negi:2009,Grenard:2011,Karppinen:2012}.

In some of the above attractive systems, shear-induced structuration has been linked to the emergence of negative normal stresses \cite{LinGibson:2004,Montesi:2004,Negi:2009} and it was proposed to interpret vorticity alignment as the consequence of an elastic instability of the elastic flocs, akin to a Weissenberg effect that would occur locally within individual flocs \cite{LinGibson:2004,Montesi:2004}. However, clear experimental evidence for such an interpretation and a detailed theory to prove the link between an elastic instability and shear-induced structuration are still lacking.

Moreover, shear-induced structuration only occurs within a certain range of shear rates. On the one hand, for very low shear, wall slip becomes predominant and generally prevents the system from being sheared in the bulk so that it remains in a homogeneous solid-like state. On the other hand, structuration does not occur above some critical shear rate, most probably due to the predominance of viscous forces and particle resuspension by shear. Coming up with a theory to provide a complete physical mechanism for the present shear-induced pattern formation and to predict both their characteristics and the shear rate limits where they appear is an important future challenge. Indeed, the striking effect of confinement not only affects the interpretation of rheological measurements but may also be of prime importance in applications involving confined flows of attractive particle systems. Finally, it is still unclear whether the structural instability that leads to pattern formation in attractive, thixotropic yield stress materials is related in any way to the mechanical noise which triggers flow cooperativity in simple yield stress fluids.

\subsubsection{Origin and scaling of the yielding time scales}
\label{hottopic_time scales}

So far, emphasis has been put on the steady state achieved by yield stress materials under a given shear rate or shear stress. However, it is quite obvious that such a steady state is not reached instantaneously and that transient regimes, e.g. from solid-like behavior at rest to liquid-like behavior above yielding, convey tremendous physical information on the yielding process.

In particular, it can be expected that upon approaching the yield stress the time needed to reach a flowing steady state can grow longer and longer, possibly pointing to a divergence of some characteristic time scale. If this time scale can be reliably estimated as a function of the various control parameters (applied stress or shear rate, packing fraction, temperature), then the question is whether physically-relevant information can be inferred on the yielding transition from such scalings. The aim of this section is to review recent work focusing on the time scales associated with yielding, open questions, and opportunities for the future.

\begin{figure}[t]
\centering
\includegraphics[width=8.5cm]{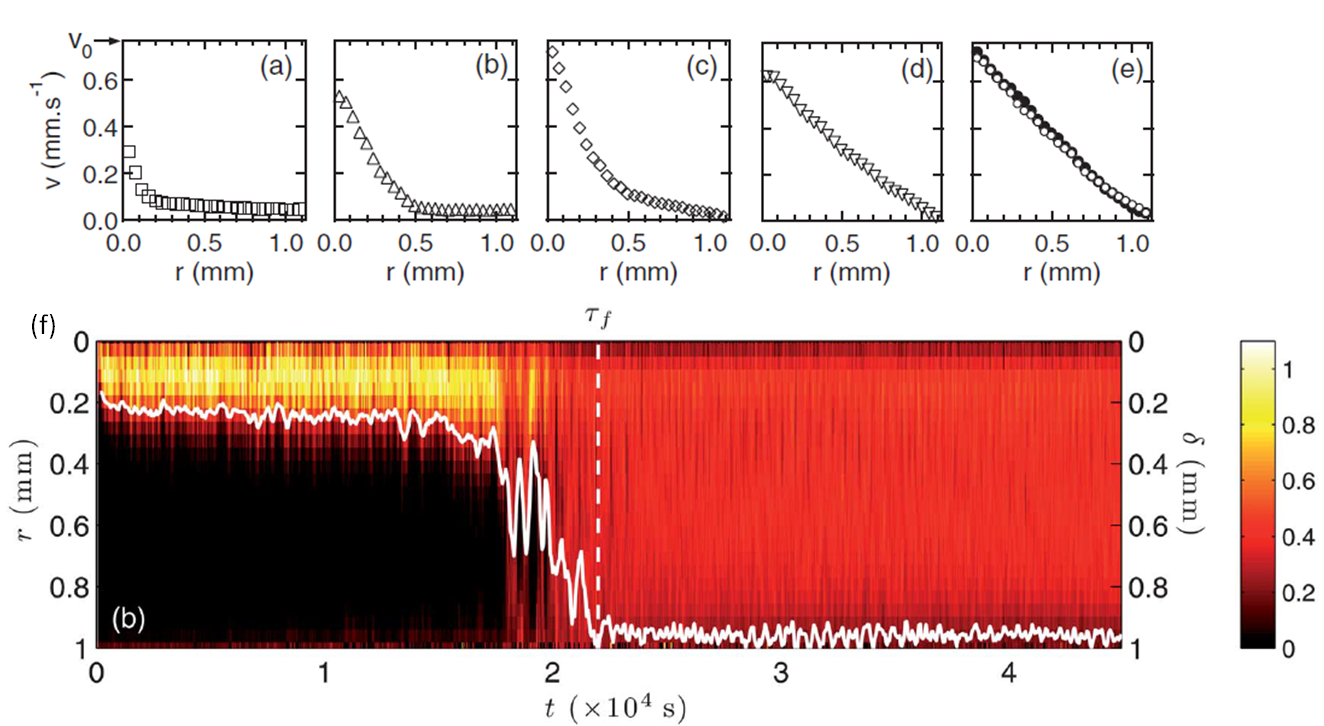}
\caption{Transient shear banding in a Carbopol microgel in the Taylor-Couette geometry. (a)--(e) Velocity profiles $v(r)$, where $r$ is the distance to the rotor, in a rough geometry at different times during the stress relaxation for an applied shear rate of 0.7~s$^{-1}$. Extracted from \cite{Divoux:2010}. (f) Spatiotemporal diagram of the local shear rate $\dot{\gamma}(r,t)$ in a smooth geometry for an applied shear rate of 0.5~s$^{-1}$. The white curve traces the position $\delta(t)$ of the interface between the fluidized band and the solid-like region. The vertical dashed line indicates the fluidization time $\tau_f$, i.e. the time at which the shear rate field becomes homogeneous. Extracted from \cite{Divoux:2012}.}
\label{Chapter3FigureTSB}
\end{figure}

\paragraph{Power-law scalings of the fluidization time and transient shear banding.}
It has been reported that transient regimes may become surprisingly long-lived in the vicinity of the yield stress. As mentioned above, it is not surprising that the dynamics becomes increasingly slow upon approaching the yield stress, which has been reported often \cite{Aral:1994,Gopalakrishnan:2007,Caton:2008,Rogers:2008}. More quantitative and local insights have been gained from recent velocimetry experiments during shear start-up and creep experiments of simple yield stress fluids, namely Carbopol microgels \cite{Divoux:2010,Divoux:2011,Divoux:2012} and, to a lesser extent, emulsions \cite{Becu:2005,Perge_thesis:2014}. These experiments have revealed that the expected homogeneous velocity profiles are reached after a transient regime that involves shear-banded velocity profiles [see Fig.~\ref{Chapter3FigureTSB}]. In Carbopol microgels, the fluidization time $\tau_f$, i.e. the duration of the transient shear banding regime, was shown to follow power-law scalings $\tau_f\sim A/\dot \gamma^\alpha$ and $\tau_f\sim B/(\sigma-\sigma_y)^\beta$ with $\alpha\simeq 2$--3 and $\beta\simeq 4$--6 under imposed shear rate and shear stress, respectively \cite{Divoux:2010,Divoux:2011b,Divoux:2012}. In all cases, the final homogeneous flow is consistent with the global steady-state rheology indicating simple yield stress behavior [see Fig.~\ref{Chapter3FigureTSB}(e)]. Interestingly, assuming that the fluidization times under imposed shear rate and shear stress are simply proportional, the above power-law scalings naturally lead to a constitutive equation $\sigma (\dot \gamma)$ that coincides with the Herschel-Bulkley equation [Eq.~(\ref{eq:HB})] in which the phenomenological exponent $n$ is given by $n=\alpha/\beta\simeq 1/2$ \cite{Divoux:2011b}. Therefore, the exponents governing the transient regimes possess a striking link with the exponent that characterizes the steady-state behavior. Such a link has been interpreted in terms of a critical-like phenomenon \cite{Divoux:2012,Chaudhuri:2013}.

It is important to clearly distinguish the transient shear-banding phenomenology from the time-dependent behavior of thixotropic materials. Here, rather than a competition between aging and shear rejuvenation, the transition from a solid-like to a liquid-like state seems to involve plastic events and damage accumulation in a way that resembles hard solids. Indeed, the flowing band can be observed to slowly ``erode'' the material at rest before the whole material experiences a rather sudden fluidization. This induction phase suggests that erosion by the fluidized band somehow fragilizes the bulk-arrested microgel, bringing it to a critical state before complete, sudden fluidization occurs. Such a critical state could be analogous to the one reached by a colloidal gel experiencing ``delayed sedimentation" right before its collapse  \cite{Buscall:2009,Teece:2011,Barlett:2012}. However, more experiments that provide access to the structure of the band at rest are needed to confirm such a picture. Moreover, a systematic comparison with recent molecular dynamic simulations of disordered systems could help to bridge the gap between yield stress fluids and amorphous solids \cite{Fusco:2014}.

From a theoretical point of view, a general criterion for the formation of transient shear bands has been proposed \cite{Moorcroft:2011,Moorcroft:2013}, providing a connection with either the stress overshoot under an imposed shear rate or the delayed yielding under creep in yield stress fluids or viscoelastic fluids. In this approach the power-law scaling for the fluidization time under creep is recovered, but with a smaller exponent $\beta\simeq 1$ \cite{Moorcroft:2013}. Another promising approach consists of a structural model of colloidal aggregates that incorporates viscoelasticity \cite{Illa:2013,Lehtinen:2013,Mohtaschemi:2014}. Such a phenomenological model recovers power-law scalings but only predicts trivial exponents $\alpha=\beta=1$ and thus fails to capture the link between both transients and the steady-state rheology observed in microgels.
Finally, theories at a more microscopic level, such as the STZ theory,
suggest that the transient shear banding and sudden fluidization is primarily a
result of microstructural disordering originating from structural
heterogeneities \cite{Hinkle:2016}. The latter results remain to be
confirmed and extended by supplemental measurements on purely repulsive
systems through simpler numerical approaches such as molecular dynamics
simulations.

\paragraph{Exponential scalings: activated processes and brittle-like failure.}
Whereas rather few papers have focused on transient fluidization under an applied shear rate, creep experiments have revealed long-lived transients in numerous yield stress fluids, including attractive gels such as carbon black gels \cite{Gibaud:2010,Grenard:2014}, coated silica particles \cite{Gopalakrishnan:2007,Sprakel:2011} and colloidal glasses \cite{Siebenburger:2012a}. The time at which the strain rate increases by several orders of magnitude defines a fluidization time $\tau_f$ which coincides with the establishment of homogeneous velocity profiles \cite{Gibaud:2010,Grenard:2014}. Interestingly, in attractive colloidal systems, $\tau_f$  generally decreases exponentially with the applied shear stress \cite{Gopalakrishnan:2007,Sprakel:2011,Gibaud:2010,Grenard:2014}. Such a scaling $\tau_f\sim\exp(-\sigma/\sigma_0)$ involves a characteristic stress $\sigma_0$, which has been interpreted and modeled in the framework of bond breaking through thermally activated processes \cite{Gopalakrishnan:2007,Lindstrom:2012}.

Alternative exponential scalings, such as the Griffith-like scalings $\tau_f\sim\exp(\sigma_0/\sigma)^p$ with $p=1$, 2 or 4 \cite{Griffith:1921,Lawn:1993,Pomeau:1992,Vanel:2009}, have been proposed in the context of yield stress fluids \cite{Caton:2008} and transient networks \cite{Tabuteau:2009,Mora:2011}. They hint to fracture-like dynamics, although it may be difficult to discriminate between various exponential --or even power-law-- scalings due to the limited range of experimentally accessible shear stresses \cite{Gibaud:2016}. This raises the question of the ``brittleness'' of yield stress fluids: while the physics of yielding in concentrated, jammed assemblies of soft particles such as microgels or emulsions appears to rely on (microscale) plasticity associated with (macroscale) shear banding, early studies based on direct visualization of the sample edges have shown some colloidal systems, such as Laponite suspensions \cite{Magnin:1990,Pignon:1996} and concentrated suspensions \cite{Aral:1994,Persello:1994}, to be prone to fracture-like behavior. Revisiting these pioneering works with modern temporally and spatially resolved techniques could classify such strain localization in terms of fracture, wall slip or shear banding and sort out the possible effects of the experimental geometry on the flow dynamics.

\paragraph{Dynamics induced by wall slip in transient and steady-state flows.}

In Sect.~\ref{wallslip} we have considered wall slip only through its effect on steady-state flow curves and velocity profiles. However, it has long been known, most prominently in the context of polymers, that wall slip often comes with instabilities and complex time-dependences \cite{Graham:1995,Denn:2001,Denn:2008}. In light of the above discussion on fluidization time scales, it also seems natural to ask whether slippage at the walls shows interesting variations, both during transient regimes and at steady state. Surprisingly only a handful of papers have dealt with the dynamics of wall slip in yield stress fluids. Various start-up experiments in smooth geometries, e.g. on emulsions \cite{Becu:2005}, Carbopol microgels \cite{Divoux:2010} and laponite clay suspensions \cite{Gibaud:2009}, have reported slip velocities that are strongly correlated to the fluidization dynamics and to the temporal evolution of the shear stress as well as stick-slip oscillations in the steady state  \cite{Pignon:1996,Ianni:2008,Divoux:2011b}.
Recent experiments of unsteady pipe flows reported a similar coupling between the solid-to-fluid transition and wall slip, including strongly fluctuating behaviors \cite{Poumaere:2014}.

Importantly, the work by \cite{Gibaud:2008,Gibaud:2009} on laponite suspensions illustrates that boundary conditions not only strongly affect the transient fluidization process but may also lead to totally different steady states---in this case, shear banded flows vs homogeneous flows. Recent numerical modeling suggests that the internal stress distribution prior to shear start-up affects the steady state \cite{Cheddadi:2012}. These results reveal the influence of both the boundary conditions and the initial conditions on the steady state reached after yielding, an issue that remains to be fully explored in experiments and models.

\subsection{Open questions}

We close this section by listing the open questions that represent the most pressing issues in current research into the dynamics of yield stress materials:
\begin{itemize}
\item How does nonlocality due to confinement set in during transient material response?
\item Is there any plasticity occurring during intial Andrade-like creep? If so, where is it localized?
\item What is the physical mechanism responsible for shear-induced structuration in confined attractive yield stress fluids?
\item What are the differences (if any) between the material microstructure in the transient shear band and in the rest of the sample?
\item What is the nature of the wall-fluid interactions that drive slip phenomena and how can they affect bulk flow?
\end{itemize}

\section{Summary and outlook}
\label{Conclusion}

We have reviewed recent progress in the understanding of yield stress fluids. Most of recent experimental advances are due to simultaneous measurements of flow structure and mechanical properties. Techniques that allow one to elucidate the flow structure such as magnetic resonance imaging, ultrasound or optical microscopy techniques have revealed a richness in the behavior of yield stress materials that was hitherto unsuspected, and have allowed for some novel physical insights.

On the theoretical side, much progress has been accomplished to account for the physical origin of solid behavior in amorphous materials across a broad range of interparticle interactions producing glassy, jammed and gel behaviors. Simultaneously, computer simulations have demonstrated their efficiency in producing convincing particle-based
models of yield stress materials, and allowed detailed investigations of the rheological behavior of these systems in various geometries. By construction, such simulations allow for a direct study of both the macroscopic rheological response and the microscopic dynamics. These studies have in turn allowed the development of a new family of coarse-grained elastoplastic models of yield stress materials, where exploration of larger-scale phenomena (such shear-banding and time-dependent flows) is better facilitated than through particle-resolved simulations.

For many decades it has been questioned whether the yield stress actually exists. It is now well-established that it does, in any case on experimentally relevant time scales. Different techniques of determining the yield stress produce similar values --- provided care is taken to account for wall slip, flow heterogeneity and time-dependences. Indeed, one of the novel insights is that not all yield stress materials behave ideally and that a distinction needs to made between two types of yield stress fluids: simple and thixotropic ones. Simple yield stress fluids show a flow behavior that is well reproduced by the Herschel-Bulkley law, with no significant time-dependence, while thixotropic ones show a pronounced time-dependence that arises from aging and shear rejuvenation phenomena. Adequate experimental protocols need to be employed that take into account the time evolution of these materials in order to get reproducible experimental estimates for the yield stress.

Tied in with the discussion of the yield stress is the shear localization that is generically observed in yield stress fluids. For simple yield stress fluids, shear banding is in general due to stress heterogeneity, and if not, it is only transient. For thixotropic materials, the situation is qualitatively different: due to the interplay between aging and shear rejuvenation, there exists a critical shear rate below which no stable homogeneous flow is possible. If a shear rate is then imposed macroscopically within the unstable regime, a shear band is formed in which the material flows at the critical shear rate, the rest of the material remaining motionless.

In addition, wall slip is commonly observed in yield stress materials, and needs to be accounted for.  In rheology, wall slip is usually defined as a viscosity that depends on the size of the gap of the measurement geometry, which distinguishes it from the two types of shear localization discussed above; the correction can be done by comparing measurements with different gap sizes, and extrapolating to an infinite gap. Besides complicating the interpretation of rheological measurements, wall slip also raises fascinating fundamental questions. Reaching a general understanding of the physics of slippage phenomena in yield stress materials appears as a challenging task for the future.

Recently, a different type of gap-dependent viscosity was uncovered for very small gaps, e.g. for microchannels with a size close to that of the microstructural elements. Here, gap-dependence was attributed to collective effects or ``spatial cooperativity'' when the range of shear-induced rearrangements span the whole system. There is still a lot of discussion on this topic, but it may challenge the simple view of yield stress materials sketched above.

One last burning issue concerns the full characterization and understanding of the transient flow behavior of both categories of yield stress fluids. The way such materials {\it start} to flow is indeed of great practical interest. In spite of notable recent progress this question of the yielding dynamics mostly remains to be explored both at microscopic and mesoscopic levels, and both experimentally and theoretically.

What is also crucial, especially for engineering purposes, is to have a predictive constitutive equation that allows for a general description of the flow (or not) of yield stress materials. For polymer systems a large number of such models have been derived from statistical mechanical models and are extensively used in practice.  
For yield stress fluids it is clear that three-dimensional invariant versions of the inelastic Bingham and Herschel-Bulkley models are inadequate, because they cannot reproduce the loss of foreaft flow symmetry in geometries with foreaft symmetry. Empirical models based on equations developed for polymeric liquids have shown promise in a few applications to complex flows, especially creeping flow past an isolated sphere, but, unlike the polymer counterparts, these models are not based on microstructural considerations and have not been tested against a full range of rheological measurements. Our limited understanding of plasticity is also a factor in incorporating pre-yield behavior into continuum models.
One promising research direction to solve this problem is to borrow statistical mechanical models from the glass-transition/soft-matter physics communities such as mode-coupling theory or the soft glassy rheology model. These would automatically also include ageing and shear rejuvenation as these are general features of the glass transition, and could thus in the end be the answer to the many remaining questions posed in this review.

\begin{acknowledgments}
We thank numerous colleagues for discussions on these matters
over the years. We especially thank
Catherine Barentin,
Jean-Louis Barrat,
Annie Colin,
Philippe Coussot,
Abdouaye Fall,
Marc-Antoine Fardin,
Thomas Gibaud,
Atsushi Ikeda,
Sara Jabbari,
Hamid Kellay,
Jorge Kurchan,
Anke Lindner,
Lisa Manning,
Jacques Meunier,
Jan Mewis,
Thijs Michels,
Peder Moller,
Guillaume Ovarlez,
Peter Sollich,
Hajime Tanaka,
for interactions on this topic.
The research leading to these results has
received funding from the European Research Council under
the European Union’s Seventh Framework Programme (Grant
No. FP7/2007-2013)/ERC Grant Agreement No 306845 and 258803.
\end{acknowledgments}

\bibliography{biblio_merged}
\bibliographystyle{apsrmp4-1}
\end{document}